\documentclass[preprint,12pt]{article}

\usepackage{hyperref} 
\usepackage{amssymb}
\usepackage{amsmath}
\usepackage[margin=1in]{geometry}
\usepackage[english]{babel}
\usepackage[utf8]{inputenc}
\usepackage[T1]{fontenc}
\usepackage{graphicx}
\usepackage{amsmath}
\usepackage{bm}

\usepackage{authblk}

\title{Flexel ecosystem: simulating mechanical systems from entities with arbitrarily complex mechanical responses}

\author{%
\small
Paul Ducarme$^{a,b}$, Bart Weber$^{b,c}$, Martin van Hecke$^{a,d}$, Johannes T.B. Overvelde$^{a,e}$\\
\footnotesize
$^a$AMOLF, Amsterdam, The Netherlands\\
$^b$Advanced Research Center for Nanolithography, Amsterdam, The Netherlands\\
$^c$Van der Waals–Zeeman Institute, Institute of Physics, Universiteit van  Amsterdam, Amsterdam, The Netherlands\\
$^d$Huygens-Kamerlingh Onnes Lab, Leiden Institute of Physics, Universiteit Leiden, The Netherlands\\
$^e$Institute for Complex Molecular Systems and Department of Mechanical Engineering, Technische Universiteit Eindhoven, Eindhoven, The Netherlands
}
\begin{document}

\maketitle

\begin{abstract}
 Nonlinearities and instabilities in mechanical structures have shown great promise for embedding advanced functionalities. However, simulating structures subject to nonlinearities can be challenging due to the complexity of their behavior, such as large shape changes, effect of pre-tension, negative stiffness and instabilities. While traditional finite element analysis is capable of simulating a specific nonlinear structure quantitatively, it can be costly and cumbersome to use due to the high number of degrees of freedom involved. We propose a framework to facilitate the exploration of highly nonlinear structures under quasistatic conditions. In our framework, models are simplified by introducing `flexels', elements capable of intrinsically representing the complex mechanical responses of compound structures. By extending the concept of nonlinear springs, flexels can be characterized by multi-valued response curves, and model various mechanical deformations, interactions and stimuli, e.g., stretching, bending, contact, pneumatic actuation, and cable-driven actuation. We demonstrate that the versatility of the formulation allows to model and simulate, with just a few elements, complex mechanical systems such as pre-stressed tensegrities, tape spring mechanisms, interaction of buckled beams and pneumatic soft gripper actuated using a metafluid.
With the implementation of the framework in an easy-to-use Python library, we believe that the flexel formulation will provide a useful modeling approach for understanding and designing nonlinear mechanical structures.
\end{abstract}

\section{Introduction}
These last years have garnered a significant interest in understanding how to leverage large deformation, nonlinearities and instabilities to design structures capable of complex yet functional mechanical responses \cite{reis_perspective_2015, bertoldi_flexible_2017}. For example, compliant mechanisms leverage flexibility to mimic conventional mechanism usually composed of many stiff components connected by joints \cite{howell_compliant_2012}. Soft robots utilize shape changes to be intrinsically more robust, adaptable and autonomous, often by harnessing mechanical instabilities to embed sequenced \cite{gorissen_hardware_2019}, asymmetric \cite{nagarkar_elastic-instabilityenabled_2021}, fast \cite{baumgartner_lesson_2020} or amplified \cite{tang_leveraging_2020} actuation. Recent progress in mechanical metamaterials have exploited large geometric changes and nonlinearities to achieve increasingly complex deformation pathways, enabling multistability \cite{liu_cellular_2023} or high energy dissipation \cite{yan_bio-inspired_2024}. While the advancements driven by the exploitation of nonlinearities increase the capabilities of these fields, they also introduce new challenges for simulation and design.\\

Whereas numerically studying structures in their small-deformation regime follows established approaches given their linear responses, studying a structure subject to large geometric changes is more challenging due to the nonlinearities involved. Nonlinear finite element analyses suffer from limitations that make them less applicable for conceptual design, and do not always contribute to the understanding of systems that undergo large deformation and instabilities. For example, due to the high number of degrees of freedom involved, they can have prohibitive computational costs,
fail upon facing mechanical instabilities or distorted meshes, and act as black boxes that are challenging to gain insight from.\\

To cope with these difficulties, reduced-order models that are characterized by fewer degrees of freedom offer a more practical alternative to finite element analyses. They are easier to define, solve, and interpret by promoting qualitative understanding over quantitative accuracy. They usually employ fewer elements which together are still able to capture the qualitative phenomenology. For example, the pseudo-rigid body model uses a combination of rigid links and torsional springs to represent flexible, slender parts \cite{howell_compliant_2012}. Trusses of bars can be used to model and even inverse design multistable compliant structures \cite{zhang_computational_2021}. Cosserat rods have shown great promise for modeling soft robotic \cite{caasenbrood_energy-shaping_2022} or soft living \cite{zhang_modeling_2019} systems.
Simple spring or beam models that can be expressed analytically have been valuable in gaining insight into a wide variety of geometrically nonlinear mechanical structures \cite{schioler_space_2007,rafsanjani_snapping_2015,dykstra_viscoelastic_2019, steinhardt_physical_2021, meng_bistability-based_2021, bekele_enhancing_2023, ten_wolde_single-input_2024}. 

Essentially, such existing approaches aim for a higher-level description of the mechanical system. Instead of describing the system as a mesh of many material elements, combining fewer abstract components allows to construct models that are cheaper to solve and easier to understand.
However, strongly nonlinear mechanical behaviors such as nonmonotonic or multi-valued force-deformation paths are still either modeled from the bottom-up, by combining multiple more simple components, or by using overly abstract models such as `hysterons', 
which require complex modeling to connect them to physical systems and moreover can lead to ill-defined systems \cite{van_hecke_profusion_2021, liu_controlled_2024, shohat_geometric_2025, shohat_aging_2025}. 

Here, we introduce an easy-to-use framework that defines components at a level of abstraction that allows them to single-handedly and intrinsically capture highly nonlinear static mechanical responses that are traditionally achieved using multiple components. Our formulation is based on energy potentials tuned to be stationary on prescribed, arbitrarily complex, possibly multi-valued generalized force-displacement curves. We show that this energy-based formulation can generate a broad ecosystem of mechanical components, with a wide range of geometries and intrinsic complexities. We also show that the framework allows for a relatively straightforward approach to build aggregate models where the various interactions between simple and complex elements can be simulated and explored. We conclude this work by demonstrating that our framework allows to model, simplify, and simulate a wide variety of mechanical systems studied previously  that are subject to, e.g., geometric nonlinearities, pre-stress, hysteresis, buckling, snapping, contact, cable-driven or pneumatic actuation.

\section{Flexels}
Before introducing the full ecosystem, let us stress that structures composed of linear springs can exhibit nonlinear force-displacement relations resulting from geometric nonlinearities. A classic example is the Von-Mises truss, which, despite being  composed of three linear springs, produces a non-monotonic (Fig.~1a) or even a multi-valued force-displacement curve (Fig.~1b), depending on the relative stiffness of the springs. The foundation of our approach is to represent such structures by single entities, which we call `flexels'. As shown in Fig. 1c,d, the intrinsic nonlinear behavior of a flexel can be tuned to capture, and be equivalent to, the geometric nonlinear behavior of compound structures, such as the Von-Mises trusses shown in Fig.~1a,b (Movie~S1).

More formally, a flexel is defined as a deformable element whose elastic energy directly depends on its geometric measure $\alpha$, a scalar quantity determined from the coordinates of the nodes composing the element, such as its length. The intrinsic nonlinear behavior of a flexel is defined by a generalized force-displacement curve, which relates the derivative of its elastic potential with respect to the geometric measure (that is, the generalized force $f$) to the change in geometric measure (that is, the generalized displacement $u=\Delta \alpha$). This relation encapsulates the nonlinear response and is used to construct an energy potential from which the ingredients needed for numerical simulations can be derived (SI section~1.1).

Flexels can be assembled to investigate how interactions between individual nonlinear elements give rise to more complex collective responses (SI section~1.2). For instance, coupling two flexels each characterized by a non-monotonic and multi-valued force-displacement curve in series reveals an equilibrium path with multiple turning points, indicating the presence of a snapping sequence upon loading and unloading (Fig.~1e, Movie~S1). By assembling these flexels in other configurations, the interplay between intrinsic and geometric nonlinearities can be explored. For example, the same pair of flexels, assembled now at an angle and loaded from their connection point, produces a different snapping sequence, due to the additional geometric nonlinearities (Fig.~1f, Movie~S1).
The more complex force-displacement responses exhibited by the assemblies shown in Fig.~1e,f can in turn be replicated by a single flexel (Fig.~1g,h). Our formulation enables simulation at a higher level, eliminating the need of simulating the individual components of larger structures that flexels are intended to mimic.
Still, it should be noted that while flexels provide a powerful approach to reduce the number of degrees of freedom while maintaining nonlinear behavior, the abstraction subsumes the deformation of the internal degrees of freedom into the flexel behavior, thereby hiding them from the surrounding and precluding direct coupling between internal nodes or non-actuated nodal loading direction with other flexels. For example, the deformation of the top node in Fig.~1b is absorbed within its flexel abstraction (Fig.~1d) and cannot be coupled to other flexels. Similarly, the deformation of the top node of the assembly shown in Fig.~1f along the horizontal direction cannot be coupled to other elements when using its flexel abstraction (Fig.~1h). Information about these internal deformations is however not lost; it could be retrieved by mapping the flexel state back to the assembly that it mimics.

\begin{figure}[ht]\centering
\includegraphics[width=.9\textwidth]{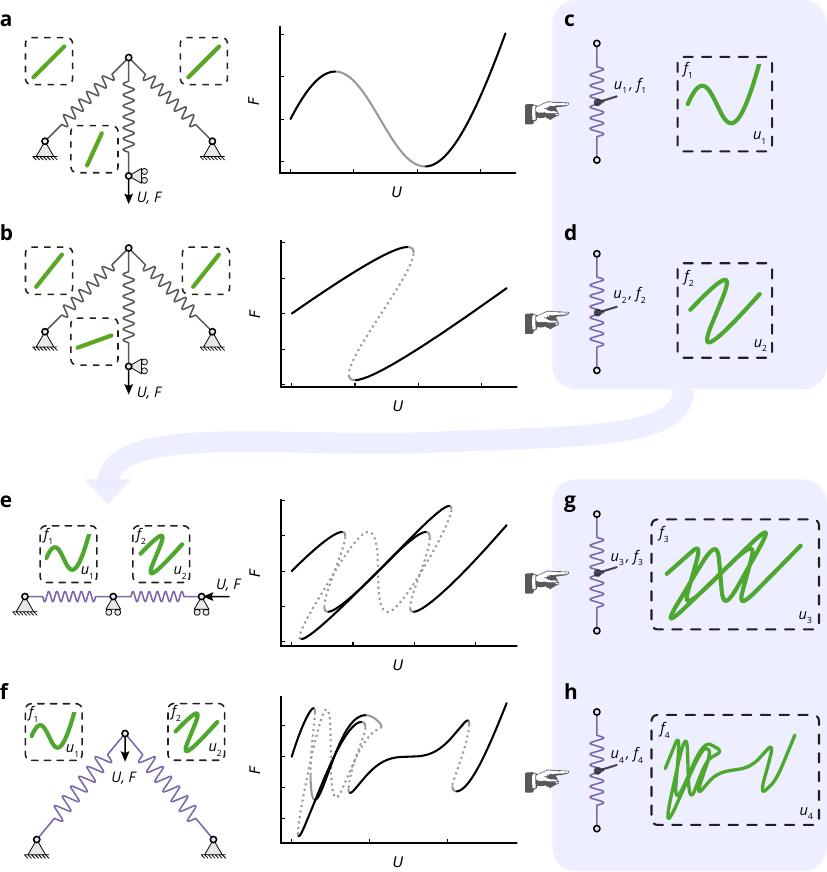}
\caption{Construction of flexels. (a-b) Von-Mises truss composed of a pair of inclined linear springs (non-dimensional stiffness of 0.6 for (a), 1.0 for (b)) driven from their connection point through a third, vertical linear spring (stiffness of 20 for (a), 0.33 for (b)). The force-displacement response of the system is either (a) non-monotonic or (b) multi-valued. (c-d) Flexel equivalents of the Von-Mises trusses shown in (a) and (b), whose mechanical behavior has been tuned to mimic their force-displacement response. (e-f) Assemblies composed of nonmonotonic and multi-valued flexels as shown in (c) and (d) loaded in series (e) or at an angle (f), exhibiting complex force-displacement curves. (g-h) Flexel equivalents of the systems shown in (e) and (f). Black (gray) lines refer to states stable (unstable) under force-controlled conditions. Solid (dashed) lines refer to states stable (unstable) under displacement-driven conditions. The full descriptions of the models are provided in SI section~7.1.}
\end{figure}
 
Simulating assemblies of flexels under quasi-static loading involves solving a system of parametrized nonlinear equations, which our toolkit achieves by implementing the arc-length method \cite{riks_incremental_1979} (SI section~2). This numerical continuation scheme retrieves an entire succession of deformed states at equilibrium even if those form a path with turning points, allowing for simulations of structures subject to snapping instabilities at either constant force or displacement. Note that if the equilibrium path splits into multiple branches at a pitchfork bifurcation, only one branch will be continued by the arclength scheme. Moreover, the arclength scheme is unable to retrieve equilibria disconnected from the initial path. Alternative methods could be implemented to cope with these two limitations \cite{farrell2016computationdisconnectedbifurcationdiagrams}.

\section{Flexel ecosystem}
The most basic flexel can be derived from the generalization of a nonlinear spring. For a nonlinear spring, the axial force is computed in two steps: the length is determined from the coordinates of its end nodes, then passed to an energy potential whose derivative yields the force. More general, a flexel extends this idea in two ways. First, the notion of length is broadened into a geometric measure (noted $\alpha$), meaning any scalar quantity computed from a list of nodes' coordinates (noted $z_i$), such as angle, area, total path length or the distance between a point and a line (Fig.~2a). Second, the class of energy potentials is widened by supporting tunable, potentially multi-valued generalized force-displacement curves (Fig.~2b). They can have an arbitrary number of turning points and intersections, allowing flexels to encode information about their loading history \cite{liu_controlled_2024}, or capture snapping or countersnapping phenomenona \cite{ducarme_exotic_2025}, for example. This is achieved by defining a custom energy potential $v(\alpha, t)$ that produces equilibria on the prescribed multi-valued path (SI section~4.2).
The only restrictions are that the curve must remain continuous and must not form loops in which the tangent vector crosses the \emph{vertical upward} direction, which prevents some structures to be abstracted into single flexels.
 
Thanks to the decoupling between geometry and intrinsic behavior, we can generate a whole flexel ecosystem by independently defining geometric measures on the one hand and generalized force-displacement curves on the other. By passing a geometric measure $\alpha$ (computed from the nodes' coordinates $z_i$) to an energy potential $v$, we pair a measure to a curve and produces a flexel that can be directly used in simulations to model complex mechanical entities (Fig.~2c, SI section~1, Movie~S2). Let us give a few examples. Pairing an angle to a multi-valued curve gives a flexel that models a flexure capable of snapping at constant angular displacement (Fig.~3a). Combining an area to a softening curve yields a flexel that mimics the behavior of a compressible fluid and can be used to model pneumatic actuation (Fig.~3b). The total length of a polygonal chain coupled to a stiffening bilinear curve gives birth to a flexel that models a rope that is yet to be taut and can be used in cable-driven systems (Fig.~3c). Pairing the point-line distance to a curve yielding a nonzero repulsion force only under a certain threshold gives a flexel that models contact (Fig.~3d).

\begin{figure}[ht]\centering
\includegraphics{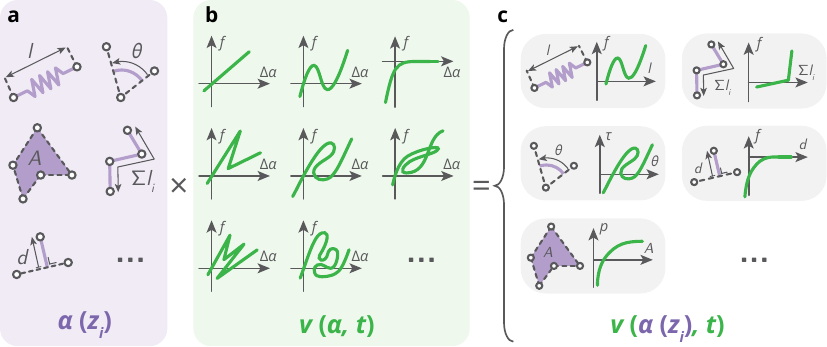}
\caption{Flexel ecosystem. (a) Various geometric measures $\alpha$ computed from a list of node coordinates $z_i$: length, angle, area, total length of a polygonal chain, distance point-line. (b) Various generalized force-displacement curves (linear, nonlinear, multi-valued) defining intrinsic mechanical behaviors via energy potentials $v(\alpha, t)$. (c) Examples of pairs of geometric measure and behavior forming the flexel ecosystem. The energy of a flexel is given by $v(\alpha(\bm{z}), t)$. More details on the definition of the geometric measures $\alpha$ and energy potentials $v$ are provided in SI section~3 and 4.}
\end{figure}

\begin{figure}[ht]\centering
\includegraphics{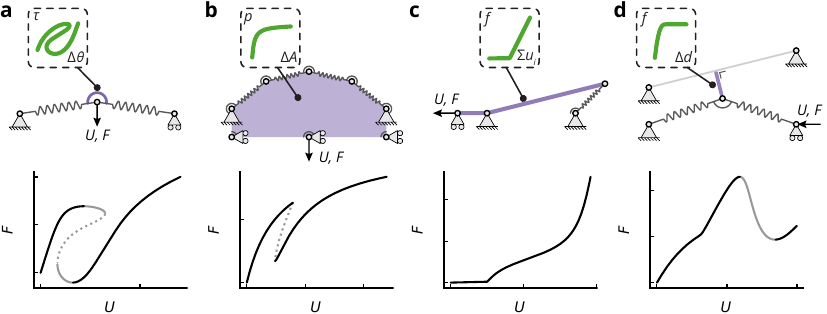}
\caption{Examples of assemblies of flexels (top) that collectively produce force-displacement responses $F(U)$ (bottom). Black (gray) lines refer to states stable (unstable) under force-controlled conditions. Solid (dashed) lines refer to states stable (unstable) under displacement-driven conditions. Gray flexels are characterized by a linear generalized force-displacement curve.
(a) An angular flexel with a multi-valued torque-angular displacement curve modeling a snapping flexure. (b) An area flexel with a softening pressure-areal displacement curve modeling pneumatic actuation. (c) A path flexel with a bilinear stiffening force-displacement curve, modeling cable-driven actuation. (d) A distance flexel with a force-displacement curve yielding a high nonzero force only for small distance values, modeling contact between a point and a line. The full descriptions of the models are provided in SI section~7.2.}
\end{figure}

\section{Workflow}

In this section we illustrate how our toolkit can be integrated to the workflow to design a structure with complex deformation pathways (Fig.~4, Movie~S3). To illustrate this, we consider two physical structures fabricated in silicone rubber (Smooth-On, Smooth-Sil 945) \cite{ducarme_exotic_2025}, shown in Fig.~4a. We set the goal of predicting the experimental tensile response of the system formed by coupling them in series.

First, a tensile test is carried out on each structure during which the force is measured while increasing and decreasing the extension. This reveals nonmonotonic force-displacement curves (Fig.~4b). Second, for each experimental test, a Bezier curve of degree~4 is fitted to the experimental data by adjusting the positions of control points (Fig.~4c-left). Fitting such a curve consists of finding the positions of control points, which form a control polygon that shapes the curve. Third, each Bezier curve is represented by a string of text listing the coordinates of the control points, which can then be used to define the generalized force-displacement curve of a longitudinal flexel, modeling each structure as a nonlinear spring (Fig.~4c-right).
Fourth, we model the assembly of the two structures by connecting the two flexels in series (Fig.~4d-right).

In practical terms, this model is written in an input file that describes the node positions, the boundary conditions, the flexels, and the loading steps (Fig.~4d-left). Note that before loading node~2, the structure is preloaded by applying a load on node~1 to account for the weight of the assembly. More details on how to interpret or compose such input file are provided in SI section~6.

Finally, we find that the simulated force-displacement curve is complex and characterized by many turning points (Fig.~4e).
We performed the physical experiment by connecting both structures in series and conducting a vertical tensile test, to validate the approach. Good agreement has been obtained both quantitatively and qualitatively (Fig.~4e). If multiple physical building blocks are available and characterized, this approach allows to build a catalog of achievable nonlinear behaviors that can be combined to quickly simulate assemblies. By scanning different nonlinear behavior combinations, the approach can help identify which blocks should be assembled within a certain network to get a specific desired response \cite{ducarme_exotic_2025}.

\begin{figure}[ht]\centering
\includegraphics{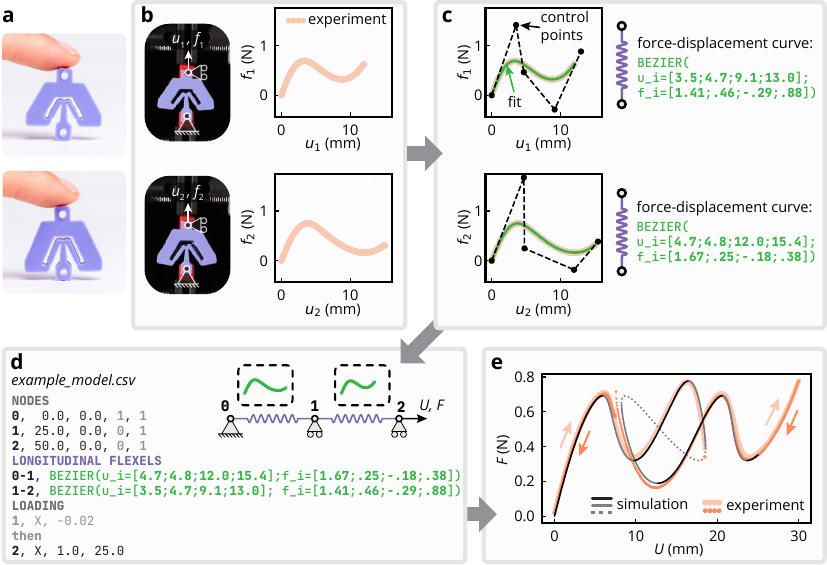}
\caption{Workflow to simulate an experimental assembly of nonlinear building blocks \cite{ducarme_exotic_2025}.
(a) Two building blocks fabricated in silicone rubber. (b) Experimental force-displacement curves obtained by performing a tensile test on the building blocks. (c) Left: Bezier curves (solid green lines) fitting the experimental force-displacement curves (orange lines). The control polygons and the control points are depicted by the black dashed lines and the black dots. Right: Equivalent flexels with tensile behaviors defined by Bezier curves, specified by a string of text listing the coordinates of the control points (green text). (d) Flexel model describing the assembly of the serially-coupled building blocks (Right) and the input text file representing the model (Left). (e) Force-displacement curve of the serially-coupled assembly. Experimental data obtained during the loading (unloading) phase is depicted by light (bright) orange dots. Simulated data is depicted by black and gray curves, where
black (gray) refers to states stable (unstable) under force-controlled conditions, and solid (dashed) curves refer to states stable (unstable) under displacement-driven conditions. Force and displacement are measured with respect to the preloaded configuration. The full descriptions of the models are provided in SI section~7.3.}
\end{figure}

\section{Use cases}
Using this workflow, we next demonstrate the versatility of the ecosystem to model and handle complex use cases (Fig.~5, Movie~S4). As a first example, we show that stress-free or prestressed tensegrity trusses \cite{obara_truth_2019} can be effectively modeled by connecting longitudinal flexels with compatible or incompatible rest lengths (SI section~7.4.1). We observe that the stress-free truss behaves as a mechanism in the small deformation regime, i.e., with zero initial stiffness, while the prestressed truss is stabilized by acquiring a nonzero initial stiffness (Fig.~5a). This confirms that this truss is a pure tensegrity structure \cite{obara_truth_2019}. Beyond small deformations, our approach is able to capture the nonlinearities originating from the rotation of the longitudinal flexels, allowing to study the behavior of such structures under larger loads. 

In the second example, we show that a tape-spring system \cite{he_grasping_2025} can be modeled using a multi-valued angular flexel and a stiff path flexel (SI section~7.4.2). By loading the system in two steps, buckling and kink formation followed by zero-stiffness deformation can be simulated (Fig.~5b), reproducing the soft deformation mode used for gripping \cite{he_grasping_2025}. The current implementation only allows to load along nodal coordinates, but the framework can naturally be extended to directly actuate the rest geometric measure of a flexel, allowing for different loading modes, e.g., angular loading, to model the various actuation modes of the mechanism \cite{he_grasping_2025}. 

Third, we show that a distance flexel can be used to model the contact between interacting buckled beams (SI section~7.4.3), reproducing the effect of the separating distance on the global buckling direction of pairs of ``bumping beams'' \cite{kwakernaak_collective_2024} (Fig.~5c). Unlike previous work studying similar systems using dynamic simulations \cite{kwakernaak_collective_2024, guerra_selfordering_2023}, our static approach allows to retrieve stable and unstable  branches,
giving insight into the various equilibrium configurations. We note that as the number of beams increases, our approach becomes costly as many stable and unstable configurations will have to be computed. 

Fourth, we show that a pneumatic gripper actuated using a metafluid \cite{djellouli_shell_2024} (medium composed of collapsable capsules surrounded by a fluid) can be modeled and simulated using a single multi-valued area flexel for the metafluid, linear angular and longitudinal flexels for the gripper, and distance flexel for the gripping contact (SI section~7.4.4). The simulation reveals that the metafluid modulates the gripping force, allowing for delicate grasping as previously demonstrated experimentally in \cite{djellouli_shell_2024}. Instead of modeling the metafluid by adding degrees of freedom for each capsules, the curve itself is directly assigned to an area flexel, drastically reducing the number of degrees of freedom. By contrast to the modeling approach shown before \cite{djellouli_shell_2024}, adding extra capsules does not cost any additional degrees of freedom. The price to pay for that improvement is that the curve should be determined in advance, via a traditional modeling approach or via an experiment \cite{djellouli_shell_2024}. But once it is known, it can be used to model systems where that fluid is used at a much cheaper cost, while still capturing the complexity.

Together, these examples show the versatility of our approach. Importantly, each example only required a few lines of code to be simulated (SI section~7).

\begin{figure}[ht]\centering
\includegraphics{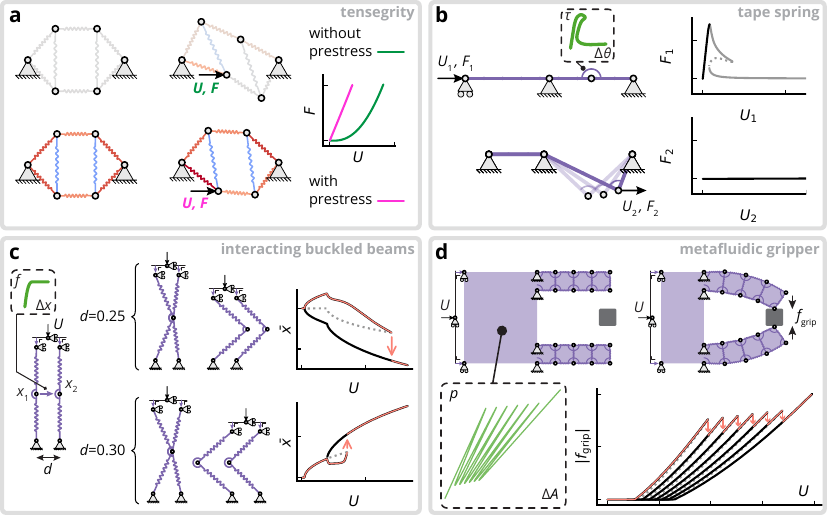}
\caption{Examples of use cases. (a) Left: Tensegrity structures without prestress (top) and with prestress (bottom) subject to an external loading. Flexels in compression, at rest and in tension are colored in blue, gray and red respectively. Right: Force-displacement curves of the tensegrity structures with and without prestress. (b) Left: Model of a tape-spring gripper \cite{he_grasping_2025} loaded in two steps. The first one applies compression to the tape spring, eventually triggering buckling and creating a kink (top). The second deforms the buckled tape spring by driving the kink (bottom). Right: Force-displacement curves corresponding to the first (top) and second (bottom) loadsteps. (c) Left: model of a pair of buckling beams separated by a distance $d$, loaded in compression and interacting through the contact of their middle node \cite{kwakernaak_counting_2023, guerra_selfordering_2023, kwakernaak_collective_2024}. Center: deformation sequence of the beams when $d=0.25$ (top) and $d=0.30$ (bottom). Right: Deformation paths when $d=0.25$ (top) and $d=0.30$ (bottom). $\bar{x}:=(x_1+x_2)/2$. (d) Top: Model of a gripper actuated using a metafluid \cite{djellouli_shell_2024} to grasp an object. Bottom: Intrinsic behavior of the area flexel modeling the metafluid (left) and deformation path of the system (right), shown as the gripping force $f_\text{grip}$ as a function of the applied displacement $U$. The full descriptions of the models are provided in SI section~7.4.}
\end{figure}

\section{Conclusion}
We introduced the concept of flexels, entities able to single-handedly capture the strongly nonlinear responses of compound systems. This facilitates the simulation of various complex mechanical systems, with a focus on nonlinearities and instabilities. To promote accessibility and adoption, the formulation is implemented in an open-source and user-friendly Python library \cite{springable}.

Key to developing this framework was the generalization of the concept of nonlinear spring along two independent axes, geometry and intrinsic mechanical behavior, which naturally generates a broad ecosystem of easy-to-combine elements. This ecosystem can be extended by defining different geometric measures or families of curves, which opens route to define custom flexels that are suitable to tackle specific mechanical problems, while still being compatible with the previously defined ones. Because our formulation is energy-based, it enables the use of well-established algorithms as well as the traditional stability analysis tools. Furthermore, even though we focused on static simulations in this work, the modularity of the implementation allows for the extension of the formulation in dynamic contexts \cite{jin_guided_2020,byun_integrated_2024}, where the module to compute the nonlinear elastic forces can be re-used.

We close by listing future challenges and perspectives. Even though each flexel can capture complexity, its response is only governed by a single geometric measure, which can be an oversimplification of reality. Being able to define a flexel with force-displacement curves along more than one dimension could help build more abstract and reduced models. This can in principle be done in the current formulation by defining energy potentials with additional dimensions. For example, flexels whose energy depends on two deformation measures could help construct 2d metamaterial models where each unit cell is a flexel \cite{bertoldi_negative_2010}. Flexels with a force-displacement curve that can be tuned via an external parameter could help model the interplay between active materials and geometric nonlinearities \cite{pal_programmable_2023}.
We believe that embracing the flexel formulation will simplify simulations, help gain insight into a large variety of nonlinear problems and strengthen the use of nonlinearities and instabilities as design paradigm for compliant mechanism, soft robots, mechanical metamaterials and nonlinear structures.

\section*{Data availability}
The `flexel' ecosystem is available on PyPI: \url{https://pypi.org/project/springable/1.0.0/} (see SI section~5 for installation instructions), source code: \url{https://github.com/ducarme/springable/releases/tag/v1.0.0}.

\section*{Acknowledgment}
We thank Michel Habets, Micha Steur, and Pavel Antonov for the useful discussions. This work is part of the Dutch Research Council (NWO) and was conducted at the AMOLF institute and the Advanced Research Center for Nanolithography. The Advanced Research Center for Nanolithography is a public–private partnership between the University of Amsterdam, Vrije Universiteit Amsterdam, Rijksuniversiteit Groningen, The Netherlands Organization for Scientifc Research (NWO), and the semiconductor-equipment manufacturer Advanced Semiconductor Materials Lithography (ASML).


\begin{thebibliography}{10}

\bibitem{reis_perspective_2015}
Pedro~M. Reis.
\newblock A {Perspective} on the {Revival} of {Structural} ({In}){Stability} {With} {Novel} {Opportunities} for {Function}: {From} {Buckliphobia} to {Buckliphilia}.
\newblock {\em Journal of Applied Mechanics}, 82(111001), September 2015.

\bibitem{bertoldi_flexible_2017}
Katia Bertoldi, Vincenzo Vitelli, Johan Christensen, and Martin van Hecke.
\newblock Flexible mechanical metamaterials.
\newblock {\em Nature Reviews Materials}, 2(11):17066, October 2017.
\newblock Publisher: Nature Publishing Group.

\bibitem{howell_compliant_2012}
Larry~L. Howell.
\newblock Compliant {Mechanisms}.
\newblock In {\em Encyclopedia of {Nanotechnology}}, pages 457--463. Springer, Dordrecht, 2012.

\bibitem{gorissen_hardware_2019}
Benjamin Gorissen, Edoardo Milana, Arne Baeyens, Eva Broeders, Jeroen Christiaens, Klaas Collin, Dominiek Reynaerts, and Michael De~Volder.
\newblock Hardware {Sequencing} of {Inflatable} {Nonlinear} {Actuators} for {Autonomous} {Soft} {Robots}.
\newblock {\em Advanced Materials}, 31(3):1804598, 2019.
\newblock \_eprint: https://onlinelibrary.wiley.com/doi/pdf/10.1002/adma.201804598.

\bibitem{nagarkar_elastic-instabilityenabled_2021}
Amit Nagarkar, Won-Kyu Lee, Daniel~J. Preston, Markus~P. Nemitz, Nan-Nan Deng, George~M. Whitesides, and L.~Mahadevan.
\newblock Elastic-instability–enabled locomotion.
\newblock {\em Proceedings of the National Academy of Sciences}, 118(8):e2013801118, February 2021.
\newblock Publisher: Proceedings of the National Academy of Sciences.

\bibitem{baumgartner_lesson_2020}
Richard Baumgartner, Alexander Kogler, Josef~M. Stadlbauer, Choon~Chiang Foo, Rainer Kaltseis, Melanie Baumgartner, Guoyong Mao, Christoph Keplinger, Soo Jin~Adrian Koh, Nikita Arnold, Zhigang Suo, Martin Kaltenbrunner, and Siegfried Bauer.
\newblock A {Lesson} from {Plants}: {High}-{Speed} {Soft} {Robotic} {Actuators}.
\newblock {\em Advanced Science}, 7(5):1903391, 2020.
\newblock \_eprint: https://advanced.onlinelibrary.wiley.com/doi/pdf/10.1002/advs.201903391.

\bibitem{tang_leveraging_2020}
Yichao Tang, Yinding Chi, Jiefeng Sun, Tzu-Hao Huang, Omid~H. Maghsoudi, Andrew Spence, Jianguo Zhao, Hao Su, and Jie Yin.
\newblock Leveraging elastic instabilities for amplified performance: {Spine}-inspired high-speed and high-force soft robots.
\newblock {\em Science Advances}, 6(19):eaaz6912, May 2020.
\newblock Publisher: American Association for the Advancement of Science.

\bibitem{liu_cellular_2023}
Zuolin Liu, Hongbin Fang, Jian Xu, and Kon-Well Wang.
\newblock Cellular {Automata} {Inspired} {Multistable} {Origami} {Metamaterials} for {Mechanical} {Learning}.
\newblock {\em Advanced Science}, 10(34):2305146, 2023.
\newblock \_eprint: https://advanced.onlinelibrary.wiley.com/doi/pdf/10.1002/advs.202305146.

\bibitem{yan_bio-inspired_2024}
Sen Yan, Wenlong Liu, Xiaojun Tan, Zhiqiang Meng, Weijia Luo, Hang Jin, Yongzheng Wen, Jingbo Sun, Lingling Wu, and Ji~Zhou.
\newblock Bio-inspired mechanical metamaterial with ultrahigh load-bearing capacity for energy dissipation.
\newblock {\em Materials Today}, 77:11--18, August 2024.

\bibitem{zhang_computational_2021}
Ran Zhang, Thomas Auzinger, and Bernd Bickel.
\newblock Computational {Design} of {Planar} {Multistable} {Compliant} {Structures}.
\newblock {\em ACM Trans. Graph.}, 40(5):186:1--186:16, October 2021.

\bibitem{caasenbrood_energy-shaping_2022}
Brandon Caasenbrood, Alexander Pogromsky, and Henk Nijmeijer.
\newblock Energy-{Shaping} {Controllers} for {Soft} {Robot} {Manipulators} {Through} {Port}-{Hamiltonian} {Cosserat} {Models}.
\newblock {\em SN Computer Science}, 3(6):494, September 2022.

\bibitem{zhang_modeling_2019}
Xiaotian Zhang, Fan~Kiat Chan, Tejaswin Parthasarathy, and Mattia Gazzola.
\newblock Modeling and simulation of complex dynamic musculoskeletal architectures.
\newblock {\em Nature Communications}, 10(1):4825, October 2019.
\newblock Publisher: Nature Publishing Group.

\bibitem{schioler_space_2007}
T.~Schioler and S.~Pellegrino.
\newblock Space {Frames} with {Multiple} {Stable} {Configurations}.
\newblock {\em AIAA Journal}, 45(7), 2007.

\bibitem{rafsanjani_snapping_2015}
Ahmad Rafsanjani, Abdolhamid Akbarzadeh, and Damiano Pasini.
\newblock Snapping {Mechanical} {Metamaterials} under {Tension}.
\newblock {\em Advanced Materials}, 27(39):5931--5935, 2015.
\newblock \_eprint: https://onlinelibrary.wiley.com/doi/pdf/10.1002/adma.201502809.

\bibitem{dykstra_viscoelastic_2019}
David Dykstra, Joris Busink, Bernard Ennis, and Corentin Coulais.
\newblock Viscoelastic {Snapping} {Metamaterials}.
\newblock {\em Journal of Applied Mechanics}, 86:1, June 2019.

\bibitem{steinhardt_physical_2021}
Emma Steinhardt, Nak-seung~P. Hyun, Je-sung Koh, Gregory Freeburn, Michelle~H. Rosen, Fatma~Zeynep Temel, S.~N. Patek, and Robert~J. Wood.
\newblock A physical model of mantis shrimp for exploring the dynamics of ultrafast systems.
\newblock {\em Proceedings of the National Academy of Sciences}, 118(33):e2026833118, August 2021.
\newblock Publisher: Proceedings of the National Academy of Sciences.

\bibitem{meng_bistability-based_2021}
Zhiqiang Meng, Weitong Chen, Tie Mei, Yuchen Lai, Yixiao Li, and C.~Q. Chen.
\newblock Bistability-based foldable origami mechanical logic gates.
\newblock {\em Extreme Mechanics Letters}, 43:101180, February 2021.

\bibitem{bekele_enhancing_2023}
Adam Bekele, M.~Ahmer Wadee, and Andrew T.~M. Phillips.
\newblock Enhancing energy absorption through sequential instabilities in mechanical metamaterials.
\newblock {\em Royal Society Open Science}, 10(8):230762, August 2023.
\newblock Publisher: Royal Society.

\bibitem{ten_wolde_single-input_2024}
Malte~A. ten Wolde and Davood Farhadi.
\newblock A single-input state-switching building block harnessing internal instabilities.
\newblock {\em Mechanism and Machine Theory}, 196:105626, June 2024.

\bibitem{van_hecke_profusion_2021}
Martin van Hecke.
\newblock Profusion of transition pathways for interacting hysterons.
\newblock {\em Physical Review E}, 104(5):054608, November 2021.
\newblock Publisher: American Physical Society.

\bibitem{liu_controlled_2024}
Jingran Liu, Margot Teunisse, George Korovin, Ivo~R. Vermaire, Lishuai Jin, Hadrien Bense, and Martin van Hecke.
\newblock Controlled pathways and sequential information processing in serially coupled mechanical hysterons.
\newblock {\em Proceedings of the National Academy of Sciences}, 121(22):e2308414121, May 2024.
\newblock Publisher: Proceedings of the National Academy of Sciences.

\bibitem{shohat_geometric_2025}
Dor Shohat and Martin van Hecke.
\newblock Geometric {Control} and {Memory} in {Networks} of {Hysteretic} {Elements}.
\newblock {\em Physical Review Letters}, 134(18):188201, May 2025.
\newblock Publisher: American Physical Society.

\bibitem{shohat_aging_2025}
Dor Shohat, Paul Baconnier, Itamar Procaccia, Martin~van Hecke, and Yoav Lahini.
\newblock Aging of amorphous materials under cyclic strain, June 2025.
\newblock arXiv:2506.08779 [cond-mat].

\bibitem{riks_incremental_1979}
E.~Riks.
\newblock An incremental approach to the solution of snapping and buckling problems.
\newblock {\em International Journal of Solids and Structures}, 15(7):529--551, January 1979.

\bibitem{farrell2016computationdisconnectedbifurcationdiagrams}
Patrick~E. Farrell, Casper H.~L. Beentjes, and Ásgeir Birkisson.
\newblock The computation of disconnected bifurcation diagrams, 2016.

\bibitem{ducarme_exotic_2025}
Paul Ducarme, Bart Weber, Martin van Hecke, and Johannes T.~B. Overvelde.
\newblock Exotic mechanical properties enabled by countersnapping instabilities.
\newblock {\em Proceedings of the National Academy of Sciences}, 122(16):e2423301122, April 2025.
\newblock Publisher: Proceedings of the National Academy of Sciences.

\bibitem{obara_truth_2019}
Paulina Obara, Joanna Kłosowska, and Wojciech Gilewski.
\newblock Truth and {Myths} about {2D} {Tensegrity} {Trusses}.
\newblock {\em Applied Sciences}, 9(1):179, January 2019.
\newblock Number: 1 Publisher: Multidisciplinary Digital Publishing Institute.

\bibitem{he_grasping_2025}
Gengzhi He, Curtis Sparks, and Nick Gravish.
\newblock Grasping and rolling in-plane manipulation using deployable tape spring appendages.
\newblock {\em Science Advances}, 11(15):eadt5905, April 2025.
\newblock Publisher: American Association for the Advancement of Science.

\bibitem{kwakernaak_collective_2024}
Lennard~J. Kwakernaak, Arman Guerra, Douglas~P. Holmes, and Martin van Hecke.
\newblock The collective snapping of a pair of bumping buckled beams.
\newblock {\em Extreme Mechanics Letters}, 69:102160, June 2024.

\bibitem{guerra_selfordering_2023}
Arman Guerra, Anja~C. Slim, Douglas~P. Holmes, and Ousmane Kodio.
\newblock Self-ordering of buckling, bending, and bumping beams.
\newblock {\em Phys. Rev. Lett.}, 130:148201, Apr 2023.

\bibitem{djellouli_shell_2024}
Adel Djellouli, Bert Van~Raemdonck, Yang Wang, Yi~Yang, Anthony Caillaud, David Weitz, Shmuel Rubinstein, Benjamin Gorissen, and Katia Bertoldi.
\newblock Shell buckling for programmable metafluids.
\newblock {\em Nature}, 628(8008):545--550, April 2024.
\newblock Publisher: Nature Publishing Group.

\bibitem{kwakernaak_counting_2023}
Lennard~J. Kwakernaak and Martin van Hecke.
\newblock Counting and sequential information processing in mechanical metamaterials.
\newblock {\em Phys. Rev. Lett.}, 130:268204, Jun 2023.

\bibitem{springable}
Paul Ducarme.
\newblock springable.
\newblock available on PyPI: \url{https://pypi.org/project/springable/1.0.0/}, source code: \url{https://github.com/ducarme/springable}, 2025.
\newblock Accessed: 2025-08-27.

\bibitem{jin_guided_2020}
Lishuai Jin, Romik Khajehtourian, Jochen Mueller, Ahmad Rafsanjani, Vincent Tournat, Katia Bertoldi, and Dennis~M. Kochmann.
\newblock Guided transition waves in multistable mechanical metamaterials.
\newblock {\em Proceedings of the National Academy of Sciences}, 117(5):2319--2325, 2020.

\bibitem{byun_integrated_2024}
Junghwan Byun, Aniket Pal, Jongkuk Ko, and Metin Sitti.
\newblock Integrated mechanical computing for autonomous soft machines.
\newblock {\em Nature Communications}, 15(1), April 2024.

\bibitem{bertoldi_negative_2010}
Katia Bertoldi, Pedro~M. Reis, Stephen Willshaw, and Tom Mullin.
\newblock Negative {Poisson}'s ratio behavior induced by an elastic instability.
\newblock {\em Advanced Materials (Deerfield Beach, Fla.)}, 22(3):361--366, January 2010.

\bibitem{pal_programmable_2023}
Aniket Pal and Metin Sitti.
\newblock Programmable mechanical devices through magnetically tunable bistable elements.
\newblock {\em Proceedings of the National Academy of Sciences}, 120(15):e2212489120, 2023.

\end{thebibliography}
\end{document}


\maketitle

\tableofcontents

\newpage
\section{Flexel formulation}

To simulate a mechanical structure quasi-statically, the strategy consists of minimizing the total potential energy at varying load levels, as more formally described in section~2. Essentially, this is done by iteratively finding configurations wherein internal elastic forces balance external ones. The process requires computing both the elastic forces and the way they vary with structural deformations. The flexel formulation that we present in this section is divided in two parts. First, we derive the equations for computing the elastic force and its variation for an individual flexel. Then, we show how to combine these individual contributions to yield the global force and variation expressions for an assembly of flexels.

\subsection{Individual flexel}
The elastic force of a flexel and how this force varies can be derived from its potential energy $e$. For a flexel, the potential energy is the stored elastic energy, which depends on its the geometric measure $\alpha$, that is, its length if it is a longitudinal flexel, its angle if an angular flexel, its area if an area flexel, etc. The geometric measure $\alpha$ is computed from the spatial coordinates $\bm{z}$ of the nodes coupled by the flexel, using the equations derived in section~\ref{section:geometric_measures}.

More formally, the energy $e$ of a flexel is described using an energy potential $v$ which is either a univariate $v(\cdot)$ or bivariate $v(\cdot, \cdot)$ function:
\begin{equation}
\label{eq:energy_flexel}
e(\bm{q}) = 
\begin{cases}
 v(\alpha(\bm{z}))&\text{if $v$ is univariate,}\\
 v(\alpha(\bm{z}), t)&\text{if $v$ is bivariate,}
\end{cases}
\end{equation}
where $\bm{q}$ is the subset of the structural coordinates $\bm{Q}$ (that describe the state of an entire assembly) used by the flexel and $\alpha$ is the geometric measure computed from the spatial coordinates $\bm{z} \subseteq \bm{q}$. For a flexel based on a univariate potential, its energy is solely determined by its geometric measure and its coordinates $\bm{q}$ are all spatial: $\bm{q}=\bm{z}$. For a flexel based on a bivariate potential, the energy depends on the geometric measure $\alpha$ and an additional (non-spatial) coordinate $t$: $\bm{q}=[\bm{z}^\top, t]^\top$.

This extra coordinate $t$ describes a degree of freedom internal to the flexel that has no direct physical interpretation. It is introduced to define flexels with multi-valued generalized force-displacement curves: equilibrium curves such that to a given geometric measure $\alpha$ (length, angle, area, etc) can correspond multiple generalized force values (force, torque, 2d-pressure, or etc). This extra coordinate $t$ disambiguates the state of the flexel, as the value of the geometric measure $\alpha$ alone would not be enough to determine what the generalized force in the flexel is. At equilibrium, the coordinate $t$ can be seen as the \emph{curve parameter} of the parametric description of the multi-valued force-displacement curve (Eq.~(\ref{eq:multi-valued_curve})). More details on bivariate potentials and how to construct them from multi-valued force-displacement curves are provided in subsection~\ref{subsection:bivariate_behaviors}.

Different types of univariate and bivariate energy potentials are implemented in \springable{}. They are defined in subsection~\ref{subsection:univariate_behaviors} and \ref{subsection:bivariate_behaviors} respectively.

\paragraph{Energy gradient (elastic force vector)}
The elastic force is described by the gradient of the energy $e$ with respect to the coordinates $\bm{q}$. We will denote this vector as $\bm{f}$, and its expression is given by
\begin{align}
\label{eq:force_bivariate_flexel}
\bm{f}& := \dfrac{\partial e}{\partial \bm{q}} = [f_i] =\left[\dfrac{\partial e}{\partial q_i}\right] = \dfrac{\partial v}{\partial \alpha}\left[\dfrac{\partial \alpha}{\partial q_i}\right] + \dfrac{\partial v}{\partial t}\left[\dfrac{\partial t}{\partial q_i}\right] \nonumber \\
&=\dfrac{\partial v}{\partial \alpha}
\begin{bmatrix}
    \partial \alpha / \partial z_0\\
    \vdots\\
    \partial \alpha / \partial z_{n-2}\\
    0
\end{bmatrix}
+ \dfrac{\partial v}{\partial t}
\begin{bmatrix}
    0\\
    \vdots\\
    0\\
    1
\end{bmatrix} \nonumber\\
&=\dfrac{\partial v}{\partial \alpha}
\begin{bmatrix}
    \partial \alpha / \partial \bm{z}\\
    0
\end{bmatrix}
+ \dfrac{\partial v}{\partial t}
\begin{bmatrix}
    \bm{0}^{(n-1) \times 1}\\
    1
\end{bmatrix}
\end{align}
where $v$ is a bivariate potential (where the $n$ coordinates used by the flexel are the spatial ones and the internal one, i.e.: $\bm{q}=[\bm{z}^\top, t]^\top$). In case of univariate potential (where $\partial v/\partial t=0$ and all the $n$ coordinates used by the flexel are spatial, i.e.: $\bm{q}=\bm{z}$), Eq.~(\ref{eq:force_bivariate_flexel}) simplifies to
\begin{align}
    \label{eq:force_univariate_flexel}
    \bm{f} = \dfrac{\partial v}{\partial \alpha}\left[\dfrac{\partial \alpha}{\partial z_i}\right] =\dfrac{\partial v}{\partial \alpha}\dfrac{\partial \alpha}{\partial \bm{z}}
\end{align}

To compute the energy gradient of an individual flexel, we therefore need to separately compute two types of quantities:
\begin{itemize}
    \item the measure $
    \alpha$ and its gradient $\partial \alpha / \partial \bm{z} =[\partial \alpha / \partial z_i]$, purely geometric quantities which reflects how the \emph{shape} of the flexel (longitudinal, angular, area, etc) couples to the nodes of the structure, and whose equations are derived in section~\ref{section:geometric_measures};
    \item the gradient of the potential $v$, i.e. $\partial v / \partial \alpha$ and $\partial v / \partial t$, quantities which reflects the \emph{intrinsic nonlinear mechanical behavior} of the flexel, and whose equations are derived in section~\ref{section:intrinsic_mechanical_behaviors}.
\end{itemize}

\paragraph{Energy hessian (stiffness matrix)}
How the elastic force varies with respect to the state of the flexel is described by the hessian of the energy $e$ with respect to the coordinates $\bm{q}$ (that is, the gradient of the gradient, or the matrix of the second-order derivatives). This matrix is commonly known as the \emph{stiffness matrix}, and we will denote it as $\bm{k}$. Its expression is given by
 \begin{align}
  \label{eq:stiffness_matrix_bivariate_flexel}
\bm{k} =& \dfrac{\partial^2 e}{\partial \bm{q} \partial\bm{q}^\top}= \left[k_{ij}\right] = \left[\dfrac{\partial^2 e}{\partial q_i\partial q_j}\right]=\left[\dfrac{\partial f_i}{\partial q_j}\right] \nonumber\\
=& \dfrac{\partial^2 v}{\partial \alpha^2}\left[\dfrac{\partial \alpha}{\partial q_i}\dfrac{\partial \alpha}{\partial q_j}\right] + \dfrac{\partial^2 v}{\partial \alpha \partial t}\left[\dfrac{\partial \alpha}{\partial q_i}\dfrac{\partial t}{\partial q_j} + \dfrac{\partial t}{\partial q_i}\dfrac{\partial \alpha}{\partial q_j}\right] + \dfrac{\partial v}{\partial \alpha}\left[\dfrac{\partial ^2 \alpha}{\partial q_i \partial q_j}\right]\nonumber\\
&+ \dfrac{\partial^2 v}{\partial t^2}\left[\dfrac{\partial t}{\partial q_i}\dfrac{\partial t}{\partial q_j}\right] + \dfrac{\partial v}{\partial t}\left[\dfrac{\partial^2 t}{\partial q_i \partial q_j}\right]\nonumber\\
=&
\dfrac{\partial^2 v}{\partial \alpha^2}
\begin{bmatrix}
    \dfrac{\partial \alpha}{\partial z_0}\dfrac{\partial \alpha}{\partial z_0}&\dots&\dfrac{\partial \alpha}{\partial z_0}\dfrac{\partial \alpha}{\partial z_{n-2}}&0\\
    \vdots&\ddots&\vdots&\vdots\\
    \dfrac{\partial \alpha}{\partial z_{n-2}}\dfrac{\partial \alpha}{\partial z_0}&\dots&\dfrac{\partial \alpha}{\partial z_{n-2}}\dfrac{\partial \alpha}{\partial z_{n-2}}&0\\
    0&\dots&0&0
\end{bmatrix}
+
\dfrac{\partial v}{\partial \alpha}
\begin{bmatrix}
    \dfrac{\partial^2 \alpha}{\partial z_0 \partial z_0}&\dots&\dfrac{\partial^2 \alpha}{\partial z_0 \partial z_{n-2}}&0\\
    \vdots&\ddots&\vdots&\vdots\\
    \dfrac{\partial^2 \alpha}{\partial z_{n-2}\partial z_0}&\dots&\dfrac{\partial^2 \alpha}{\partial z_{n-2}\partial z_{n-2}}&0\\
    0&\dots&0&0
\end{bmatrix}\nonumber\\
&+
\dfrac{\partial^2 v}{\partial \alpha \partial t}
\begin{bmatrix}
    0&\dots&0&\dfrac{\partial \alpha}{\partial z_{0}}\\
    \vdots&\ddots&\vdots&\vdots\\
    0&\dots&0&\dfrac{\partial \alpha}{\partial z_{n-2}}\\
    \dfrac{\partial \alpha}{\partial z_{0}}&\dots&\dfrac{\partial \alpha}{\partial z_{n-2}}&0
\end{bmatrix}
+
\dfrac{\partial^2 v}{ \partial t^2}
\begin{bmatrix}
    0&\dots&~&~&0\\
    \vdots&\ddots&~&~&\vdots\\
    ~&~&~&~&~\\
    ~&~&~&0&0\\
    0&\dots&~&0&1
\end{bmatrix}\nonumber\\
~&~\nonumber\\
=&
\dfrac{\partial^2 v}{\partial \alpha^2}
\begin{bmatrix}
    \dfrac{\partial \alpha}{\partial \bm{z}}\dfrac{\partial \alpha}{\partial \bm{z}^\top}&\bm{0}^{(n-1)\times1}\\
    ~&~\\
    \bm{0}^{1\times(n-1)}&0
\end{bmatrix}
+
\dfrac{\partial v}{\partial \alpha}
\begin{bmatrix}
    \dfrac{\partial^2 \alpha}{\partial \bm{z} \partial \bm{z}^\top}&\bm{0}^{(n-1)\times1}\\
    ~&~\\
    \bm{0}^{1\times(n-1)}&0
\end{bmatrix}\nonumber\\
&+
\dfrac{\partial^2 v}{\partial \alpha \partial t}
\begin{bmatrix}
    \bm{0}^{(n-1)\times(n-1)}&\dfrac{\partial \alpha}{\partial \bm{z}}\\
    ~&~\\
    \dfrac{\partial \alpha}{\partial \bm{z}^\top}&0
\end{bmatrix}
+
\dfrac{\partial^2 v}{ \partial t^2}
\begin{bmatrix}
    \bm{0}^{(n-1)\times(n-1)}&\bm{0}^{(n-1)\times1}\\
    ~&~\\
    \bm{0}^{1\times(n-1)}&1
\end{bmatrix}
\end{align}
where $v$ is a bivariate potential (where the $n$ coordinates used by the flexel are the spatial ones and the internal one, i.e.: $\bm{q}=[\bm{z}^\top, t]^\top$). In case of a univariate potential (where $\partial^2 v/\partial \alpha\partial t=\partial v^2/\partial t^2=0$ and all the $n$ coordinates used by the flexel are spatial, i.e.: $\bm{q}=\bm{z}$), Eq.~(\ref{eq:stiffness_matrix_bivariate_flexel}) simplifies to
\begin{align}
 \label{eq:stiffness_matrix_univariate_flexel}
\bm{k} &= \dfrac{\partial^2 v}{\partial \alpha^2}\left[\dfrac{\partial \alpha}{\partial z_i}\dfrac{\partial \alpha}{\partial z_j}\right] + \dfrac{\partial v}{\partial \alpha}\left[\dfrac{\partial^2 \alpha}{\partial z_i\partial z_j}\right]=
\dfrac{\partial^2 v}{\partial \alpha^2}\dfrac{\partial \alpha}{\partial \bm{z}}\dfrac{\partial \alpha}{\partial \bm{z}^\top} + \dfrac{\partial v}{\partial \alpha}\dfrac{\partial^2 \alpha}{\partial \bm{z} \partial \bm{z}^\top}.
\end{align}

Similarly to the energy gradient, to compute the energy hessian, we need to separately compute quantities describing the geometry and quantities describing the intrinsic nonlinear behavior:
\begin{itemize}
    \item the geometric measure, its gradient and its hessian, $
    \alpha$, $\partial \alpha / \partial \bm{z} =[\partial \alpha / \partial z_i]$ and $\partial^2 \alpha / \partial \bm{z}\partial \bm{z}^\top=[\partial^2 \alpha / \partial z_i \partial z_j]$, purely geometric quantities which reflects how the \emph{shape} of the flexel (longitudinal, angular, area, etc) couples to the rest of the structure, and whose equations are derived in section~\ref{section:geometric_measures};
    \item the gradient and hessian of the potential, $\partial v / \partial \alpha$, $\partial^2 v / \partial \alpha^2$, $\partial^2 v / \partial \alpha \partial t$ and  $\partial^2 v / \partial t^2$, which reflects the \emph{intrinsic nonlinear mechanical behavior} of the flexel, and whose equations are derived in section~\ref{section:intrinsic_mechanical_behaviors}.
\end{itemize}

\subsection{Assembly of flexels}
The elastic force of an assembly of flexels and how this force varies can be derived from its elastic energy $E$. The total elastic energy $E$ stored in the structure is the sum of each flexel's elastic energy:

\begin{equation}
\label{eq:total_elastic_energy}
    E(\bm{Q}) = \sum_{s=0}^{N_\text{f}-1} e_{s}(\bm{q}^s),
\end{equation}
where $\bm{Q}$ is the set of all the structural coordinates, which describe the state of the assembly, $N_\text{f}$ is the number of flexels in the assembly, $e_s$ is the energy of flexel $s$ (Eq.~(\ref{eq:energy_flexel})) and $\bm{q}^s \subseteq \bm{Q}$ is the subset of structural coordinates used by flexel $s$.

\paragraph{Energy gradient (elastic force vector)} The elastic force is described by the gradient of the energy $E$ with respect to the coordinates $\bm{Q}$. We will denote this vector as $\bm{F}$, and its expression is given by
\begin{equation}
    \bm{F}=\dfrac{\partial E}{\partial \bm{Q}} = \left[ \dfrac{\partial E}{\partial Q_k}\right]=\sum_{s=0}^{N_\text{f}-1} \left[\dfrac{\partial e_s(\bm{q}^s)}{\partial Q_k}\right]= \sum_{s=0}^{N_\text{f}-1}\left(\sum_{i=0}^{n_s-1}\dfrac{\partial e_s}{\partial q^s_i}\left[\dfrac{\partial q^s_i}{\partial Q_k}\right]\right) = \sum_{s=0}^{N_\text{f}-1}\left(\sum_{i=0}^{n_s-1}\dfrac{\partial e_s}{\partial q^s_i}\left[L^s_{ik}\right]\right),
    \label{eq:force_vector_entire_structure}
\end{equation}
where $n_s$ is the number of coordinates in $\bm{q}^s$, $\partial e_s/\partial q^s_i$ are the components of the energy gradient $\bm{f}^s$ of flexel $s$ (Eq.~(\ref{eq:force_bivariate_flexel})),
and where
\begin{equation}
    \label{eq:Lik}
    L^s_{ik}=
    \begin{cases}
        1&\text{if $q^s_i = Q_k$},\\
        0&\text{else.}
    \end{cases}
\end{equation} In practical terms, Eq.~(\ref{eq:force_vector_entire_structure}) means that the gradient $\bm{F}=[F_k]$ is constructed by adding together the components of each individual flexel's energy gradient $\bm{f}^s=[f^s_i]=\left[\partial e_s / \partial q^s_i\right]$ that correspond to the same coordinate $Q_k$.

\paragraph{Energy hessian (stiffness matrix)}How the elastic force varies with respect to the state of the assembly is described by the hessian of the energy $E$ with respect to the coordinates $\bm{Q}$ (that is, the gradient of the gradient, or the matrix of the second-order derivatives). This matrix is commonly known as the \emph{stiffness matrix}, and we will denote it as $\bm{K}$. Its expression is given by:
\begin{align}
    \label{eq:stiffness_matrix_entire_structure}
    \bm{K}&=\dfrac{\partial^2 E}{\partial \bm{Q}\partial\bm{Q}^\top} = \left[K_{kl}\right]=\left[\dfrac{\partial^2 E}{\partial Q_k \partial Q_l}\right]= \left[\dfrac{\partial F_k}{\partial Q_l}\right]=\sum_{s=0}^{N_\text{f}-1} \left[\dfrac{\partial^2 e_s(\bm{q}^s)}{\partial Q_k \partial Q_l}\right] \nonumber\\
    &=\sum_{s=0}^{N_\text{f}-1}\left(\sum_{i=0}^{n_s-1} \sum_{j=0}^{n_s-1}\dfrac{\partial^2 e_s}{\partial q^s_i \partial q^s_j}\left[\dfrac{\partial q^s_i}{\partial Q_k}\dfrac{\partial q^s_j}{\partial Q_l}\right]\right) = \sum_{s=0}^{N_\text{f}-1}\left(\sum_{i=0}^{n_s-1}\sum_{j=0}^{n_s-1}\dfrac{\partial^2 e_s}{\partial q^s_i \partial q^s_j}\left[L^s_{ik}L^s_{jl}\right]\right),
\end{align}
where $\partial^2 e_s/\partial q^s_i \partial q^s_j$ are the components of the energy hessian of flexel $s$ (Eq.~(\ref{eq:stiffness_matrix_bivariate_flexel})),
and where $[L_{ik}L_{jl}]$ is determined using Eq.~(\ref{eq:Lik}). In practical terms, Eq.~(\ref{eq:stiffness_matrix_entire_structure}) means that the energy hessian $\bm{K}=[K_{kl}]$ is constructed by adding together the components of each individual flexel's energy hessian $\bm{k}^s=[k^s_{ij}]=[\partial e_s / \partial q^s_i \partial q^s_j]$ that correspond to the same pair of coordinates $(Q_k, Q_l)$.

\section{Path-following algorithm}
\label{section:pathfollowing_algo}
First, we provide a general mathematical description of the mechanical systems that we aim to model and simulate, and express the governing equations that need to be solved by the algorithm to simulate such systems. Using that formalism, we then describe the algorithm itself.

\subsection{Model descriptions}
Let us consider a mechanical system (or structure), defined as a set of nodes coupled by flexels. The state of the system is defined by $N$ coordinates gathered in the array $\bm{Q}$. Among these $N$ coordinates, $M$ are free, while the other are fixed. The free coordinates are denoted by $\tilde{\bm{Q}}$, which is the array $\bm{Q}$ without the fixed coordinates. The objective of the simulation is to calculate the equilibrium states of the structure from a given initial load up to a final prescribed load, acting on the free coordinates $\tilde{\bm{Q}}$. To this end, we first consider the total potential energy $\Pi$ of the system:
\begin{equation}
\Pi = E - \tilde{\bm{F}}^\text{ext} \cdot \tilde{\bm{U}} = E - \sum_{k=0}^{M-1} \tilde{F}^\text{ext}_k\tilde{U}_k,
\end{equation}
where $E$ is the elastic energy stored in the system and $\tilde{\bm{F}}^\text{ext} \cdot \tilde{\bm{U}}$ is the combined work done by each external force $\tilde{F}^\text{ext}_k$ applied along the displacement $\tilde{U}_k=\Delta \tilde{Q}_k$ of the free coordinate $\tilde{Q}_k$. Since equilibrium states under the external load $\tilde{\bm{F}}^\text{ext}$ are stationary points of the total potential energy with respect to the free coordinates $\tilde{\bm{Q}}$, they are solutions of the following system of $M$ equations
\begin{align}
    &\dfrac{\partial E}{\partial \tilde{\bm{Q}}}(\bm{Q}) - \tilde{\bm{F}}^\text{ext} =0\\
    &\Leftrightarrow\nonumber\\
    &\begin{cases}
        \dfrac{\partial E}{\partial \tilde{Q}_0}(\bm{Q}) -  \tilde{F}^\text{ext}_0 = 0&\\
        \vdots&\\
        \dfrac{\partial E}{\partial \tilde{Q}_k}(\bm{Q}) -  \tilde{F}^\text{ext}_k = 0&\\
        \vdots&\\
        \dfrac{\partial E}{\partial \tilde{Q}_{M-1}}(\bm{Q}) -  \tilde{F}^\text{ext}_{M-1} = 0.
    \end{cases}
\end{align}
Each equation in this system represents the balance between the internal elastic forces and the external forces on coordinate $\tilde{Q}_k$ that must be achieved by a configuration to be in a static equilibrium under the external forces $\tilde{\bm{F}}^\text{ext}$. Since we are interested in solving this system for various levels of loading from the initial given load to the final prescribed load (that we will now denote as $\tilde{\bm{F}}^{\text{ext}, 0}$ and $\tilde{\bm{F}}^\text{ext, final}$), the system can be rewritten in the following parametrized form
\begin{align}
    \label{eq:parametrized_sys}
    &\dfrac{\partial E}{\partial \tilde{\bm{Q}}}(\bm{Q}) - \tilde{\bm{F}}^\text{ext} = \dfrac{\partial E}{\partial \tilde{\bm{Q}}}(\bm{Q}) - (\tilde{\bm{F}}^{\text{ext}, 0} +\lambda \tilde{\bm{F}}^\text{dir}) =0\\
    &\Leftrightarrow\nonumber\\
    &\begin{cases}
        \dfrac{\partial E}{\partial \tilde{Q}_0}(\bm{Q}) - (\tilde{F}_0^{\text{ext}, 0} +\lambda \tilde{F}_0^\text{dir}) = 0&\\
        \vdots&\\
        \dfrac{\partial E}{\partial \tilde{Q}_k}(\bm{Q}) - (\tilde{F}_k^{\text{ext}, 0} +\lambda \tilde{F}_k^\text{dir}) = 0&\\
        \vdots&\\
        \dfrac{\partial E}{\partial \tilde{Q}_{M-1}}(\bm{Q}) - (\tilde{F}_{M-1}^{\text{ext}, 0} +\lambda \tilde{F}_{M-1}^\text{dir}) = 0.
    \end{cases}\nonumber
\end{align}
where
\begin{equation}
\tilde{\bm{F}}^{\text{dir}}:= \dfrac{\tilde{\bm{F}}^\text{step}}{||\tilde{\bm{F}}^\text{step}||}
\end{equation}
is the direction of the step load vector $\tilde{\bm{F}}^\text{step}:= \tilde{\bm{F}}^\text{ext, final} - \tilde{\bm{F}}^{\text{ext}, 0}$ (which represents the additional load needed to reach the final prescribed load from the initial load), and where $\lambda$
is the load parameter, which is initially 0 and reaches $||\tilde{\bm{F}}^\text{step}||$ when the load step is completed. Eq.~(\ref{eq:parametrized_sys}) represents a system of parametrized nonlinear equations, which can be solved using numerical continuation algorithms.

\subsection{Algorithm}
In \springable, we solve Eq.~(\ref{eq:parametrized_sys}) using the arc-length continuation scheme introduced in \cite{riks_incremental_1979, crisfield1981fast}, as it is able to reliably calculate the entire succession of equilibrium states even if those form a branch with turning points, unlike more naive approaches that consists of iteratively increasing the parameter $\lambda$ and applying Newton's method. The algorithm consists of computing successive increments, each performing one prediction, followed by multiple corrections to converge towards an equilibrium state. The algorithm proceeds as follows. Note: $x \leftarrow y$ means that $y$ is assigned to the variable $x$.

\begin{enumerate}
\item The inputs of the problem are provided: initial coordinates $\bm{Q}^0$, initial external forces $\tilde{\bm{F}}^{\text{ext}, 0}$, step load vector $\tilde{\bm{F}}^\text{step}$ (defined as $\tilde{\bm{F}}^\text{ext, final} - \tilde{\bm{F}}^{\text{ext}, 0}$).
\item The settings of the solver are provided: initial radius $r_0$, convergence tolerance $\varepsilon$.
\item The direction of of the step load vector is computed: $\tilde{\bm{F}}^\text{dir}:= \tilde{\bm{F}}^\text{step} / ||\tilde{\bm{F}}^\text{step}||$.
\item The variables that will be used and updated by the algorithm are initialized:
\begin{itemize}
    \item current radius $r \leftarrow r_0$,
    \item load parameter $\lambda \leftarrow 0$,
    \item current coordinates $\bm{Q}\leftarrow\bm{Q}^0$,
    \item current external force vector $\tilde{\bm{F}}^\text{ext}\leftarrow \tilde{\bm{F}}^{\text{ext}, 0}$,
    \item list of previously computed equilibrium states $\mathcal{Q}^\text{eq} \leftarrow [\bm{Q}^0]$,
    \item list of external force vectors corresponding to the previously computed equilibrium states $\tilde{\mathcal{F}}^\text{eq} \leftarrow [\tilde{\bm{F}}^{\text{ext}, 0}]$,
    \item the current hessian matrix $\bm{K}$ is computed based on the coordinates $\bm{Q}^0$ (Eq.~(\ref{eq:stiffness_matrix_entire_structure})), the current reduced hessian matrix $\tilde{\bm{K}}$ is formed by removing the rows and columns corresponding to the fixed coordinates.
\end{itemize}
\item A new increment is started.
\begin{enumerate} 
\item The prediction step is started.
\begin{enumerate}
    \item The increment is initialized: $\Delta \bm{U}^\text{inc} = \bm{0}$, $\Delta \lambda^\text{inc} = 0$.
    \item Eq.~(\ref{eq:uhat_prediction}) is solved for $\Delta \hat{\bm{U}}$:
    \begin{equation}
        \tilde{\bm{K}}\Delta \hat{\bm{U}}=\tilde{\bm{F}}^\text{dir}.
        \label{eq:uhat_prediction}
    \end{equation}
    \item The quantity $\Delta \lambda^\text{ite}$ is computed  \cite{ritto2008arclength}:
    \begin{equation}
        \Delta \lambda^\text{ite} = s\dfrac{r}{\sqrt{\Delta \hat{\bm{U}} \cdot \Delta \hat{\bm{U}}}},
    \end{equation}
    with
    \begin{equation}
    s=
    \begin{cases}
        1&\text{if first increment, or if $\Delta \hat{\bm{U}}\cdot \Delta \bm{U}^\text{prev. inc} \ge 0$}\\
        -1&\text{else.}
    \end{cases}
    \end{equation}
    \item The following quantities are updated: $\Delta \lambda^\text{inc} \leftarrow \Delta \lambda^\text{inc} + \Delta \lambda^\text{ite}$, and $\Delta \bm{U}^\text{inc} \leftarrow  \Delta \bm{U}^\text{inc} + \Delta\bm{U}^\text{ite}$, where 
    \begin{equation}
        \Delta\bm{U}^\text{ite} = \Delta \lambda^\text{ite}\Delta \hat{\bm{U}}.
    \end{equation}
    \item The external force vector is updated: $\tilde{\bm{F}}^\text{ext} \leftarrow \tilde{\bm{F}}^\text{ext} + \Delta\lambda^\text{ite}\tilde{\bm{F}}^\text{dir}$.
    \item The free coordinates in $\bm{Q}$ are updated:  $\tilde{\bm{Q}} \leftarrow \tilde{\bm{Q}} + \Delta \bm{U}^\text{ite}$.
    \item The residual $\tilde{\bm{R}}$ of the force balance based on the updated coordinates is computed:
    \begin{equation}
        \tilde{\bm{R}} = \tilde{\bm{F}} - \tilde{\bm{F}}^\text{ext},
    \end{equation}
    where $\tilde{\bm{F}}$ is the array $\bm{F}:=\partial E / \partial \bm{Q}$ (Eq.~(\ref{eq:force_vector_entire_structure})) from which the components corresponding to fixed coordinates have been removed. The array $\bm{F}$ represents the internal net force resulting on each coordinate due the elastic forces of the flexels, 
\item The next action is decided based on the convergence criterion:
\begin{equation}
    \begin{cases}
    \text{start a correction step}&\text{if }\quad\dfrac{||\tilde{\bm{R}}||}{||\tilde{\bm{F}}^\text{step}||}> \varepsilon,\\
    ~&~\\
    \text{prepare for the next increment}&\text{else.}
    \end{cases}
    \nonumber
\end{equation}
\end{enumerate}
\item A correction step is started (if no convergence):
\begin{enumerate}
    \item The current hessian matrix $\bm{K}$ is computed based on the updated coordinates $\bm{Q}$ (see Eq.~\ref{eq:stiffness_matrix_entire_structure}), and the current reduced hessian matrix $\tilde{\bm{K}}$ is formed by removing the rows and columns corresponding to the fixed coordinates.
    \item Eqs.~(\ref{eq:uhat_correction}) and (\ref{eq:ubar_correction}) are solved for $\Delta \hat{\bm{U}}$ and $\Delta \bar{\bm{U}}$ respectively:
    \begin{align}
        \tilde{\bm{K}}\Delta\hat{\bm{U}} &= \tilde{\bm{F}}^\text{dir}, \label{eq:uhat_correction}\\
        \tilde{\bm{K}}\Delta\bar{\bm{U}} &= -\tilde{\bm{R}}.
        \label{eq:ubar_correction}
    \end{align}
    \item The following quantities are computed \cite{ritto2008arclength}:
    \begin{itemize}
        \item $a_0=\Delta \hat{\bm{U}}\cdot \Delta \hat{\bm{U}}$,
        \item $b_0=2\Delta\bm{U}^\text{inc}\cdot\Delta \hat{\bm{U}}$,
        \item $b_1=2\Delta \bar{\bm{U}} \cdot \Delta \hat{\bm{U}}$,
        \item $c_0 =\Delta \bm{U}^\text{inc}\cdot \Delta \bm{U}^\text{inc} - r^2$,
        \item $c_1=2\Delta \bm{U}^\text{inc}\cdot \Delta \bar{\bm{U}}$,
        \item $c_2=\Delta\bar{\bm{U}}\cdot\Delta\bar{\bm{U}}$, 
        \item $a=a_0$,
        \item $b=b_0+b_1$,
        \item $c=c_0+c_1+c_2$,
        \item $\rho=b^2 - 4ac$.
    \end{itemize}

    \item The correction feasibility is assessed, and the next action is decided:
    \begin{equation}
    \begin{cases}
    \text{continue}&\text{if $\rho > 0$},\\
    ~&~\\
    \text{prepare for increment restart}&\text{else}.
    \end{cases}
    \nonumber
\end{equation}
\item The quantity $\Delta \lambda^\text{ite}$ is computed:
\begin{equation}
\Delta \lambda^\text{ite}=
    \begin{cases}
        \dfrac{-b + \sqrt{\rho}}{2a}&\text{if $b_0 > 0$},\\
        ~&~\\
        \dfrac{-b - \sqrt{\rho}}{2a}&\text{else}.
    \end{cases}
\end{equation}
\item The following quantities are updated: $\Delta \lambda^\text{inc} \leftarrow \Delta \lambda ^\text{inc} + \Delta\lambda^\text{ite}$, and $\Delta \bm{U}^\text{inc} \leftarrow \Delta \bm{U}^\text{inc} + \Delta \bm{U}^\text{ite}$, where
\begin{equation}
\Delta \bm{U}^\text{ite} \leftarrow \Delta\bar{\bm{U}} + \Delta\lambda^\text{ite}\Delta\hat{\bm{U}}.
\end{equation}
\item The external force vector is updated: $\tilde{\bm{F}}^\text{ext} \leftarrow \tilde{\bm{F}}^\text{ext} + \Delta\lambda^\text{ite}\tilde{\bm{F}}^\text{dir}$.
\item The free coordinates in $\bm{Q}$ are updated:  $\tilde{\bm{Q}} \leftarrow \tilde{\bm{Q}} + \Delta \bm{U}^\text{ite}$.
\item The residual $\tilde{\bm{R}}$ of the force balance based on the updated coordinates is computed:
\begin{equation}
    \tilde{\bm{R}} = \tilde{\bm{F}} - \tilde{\bm{F}}^\text{ext},
\end{equation}
where $\tilde{\bm{F}}$ is the array $\bm{F}:=\partial E / \partial \bm{Q}$ (Eq.~(\ref{eq:force_vector_entire_structure})) from which the components corresponding to the fixed coordinates have been removed.
\item The next action is decided based on the convergence criterion:
    \begin{equation}
        \begin{cases}
        \text{start an additional correction step}&\text{if }\quad\||\tilde{\bm{R}}|| > \varepsilon,\\
        ~&~\\
        \text{prepare for the next increment}&\text{else.}
        \end{cases}
        \nonumber
    \end{equation}
\end{enumerate}
\item The next increment is prepared.
\begin{enumerate}
    \item The current hessian matrix $\bm{K}$ is computed based on the coordinates $\bm{Q}$ (Eq.~(\ref{eq:stiffness_matrix_entire_structure})), the current reduced hessian matrix $\tilde{\bm{K}}$ is formed by removing the rows and columns corresponding to the fixed coordinates.
    \item The following quantities are updated: $\Delta \bm{U}^\text{prev. inc}\leftarrow \Delta \bm{U}^\text{inc}$, and $r \leftarrow \min(r_0, 2r)$.
    \item The load parameter is updated: $\lambda \leftarrow \lambda + \Delta \lambda_\text{inc}$.
    \item The new equilibrium state $\bm{Q}$ is appended to the list $\mathcal{Q}^\text{eq}$ of previously computed equilibrium states: $\mathcal{Q}^\text{eq} \leftarrow \mathcal{Q}^\text{eq} + [\bm{Q}]$.
    \item The new external force vector $\tilde{\bm{F}}^\text{ext}$ is appended to the list $\tilde{\mathcal{F}}^\text{eq}$ of external force vectors corresponding to previously computed equilibrium states: $\tilde{\mathcal{F}}^\text{eq} \leftarrow \tilde{\mathcal{F}}^\text{eq} + [\tilde{\bm{F}}^\text{ext}]$.
    \item The radius is updated: $r \leftarrow \min(2r, r_0)$.
    \item The next action is decided based on whether the final load has been reached:
    \begin{equation}
        \begin{cases}
        \text{stop (load step has successfully finished)}&\text{if}\quad \lambda  \ge ||\tilde{\bm{F}}^\text{step}||,\\
        ~&~\\
        \text{start new increment}&\text{else.}
        \end{cases}
        \nonumber
    \end{equation}
\end{enumerate}
\item An increment restart is prepared (only after the correction phase fails).
\begin{enumerate}
    \item The array of coordinates $\bm{Q}$ is reset to the last item in the list $\mathcal{Q}^\text{eq}$ of previously computed equilibrium states: $\bm{Q} \leftarrow \mathcal{Q}^\text{eq}[\text{last}]$.
    \item The external force vector is reset the last item in the list $\tilde{\bm{F}}^\text{ext}$ of external force vectors corresponding to the previously computed equilibrium states : $\tilde{\bm{F}}^\text{ext} \leftarrow \tilde{\mathcal{F}}^\text{eq}[\text{last}]$.
     \item The current hessian matrix $\bm{K}$ is computed based on the coordinates $\bm{Q}$ (Eq.~(\ref{eq:stiffness_matrix_entire_structure})), the current reduced hessian matrix $\tilde{\bm{K}}$ is formed by removing the rows and columns corresponding to the fixed coordinates.
    \item The current radius is reduced: $r \leftarrow r / 2$.
    \item A new increment is started.
     
\end{enumerate}
\end{enumerate}
\end{enumerate}

\newpage
\section{Geometric measures, gradients and hessians}
\label{section:geometric_measures}

\begin{figure}[!ht]
    \centering
    \includegraphics{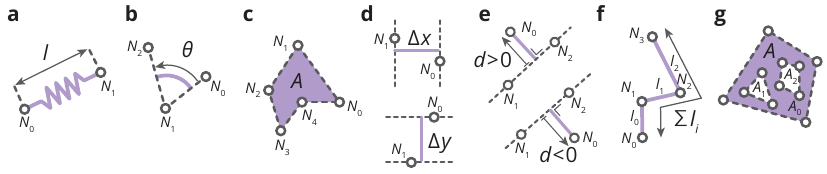}
    \caption{Geometric measures. (a) length, (b) angle, (c) area, (d) signed $x$- (top) and signed $y$- (bottom) distances, (e) signed point-line distance, which can be positive (top) or negative (bottom), (f) path length, (g) area with holes.}
    \label{fig:geometric_measures}
\end{figure}

When computing the gradient and hessian of the elastic energy, necessary for the arc-length method, the measure of the flexel (length, angle, area, etc) and its derivatives (gradient, hessian) with respect to the spatial coordinates $z_i$ must be computed. In this section, the equations used to calculate those quantities are derived. An overview of the different geometric measure that can be used in \springable{} is shown in Fig.~\ref{fig:geometric_measures}.\\

Notation:
\begin{itemize}
    \item $\alpha$: the geometric measure,
    \item $x_k$: horizontal coordinate of node $k$,
    \item $y_k$: vertical coordinate of node $k$,
    \item $N_k$: node $k$,
    \item $\overrightarrow{N_kN_l}$: vector going from node $N_k$ to node $N_l$,
    \item $\bm{z}$: array of spatial nodal coordinates,
    \item $z_i$: $(i+1)$th component in the array $\bm{z}$ (so, $z_0$ is the first component).
\end{itemize}

\subsection{Length (longitudinal flexel)}
\label{subsection:length}
A longitudinal flexel is defined by two nodes, node~$N_0$ and node~$N_1$ (Fig.~\ref{fig:geometric_measures}a). The measure $\alpha$ of a longitudinal flexel is the length $l$ of the segment connecting the two nodes. In 2d, this length depends on the four coordinates $\bm{z}=[x_0, y_0, x_1, y_1]$.

\paragraph{Measure $\alpha$}
\begin{equation}
    \label{eq:length}
    \alpha = l(\bm{z}) = \sqrt{(x_0-x_1)^2 + (y_0-y_1)^2}\quad.
\end{equation}

\paragraph{Gradient $\partial \alpha / \partial \bm {z}$}
\begin{align}
    \label{eq:length_gradient}
    \dfrac{\partial \alpha}{\partial \bm{z}} &=
    \begin{bmatrix}\dfrac{\partial \alpha}{\partial z_i}\end{bmatrix}
    =
    \begin{bmatrix}
    \dfrac{\partial l}{\partial t_0}&
    \dfrac{\partial l}{\partial y_0}&
    \dfrac{\partial l}{\partial t_1}&
    \dfrac{\partial l}{\partial y_1}
    \end{bmatrix}^\top
    =\dfrac{1}{l}\begin{bmatrix}
                    x_0-x_1\\
                    y_0-y_1\\
                    x_1-x_0\\
                    y_1-y_0\\
                    \end{bmatrix}.
\end{align}
\paragraph{Hessian $\partial^2 \alpha / \partial \bm{z} \partial \bm{z}^\top$}
\begin{align}
    \label{eq:length_hessian}
    \dfrac{\partial^2 \alpha}{\partial \bm{z} \partial \bm{z}^\top} &=
    \begin{bmatrix}
    \dfrac{\partial^2 \alpha}{\partial z_i \partial z_j}
    \end{bmatrix}\nonumber\\
    &= \dfrac{1}{l}\begin{bmatrix}
                    1&0&-1&0\\
                    0&1&0&-1\\
                    -1&0&1&0\\
                    0&-1&0&1
                    \end{bmatrix}-\dfrac{1}{l^2}\dfrac{1}{l}\begin{bmatrix}
                    x_0-x_1\\
                    y_0-y_1\\
                    x_1-x_0\\
                    y_1-y_0\\
                    \end{bmatrix}
                    \begin{bmatrix}
                    x_0-x_1\\
                    y_0-y_1\\
                    x_1-x_0\\
                    y_1-y_0\\
                    \end{bmatrix}^\top\nonumber\\
                    &=\dfrac{1}{l}\begin{bmatrix}
                    1&0&-1&0\\
                    0&1&0&-1\\
                    -1&0&1&0\\
                    0&-1&0&1
                    \end{bmatrix}
                    -\dfrac{1}{l^3}\begin{bmatrix}
                    \Delta x \Delta x&\Delta x \Delta y&-\Delta x \Delta x&-\Delta x \Delta y\\
                    \Delta x \Delta y&\Delta y \Delta y&-\Delta x \Delta y&-\Delta y \Delta y\\
                    -\Delta x \Delta x&-\Delta x \Delta y&\Delta x \Delta x&\Delta x \Delta y\\
                    -\Delta x \Delta y&-\Delta y \Delta y&\Delta x \Delta y&\Delta y \Delta y
                    \end{bmatrix},
\end{align}
where $\Delta x:= x_0-x_1$ and $\Delta y:= y_0-y_1$.

\subsection{Angle (angular flexel)}
\label{subsection:angle}
An angular flexel is defined by three nodes: node~$N_0$, node~$N_1$ and node~$N_2$ (Fig.~\ref{fig:geometric_measures}b). The measure $\alpha$ of an angular flexel is the angle $\theta$ defined by the three nodes, where node~$N_1$ is the vertex. More precisely, the angle $\theta \in [0, 2\pi[$ is the angle by which the vector $\overrightarrow{N_1 N_0}$ must rotate counter-clockwise about $N_1$ to align with the vector $\overrightarrow{N_1 N_2}$. This angle depends on the six coordinates $\bm{z}=[x_0, y_0, x_1, y_1, x_2, y_2]^\top$.

\paragraph{Measure $\alpha$}
\begin{equation}
    \alpha = \theta(\bm{z}) = \atantwo{\left(Y(\bm{z}) , X(\bm{z})\right)} \mod 2\pi,
\end{equation}
where
\begin{equation}
    Y(\bm{z}) = (x_1 - x_0) (y_1 - y_2) - (y_1 - y_0) (x_1 - x_2),
\end{equation}
and
\begin{equation}
    X(\bm{z}) = (x_1 - x_0) (x_1 - x_2) + (y_1 - y_0) (y_1 - y_2).
\end{equation}

\paragraph{Gradient $\partial \alpha / \partial \bm {z}$}
\begin{align}
    \dfrac{\partial \alpha}{\partial \bm{z}} &=
    \begin{bmatrix}\dfrac{\partial \alpha}{\partial z_i}\end{bmatrix}
    =\begin{bmatrix}\dfrac{\partial \theta}{\partial z_i}\end{bmatrix} =
    \begin{bmatrix}\dfrac{\partial \theta}{\partial Y}\dfrac{\partial Y}{\partial z_i} + \dfrac{\partial \theta}{\partial X}\dfrac{\partial X}{\partial z_i}\end{bmatrix}
    =\dfrac{\partial \theta}{\partial Y}
    \begin{bmatrix}
        \dfrac{\partial Y}{\partial z_i}
    \end{bmatrix}+
    \dfrac{\partial \theta}{\partial X}
    \begin{bmatrix}
        \dfrac{\partial X}{\partial z_i}
    \end{bmatrix},
\end{align}
where
\begin{equation}
    \begin{bmatrix}
        \dfrac{\partial Y}{\partial z_i}
    \end{bmatrix}=
    \begin{bmatrix}
        y_2-y_1\\
        x_1-x_2\\
        y_0-y_2\\
        x_2-x_0\\
        y_1-y_0\\
        x_0-x_1
    \end{bmatrix},
\end{equation} 
\begin{equation}
    \begin{bmatrix}
        \dfrac{\partial X}{\partial z_i}
    \end{bmatrix}=
        \begin{bmatrix}
        x_2-x_1\\
        y_2-y_1\\
        2x_1-x_0-x_2\\
        2y_1-y_0-y_2\\
        x_0-x_1\\
        y_0-y_1
    \end{bmatrix},
\end{equation}
\begin{equation}
\dfrac{\partial \theta}{\partial Y} = \dfrac{X}{ X^2 + Y^2}\quad,
\end{equation}
and
\begin{equation}
\dfrac{\partial \theta}{\partial X} = -\dfrac{Y}{ X^2 + Y^2}\quad.
\end{equation}
\paragraph{Hessian $\partial^2 \alpha / \partial \bm{z} \partial \bm{z}^\top$}
\begin{align}
    \dfrac{\partial^2 \alpha}{\partial \bm{z} \partial \bm{z}^\top} =&
    \begin{bmatrix}
    \dfrac{\partial^2 \alpha}{\partial z_i \partial z_j}
    \end{bmatrix} \nonumber \\
    =&
\dfrac{\partial \theta}{\partial Y}\left[\dfrac{\partial^2 Y}{\partial z_i \partial z_j}\right]
+ \dfrac{\partial \theta}{\partial X}\left[\dfrac{\partial^2 X}{\partial z_i \partial z_j}\right]\nonumber\\
&+ \dfrac{\partial^2 \theta}{\partial Y^2}\left[\dfrac{\partial Y}{\partial z_i} \dfrac{\partial Y}{\partial z_j}\right]
+ \dfrac{\partial^2 \theta}{\partial Y \partial X}\left[\dfrac{\partial Y}{\partial z_i}\dfrac{\partial X}{\partial z_j} + \dfrac{\partial X}{\partial z_i}\dfrac{\partial Y}{\partial z_j}
\right] + \dfrac{\partial^2 \theta}{\partial X^2}\left[\dfrac{\partial X}{\partial z_i} \dfrac{\partial X}{\partial z_j}\right],
\end{align}
where
\begin{equation}
    \left[\dfrac{\partial^2 Y}{\partial z_i \partial z_j}\right]=
    \begin{bmatrix}
        0&0&0&-1&0&1\\
        0&0&1&0&-1&0\\
        0&1&0&0&0&-1\\
        -1&0&0&0&1&0\\
        0&-1&0&1&0&0\\
        1&0&-1&0&0&0
    \end{bmatrix},
\end{equation}
\begin{equation}
    \left[\dfrac{\partial^2 X}{\partial z_i \partial z_j}\right]=
    \begin{bmatrix}
        0&0&-1&0&1&0\\
        0&0&0&-1&0&1\\
        -1&0&2&0&-1&0\\
        0&-1&0&2&0&-1\\
        1&0&-1&0&0&0\\
        0&1&0&-1&0&0
    \end{bmatrix},
\end{equation}
\begin{equation}
\dfrac{\partial^2 \theta}{\partial Y^2} = -\dfrac{2XY}{\left(X^2 +Y^2\right)^2}\quad,
\end{equation}
\begin{equation}
\dfrac{\partial^2 \theta}{\partial Y \partial X} = \dfrac{Y^2-X^2}{\left(X^2 +Y^2\right)^2}\quad,
\end{equation}
and
\begin{equation}
\dfrac{\partial^2 \theta}{\partial X^2} = \dfrac{2XY}{\left(X^2 +Y^2\right)^2}\quad.
\end{equation}

\subsection{Signed area}
\label{subsection:signed_area}
Currently, the measure `signed area' is not directly used as the measure of a flexel. Yet it is used indirectly to define area flexels and `signed distance' flexels (see subsection~\ref{subsection:area} and \ref{subsection:signed_distance}).
A signed area $A_\text{s}$ is defined by $n \ge 3$ nodes: $N_k$, $k=0,1,\dots, n-1$, which are the vertices forming the boundary of a simple polygon. If the sequence $N_k$ are the vertices listed sequentially counter-clockwise, the `signed area' is the (positive) area of the polygon. Conversely, if the sequence $N_k$ are the vertices listed sequentially clockwise, the `signed area' is the negative of the (positive) area of the polygon. The polygon is assumed to be not self-intersecting. The signed area depends on $2n$ coordinates $\bm{z}=[x_0, y_0, \dots, x_{n-1}, y_{n-1}]^\top$

\paragraph{Measure $\alpha$}
\begin{equation}
    \label{eq:signed_area}
    \alpha = A_\text{s}(\bm{z}) = \dfrac{1}{2} \sum_{k=0}^{n-1} \left( x_k y_{\nextt{k}} -  y_kx_{\nextt{k}} \right)
\end{equation}
where
$\nextt{k} =  (k + 1) \mod n$.

\paragraph{Gradient $\partial \alpha / \partial \bm {z}$}
\begin{align}
\label{eq:signed_area_gradient}
\dfrac{\partial \alpha}{\partial \bm{z}} &=
    \begin{bmatrix}\dfrac{\partial \alpha}{\partial z_i}\end{bmatrix}=\left[\dfrac{\partial A_\text{s}}{\partial z_i}\right]=\dfrac{1}{2}
    \begin{bmatrix}
        S_i
    \end{bmatrix},
\end{align}
where
\begin{equation}
    \label{eq:signed_area_Si}
    S_i = \begin{cases}
            y_{\nextt{k}} - y_{\prev{k}}&\text{if $i$ is even $\Leftrightarrow z_i$ is a $x$-coordinate}\\
            \text{with $k=i/2$}&\\
            ~&~\\
            x_{\prev{k}} - x_{\nextt{k}}&\text{if $i$ is odd $\Leftrightarrow z_i$ is a $y$-coordinate}\\
            \text{with $k=i/2 - 1/2$},&
        \end{cases}
\end{equation}
with $\prev{k} =  (k - 1) \mod n$. Put differently,
\begin{equation}
    \left[S_i\right] =
    \begin{bmatrix}
        y_{1} - y_{n-1}\\
        x_{n-1} - x_{1}\\
        y_{2} - y_{0}\\
        x_{0} - x_{2}\\
        \vdots\\
        y_0 - y_{n-2}\\
        x_{n-2} - x_{0}\\
    \end{bmatrix}.
\end{equation}

\begin{figure}[ht]
    \centering
    \includegraphics{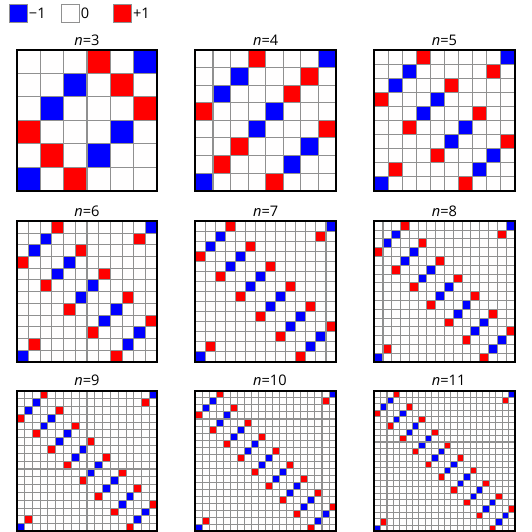}
    \caption{$\left[T_{ij}\right]$ matrices for $n=3$ to 11.}
    \label{fig:area_hessian_matricies}
\end{figure}

\paragraph{Hessian $\partial^2 \alpha / \partial \bm{z} \partial \bm{z}^\top$}
\begin{align}
    \label{eq:signed_area_hessian}
    \dfrac{\partial^2 \alpha}{\partial \bm{z} \partial \bm{z}^\top} &=
    \begin{bmatrix}
    \dfrac{\partial^2 \alpha}{\partial z_i \partial z_j}
    \end{bmatrix}
    =\dfrac{1}{2}
    \begin{bmatrix}
        T_{ij}
    \end{bmatrix},
\end{align}
where
\begin{equation}
    \label{eq:signed_area_Tij}
    T_{ij}=
    \begin{cases}
     +1&\text{if $i$ is even and $j=(i+3) \mod 2n$}\\
    -1&\text{if $i$ is even and $j=(i-1) \mod 2n$}\\
    -1&\text{if $i$ is odd and $j=(i+1) \mod 2n$}\\
     +1&\text{if $i$ is odd and $j=(i-3) \mod 2n$}\\
    0&\text{else}.
    \end{cases}
\end{equation} Matrices $\left[T_{ij}\right]$ for $n \in [3, 11]$ are shown in Fig.~\ref{fig:area_hessian_matricies}.

\subsection{Area (area flexel)}
\label{subsection:area}
An area flexel is defined by $n\ge3$ nodes: $N_0, N_1, \dots, N_{n-1}$ (Fig.~\ref{fig:geometric_measures}c). The measure $\alpha$ of an area flexel is the area $A\ge0$ of the polygon whose vertices is the sequence of nodes $N_k$. The area $A$ is always positive regardless of whether the vertices are sequentially listed clockwise or counter-clockwise. The area depends on $2n$ coordinates $\bm{z}=[x_0, y_0, \dots, x_{n-1}, y_{n-1}]$.

\paragraph{Measure $\alpha$}
\begin{equation}
    \label{eq:area}
    \alpha=A(\bm{z})=\left|A_\text{s}(\bm{z})\right|,
\end{equation}
where $A_\text{s}$ is the `signed area' measure, defined in Eq.~(\ref{eq:signed_area}).

\paragraph{Gradient $\partial \alpha / \partial \bm {z}$}
\begin{align}
    \label{eq:area_gradient}
     \dfrac{\partial \alpha}{\partial \bm{z}} &=
    \begin{bmatrix}\dfrac{\partial \alpha}{\partial z_i}\end{bmatrix}=\sgn{A_\text{s}}\left[\dfrac{\partial A_\text{s}}{\partial z_i}\right]=\dfrac{1}{2}\sgn{A_\text{s}}\left[S_i\right],
\end{align}
where $\left[S_i\right]$ is the array defined in Eq.~(\ref{eq:signed_area_Si}).

\paragraph{Hessian $\partial^2 \alpha / \partial \bm{z} \partial \bm{z}^\top$}
\begin{align}
    \label{eq:area_hessian}
    \dfrac{\partial^2 \alpha}{\partial \bm{z} \partial \bm{z}^\top} &=
    \begin{bmatrix}
    \dfrac{\partial^2 \alpha}{\partial z_i \partial z_j}
    \end{bmatrix}=\sgn{A_\text{s}}\left[\dfrac{\partial^2 A_\text{s}}{\partial z_i \partial z_j}\right]
    =\dfrac{1}{2}\sgn{A_\text{s}}
    \begin{bmatrix}
        T_{ij}
    \end{bmatrix},
\end{align}
where $\left[T_{ij}\right]$ is the matrix defined in Eq.~(\ref{eq:signed_area_Tij}).

\subsection{Signed $x$-distance and $y$-distance ($x$-distance flexel and $y$-distance flexel)}
\label{subsection:signed_xy_distance}
An $x$ (resp. $y$) -distance flexel is defined by two nodes $N_0$ and $N_1$ (Fig.~\ref{fig:geometric_measures}d). The measure of such flexel is the signed horizontal (resp. vertical) distance between both nodes. If $N_0$ is on the right of (resp. above) $N_1$, the signed distance is positive, else negative. The signed distance depends on four coordinates $\bm{z}=[x_0, y_0, x_1, y_1]^\top$.

\paragraph{Measure $\alpha$}
\begin{equation}
\alpha = \Delta x(\bm{z}) =x_0 - x_1\quad\text{(resp. $\quad \alpha=\Delta y(\bm{z})=y_0-y_1$)}.
\end{equation}

\paragraph{Gradient $\partial \alpha / \partial \bm{z}$}
\begin{equation}
\dfrac{\partial \alpha}{\partial \bm{z}}=[1, 0, -1, 0]^\top\quad\text{(resp. $\quad \dfrac{\partial \alpha}{\partial \bm{z}}=[0, 1, 0, -1]^\top$)}.
\end{equation}

\paragraph{Hessian $\partial^2 \alpha / \partial \bm{z} \partial \bm{z}^\top$}
\begin{equation}
\dfrac{\partial^2 \alpha}{\partial \bm{z} \partial \bm{z}^\top}=\bm{0}^{4\times 4}.
\end{equation}

\subsection{Compound measure}
\label{subsection:compound_measure}
A flexel's measure $\alpha$ can be expressed as a function $C$ of $p$ other measures $\alpha_k$, $k=0, 1, \dots, p-1$. Each measure $\alpha_k$ itself depends on $m_k$ coordinates $\bm{z}^k$. In general, it is possible that measures $\alpha_k$ share some of their coordinates; therefore the number $m$ of coordinates that the measure $\alpha$ depends on is less than $\sum_{k=0}^{p-1} m_k$. The array $\bm{z}$ gathers the set of unique coordinates collected across each $\bm{z}^k$.


\paragraph{Measure $\alpha$}
\begin{equation}
    \alpha=C(\alpha_0, \alpha_1, \dots, \alpha_{p-1}).
\end{equation}

\paragraph{Gradient $\partial \alpha / \partial \bm {z}$}
\begin{align}
    \label{eq:compound_gradient}
     \dfrac{\partial \alpha}{\partial \bm{z}} &=\left[\dfrac{\partial \alpha}{ \partial z_i}\right]=\sum_{k=0}^{p-1}\dfrac{\partial C}{\partial \alpha_k}\left[\dfrac{\partial \alpha_k(\bm{z}^k)}{\partial z_i}\right]=\sum_{k=0}^{p-1}\sum_{r=0}^{m_k-1}\dfrac{\partial C}{\partial \alpha_k}\dfrac{\partial \alpha_k}{\partial z_r^k}\left[\dfrac{\partial z_r^k}{\partial z_i}\right]\nonumber\\
     &=\sum_{k=0}^{p-1}\sum_{r=0}^{m_k-1}\dfrac{\partial C}{\partial \alpha_k}\dfrac{\partial \alpha_k}{\partial z_r^k}\left[L^k_{ir}\right],
\end{align}
where
$\left[\partial \alpha_k / \partial z^k_{r}\right]=\partial \alpha_k / \partial \bm{z}^k$ is the gradient of the measure $\alpha_k$ in its `own coordinates' $\bm{z}^k$,
and
\begin{equation}
\label{eq:Lirk}
L_{ir}^k=
\begin{cases}
   1&\text{if $z_i=z^k_r$}\\
   0&\text{else.}
\end{cases}
\end{equation}
In practical terms, Eq.~(\ref{eq:compound_gradient}) means that the gradient $\partial \alpha / \partial \bm{z}=[\partial \alpha / \partial z_i]$ of a compound measure is constructed by summing together the components of each individual measure's gradient $\partial \alpha_k / \partial \bm{z}^k$ that correspond to the same coordinate $z_i$, weighted by $\partial C_k / \partial \alpha_k$.

\paragraph{Hessian $\partial^2 \alpha / \partial \bm{z} \partial \bm{z}^\top$}

\begin{align}
    \label{eq:compound_hessian}
    \dfrac{\partial^2 \alpha}{\partial \bm{z} \partial \bm{z}^\top} &=
    \begin{bmatrix}
    \dfrac{\partial^2 \alpha}{\partial z_i \partial z_j}
    \end{bmatrix}=\sum_{k=0}^{p-1}\sum_{l=0}^{p-1}\dfrac{\partial^2 C}{\partial \alpha_k \partial \alpha_l}\left[\dfrac{\partial \alpha_k(\bm{z}^k)}{\partial z_i}\dfrac{\partial \alpha_l(\bm{z}^l)}{\partial z_j}\right] +\sum_{k=0}^{p-1} \dfrac{\partial C}{\partial \alpha_k} \dfrac{\partial^2\alpha_k(\bm{z}^k)}{\partial z_i \partial z_j}\nonumber\\
    &=\sum_{k=0}^{p-1}\sum_{l=0}^{p-1}\sum_{r=0}^{m_k-1}\sum_{s=0}^{m_l-1}\dfrac{\partial^2 C}{\partial \alpha_k \partial \alpha_l}\dfrac{\partial \alpha_k}{\partial z^k_r}\dfrac{\partial \alpha_l}{\partial z^l_s}\left[L^k_{ir}L^l_{js}\right] +\sum_{k=0}^{p-1}\sum_{r=0}^{m_k-1}\sum_{s=0}^{m_k-1} \dfrac{\partial C}{\partial \alpha_k} \dfrac{\partial^2\alpha_k}{\partial z^k_r \partial z^k_s}\left[L^k_{ir}L^k_{js}\right],
\end{align}
where $\left[\partial \alpha_k / \partial z^k_{r}\right]=\partial \alpha_k / \partial \bm{z}^k$ and $\left[\partial \alpha_l / \partial q^l_{s}\right]=\partial \alpha_l / \partial \bm{z}^l$ are the gradients of the individual measures in their own coordinates; $\left[\partial^2\alpha_k/\partial z^k_r \partial z^k_s\right]=\partial^2\alpha_k/\partial \bm{z}^k\partial {\bm{z}^k}^\top$ are the hessian matrices of the individual measures in their own coordinates; and where $\left[L^k_{ir}L^l_{js}\right]$ and $\left[L^k_{ir}L^k_{js}\right]$ are computed using Eq.~(\ref{eq:Lirk}). In practical terms, Eq.~(\ref{eq:compound_hessian}) means that the hessian matrix $\partial^2 \alpha / \partial \bm{z} \partial \bm{z}^\top$ is constructed by summing together the components
\begin{itemize}
    \item of the outer-product matrices $\partial \alpha_k / \partial \bm{z}^k \otimes \partial \alpha_l / \partial \bm{z}^l$ for every possible pair of individual gradients (weighted by $\partial^2 C / \partial \alpha_k \partial \alpha_l)$; and
    \item of the individual hessian matrices $\partial^2 \alpha_k / \partial \bm{z}^k \partial {\bm{z}^k}^\top$ (weighted by $\partial C / \partial \alpha_k$)
\end{itemize} that correspond to the same pair of coordinates $(z_i, z_j)$.

\subsection{Signed point-line distance (distance flexel)}
\label{subsection:signed_distance}
 A distance flexel is defined by 3 nodes: $N_0$, $N_1$ and $N_2$ (Fig.~\ref{fig:geometric_measures}e). The measure $\alpha$ of a distance flexel is the `signed distance' $d$ between $N_0$ and the infinite line passing through $N_1$ and $N_2$. If $N_0$ is on the left of that line (whose direction is defined by the vector $\overrightarrow{N_1N_2}$), the measure is the (positive) distance. If $N_0$ is on the right of the line, the measure is the negative of the (positive) distance. The `signed distance' depends on the six coordinates $\bm{z}=[x_0, y_0, x_1, y_1, x_2, y_2]^\top$.

\paragraph{Measure $\alpha$}~\\
The `signed distance' measure $\alpha$ can be regarded as a compound measure: it is twice the `signed area' of the triangle $N_0N_1N_2$ ($\alpha_0$) divided by the length of the segment $N_1N_2$ ($\alpha_1$):

\begin{equation}
    \alpha=C(\alpha_0, \alpha_1)=\dfrac{2\alpha_0}{\alpha_1}=\dfrac{2A_\text{s}(\bm{z}^0)}{l(\bm{z}^1)}=d(\bm{z}),
\end{equation}
where $A_\text{s}(\bm{z}^0)$ is the `signed area' of the triangle $N_0N_1N_2$ (Eq.~(\ref{eq:signed_area})), with $\bm{z}^0=[x_0, y_0, x_1, y_1, x_2, y_2]^\top$; and  $l(\bm{z}^1)$ is the length of the segment $N_1N_2$ (Eq.~(\ref{eq:length})), with $\bm{z}^1=[x_1, y_1, x_2, y_2]^\top$.

\paragraph{Gradient $\partial \alpha / \partial \bm {z}$}
As the signed distance measure is a compound measure, its gradient $\partial \alpha / \partial \bm {z}$ is computed using Eq.~(\ref{eq:compound_gradient}), with $\partial C / \partial \alpha_0=2/\alpha_1$ and $\partial C / \partial \alpha_1=-2\alpha_0/\alpha_1^2$, and where the individual gradients $\partial \alpha_0 / \partial \bm{z}^0$ and $\partial \alpha_1 / \partial \bm{z}^1$ are computed using Eq.~(\ref{eq:signed_area_gradient}) and Eq.~(\ref{eq:length_gradient}), respectively.

\paragraph{Hessian $\partial^2 \alpha / \partial \bm{z} \partial \bm{z}^\top$}
As the signed distance measure is a compound measure, its hessian matrix $\partial^2 \alpha / \partial \bm{z} \partial \bm{z}^\top$ is computed using Eq.~(\ref{eq:compound_hessian}), with $\partial C / \partial \alpha_0=2/\alpha_1$, $\partial C / \partial \alpha_1=-2\alpha_0/\alpha_1^2$, $\partial^2 C / \partial \alpha_0^2=0$, $\partial^2 C / \partial \alpha_0 \partial \alpha_1=\partial^2 C / \partial \alpha_1 \partial \alpha_0=-2/\alpha_1^2$ and $\partial^2 C / \partial \alpha_1^2=4\alpha_0/\alpha_1^3$, where the individual gradients and hessian matrices $\partial \alpha_0 / \partial \bm{z}^0$, $\partial \alpha_1 / \partial \bm{z}^1$, $\partial^2 \alpha_0 / \partial \bm{z}^0 {\partial \bm{z}^0}^\top$, $\partial^2 \alpha_1 / \partial \bm{z}^1 {\partial \bm{z}^1}^\top$ are computed using Eq.~(\ref{eq:signed_area_gradient}), Eq.~(\ref{eq:length_gradient}), Eq.~(\ref{eq:signed_area_hessian}) and Eq.~(\ref{eq:length_hessian}), respectively.

\subsection{Path length (path flexel)}
\label{subsection:path_length}
A path flexel is defined by an ordered sequence of $p+1$ nodes \{$N^k$\} selected from a set of $n\ge2$ distinct nodes $N_0, N_1, \dots, N_{n-1}$ (Fig.~\ref{fig:geometric_measures}f). The measure of $\alpha$ of a path flexel is the length of the polygonal chain whose vertices are defined by the sequence of nodes \{$N^k$\}. The path length depends on $2n$ coordinates $\bm{z}=[x_0, y_0, \dots, x_{n-1}, y_{n-1}]^\top$.

\paragraph{Measure $\alpha$}~\\
The path length measure $\alpha$ can be regarded as a compound measure: it is the sum of the lengths of the $p$ segments $N^kN^{k+1}$:
\begin{equation}
    \alpha=C(\alpha_0, \alpha_1, \dots, \alpha_{p-1})=\alpha_0+\alpha_1+\dots+\alpha_{p-1}=\sum_{k=0}^{p-1} l_k(\bm{z}^k),
\end{equation}
where $\alpha_k=l_k$ is the length of the segment $N^kN^{k+1}$ (Eq.~(\ref{eq:length})) and $\bm{z}^k=[x_k, y_k, x_{k+1}, y_{k+1}]^\top$.

\paragraph{Gradient $\partial \alpha / \partial \bm {z}$}
As the path length measure is a compound measure, its gradient $\partial \alpha / \partial \bm {z}$ is computed using Eq.~(\ref{eq:compound_gradient}), with $\partial C / \partial \alpha_k=1$, and where the individual gradients $\partial \alpha_k / \partial \bm{z}^k$ are computed using Eq.~(\ref{eq:length_gradient}). In case that the sequence is composed of distinct nodes ($N^k=N_k$), the gradient $\partial \alpha / \partial \bm {z}$ takes the following form:
\begin{equation}
\dfrac{\partial \alpha}{\partial \bm{z}} =\dfrac{1}{l_0}
\begin{bmatrix}
x_0-x_1\\
y_0-y_1\\
x_1-x_0\\
y_1-y_0\\
0\\
\vdots\\
~\\
~\\
~\\
~\\
0\end{bmatrix}
+
\dfrac{1}{l_1}
\begin{bmatrix}
0\\
0\\
x_1-x_2\\
y_1-y_2\\
x_2-x_1\\
y_2-y_1\\
0\\
\vdots\\
~\\
~\\
0\end{bmatrix}
+\dots
+
\dfrac{1}{l_{n-2}}
\begin{bmatrix}
0\\
\vdots\\
~\\
~\\
~\\
~\\
0\\
x_{n-2}-x_{n-1}\\
y_{n-2}-y_{n-1}\\
x_{n-1}-x_{n-2}\\
y_{n-1}-y_{n-2}
\end{bmatrix}.
\end{equation}

\paragraph{Hessian $\partial^2 \alpha / \partial \bm{z} \partial \bm{z}^\top$}
As the path length measure is a compound measure, its hessian matrix $\partial^2 \alpha / \partial \bm{z} \partial \bm{z}^\top$ is computed using Eq.~(\ref{eq:compound_hessian}), with $\partial C / \partial \alpha_k=1$ and $\partial^2 C / \partial \alpha_k \partial \alpha_l=0$, where the individual gradients and hessian matrices $\partial \alpha_k / \partial \bm{z}^k$ and $\partial^2 \alpha_k / \partial \bm{z}^k {\partial \bm{z}^k}^\top$ are computed using Eq.~(\ref{eq:length_gradient}) and Eq.~(\ref{eq:length_hessian}), respectively.

\subsection{Area with holes (area flexel)}
\label{subsection:area_with_holes}
In subsection~\ref{subsection:area}, the geometric measure measure of an area flexel was defined as the area of a simple polygon. An area flexel can also be defined by a polygon with holes, that is, a polygon with an outer boundary and one or multiple inner boundaries (Fig.~\ref{fig:geometric_measures}g). The geometric measure $\alpha$ of such polygon is the area delimited by the outer boundary $A_0$ to which is subtracted the areas delimited by the inner boundaries $A_1, A_2, \dots, A_{p-1}$. Each area $A_k$ is defined by a set of coordinates, noted $\bm{z}^k$. The array of unique coordinates $\bm{z}$ is constructed by sequentially appending each coordinate from $\bm{z}^0, \dots,\bm{z}^{p-1}$ that has not yet been encountered.

\paragraph{Measure $\alpha$}
The area-with-holes measure $\alpha$ can be regarded as a compound measure: it is outer area minus the sum of the inners areas:
\begin{equation}
    \alpha=C(\alpha_0, \alpha_1, \dots, \alpha_{p-1})=A_0 - \sum_{k=1}^{p-1} A_k(\bm{z}^k)=A(\bm{z}).
\end{equation}
where $\alpha_k=A_k(\bm{z}^k)$ is computed using Eq.~(\ref{eq:area}).

\paragraph{Gradient $\partial \alpha / \partial \bm {z}$}
As the area-with-holes measure is a compound measure, its gradient $\partial \alpha / \partial \bm {z}$ is computed using Eq.~(\ref{eq:compound_gradient}), with $\partial C / \partial \alpha_0=1$ and $\partial C / \partial \alpha_k=-1\quad(\forall k \ne 0)$, and where the individual gradients $\partial \alpha_k / \partial \bm{z}^k$ are computed using Eq.~(\ref{eq:area_gradient}). 

\paragraph{Hessian $\partial^2 \alpha / \partial \bm{z} \partial \bm{z}^\top$}
As the area-with-holes measure is a compound measure, its hessian matrix $\partial^2 \alpha / \partial \bm{z} \partial \bm{z}^\top$ is computed using Eq.~(\ref{eq:compound_hessian}), with $\partial C / \partial \alpha_0=1$, $\partial C / \partial \alpha_k=-1\quad(\forall k \ne 0)$ and $\partial^2 C / \partial \alpha_k \partial \alpha_l=0$, where the individual gradients and hessian matrices $\partial \alpha_k / \partial \bm{z}^k$ and $\partial^2 \alpha_k / \partial \bm{z}^k {\partial \bm{z}^k}^\top$ are computed using Eq.~(\ref{eq:area_gradient}) and Eq.~(\ref{eq:area_hessian}), respectively.

\section{Intrinsic nonlinear mechanical behaviors}
\label{section:intrinsic_mechanical_behaviors}
\begin{figure}[ht]
    \centering
    \includegraphics[width=\textwidth]{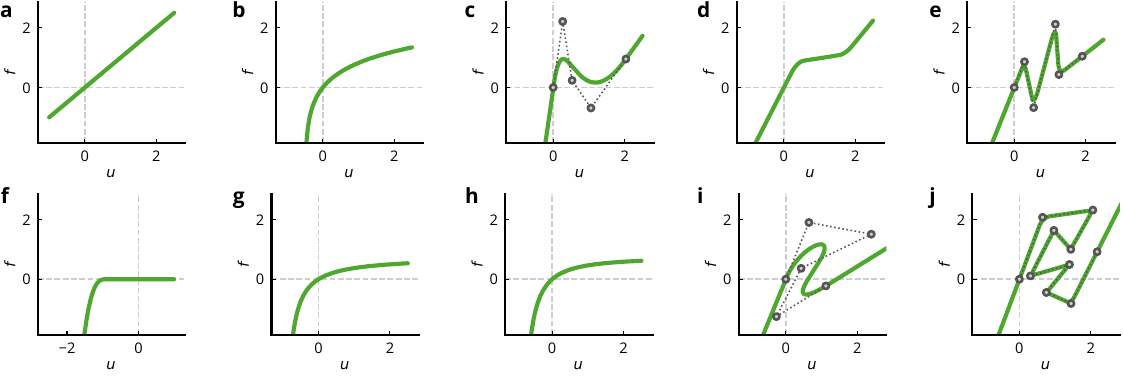}
    \caption{Nonlinear generalized force ($f$)-displacement ($u$) curves used to construct nonlinear behaviors. (a) linear, (b) logarithmic, (c) Bezier (univariate) (d) $\mathcal{C}_1$-piecewise linear, (e) zigzag (univariate), (f) contact, (g) isothermal, (h) isentropic, (i) Bezier (bivariate), (j) zigzag (bivariate). Control points and control polygons are depicted by gray markers and dotted lines.}
    \label{fig:nonlinear_behaviors}
\end{figure}

When computing the gradient and hessian of the elastic energy, necessary for the arc-length method, the gradient and hessian of the intrinsic elastic potential $v$ of the flexel with respect to the measure $\alpha$ (and the internal degree of freedom $x$ for bivariate potentials) must be computed. In this section, the equations used to calculate those quantities are derived for various kinds of potentials. In \springable, potentials are specified by providing the generalized force-displacement curve, where the generalized force $f$ is defined as $\partial v / \partial \alpha $, and the generalized displacement $u$ as $\alpha - \alpha_0$ ($\alpha_0$ is a constant representing the natural measure). An overview of the different types of nonlinear generalized force-displacement curves implemented in \springable{} is shown in Fig.~\ref{fig:nonlinear_behaviors}. In the following, we use apostrophes, $g'$ and $g''$, to denote the first and second derivatives of functions $g$ of one single argument.

\subsection{Univariate behaviors}
\label{subsection:univariate_behaviors}
A univariate behavior is characterized by a univariate energy potential, which is constructed from a generalized force-displacement curve that can be described as function: $f(u)$.

\begin{figure}[ht]
    \centering
    \includegraphics{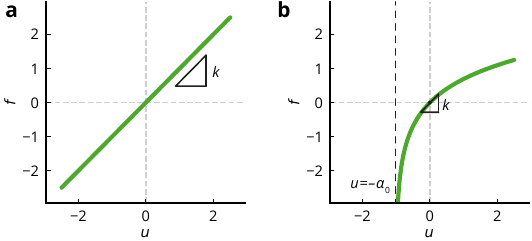}
    \caption{Linear (a) and logarithmic (b) force-displacement curves.}
    \label{fig:linear_loga}
\end{figure}

\paragraph{Linear behavior}
A linear behavior is defined by a linear generalized force-displacement curve (Fig.~\ref{fig:linear_loga}a),
\begin{equation}
\label{eq:linear_behavior}
    f(u) = ku,
\end{equation}
where $k$ is the spring constant, that is, the slope of the generalized force-displacement curve, whose unit is the unit of the generalized force $f$ divided by the unit of the generalized displacement $u$. From the curve, the potential $v$, its gradient $\partial v / \partial \alpha$  and hessian $\partial^2 v / \partial \alpha^2$ can be constructed as follows:
\begin{align}
v(\alpha) &= \int_{0}^{\alpha-\alpha_0} f(u)\mathrm{d}u = \dfrac12 k (\alpha - \alpha_0)^2,\\
\dfrac{\partial v}{\partial\alpha}(\alpha) &=f(\alpha-\alpha_0)= k (\alpha - \alpha_0),\\
\dfrac{\partial^2 v}{\partial\alpha^2}(\alpha) &= \dfrac{\mathrm{d}f}{\mathrm{d}u}(\alpha-\alpha_0)=k.
\end{align}

\paragraph{Logarithmic behavior}
A logarithmic behavior is defined by a generalized force-displacement curve given by

\begin{equation}
\label{eq:logarithmic_behavior}
    f(u) = k\alpha_0 \ln \left(\dfrac{u + \alpha_0}{\alpha_0}\right),
\end{equation}
where $k$ is a parameter that represents the slope of the generalized force-displacement curve at $u=0$, whose unit is the unit of the generalized force $f$ divided by the unit of the generalized displacement $u$. Unlike the linear behavior, a logarithmic behavior can prevent flexels from reaching a zero measure ($\alpha=0$), as the force tends to infinity when the flexel becomes `fully compressed' (Fig.~\ref{fig:linear_loga}b). The behavior is identical to the linear behavior at first order around the natural measure $\alpha_0$. From the curve, the potential $v$, its gradient $\partial v / \partial \alpha$  and hessian $\partial^2 v / \partial \alpha^2$ can be constructed as follows:
\begin{align}
v(\alpha) &= \int_{0}^{\alpha-\alpha_0} f(u)\mathrm{d}u = k\alpha \alpha_0 \left(\ln \left(\alpha/\alpha_0\right) - 1\right),\\
\dfrac{\partial v}{\partial\alpha}(\alpha) &=f(\alpha-\alpha_0)= k \alpha_0 \ln\left(\alpha / \alpha_0\right),\\
\dfrac{\partial^2 v}{\partial\alpha^2}(\alpha) &=\dfrac{\mathrm{d}f}{\mathrm{d}u}(\alpha-\alpha_0)= k \alpha_0 / \alpha.
\end{align}

\begin{figure}[ht]
    \centering
    \includegraphics{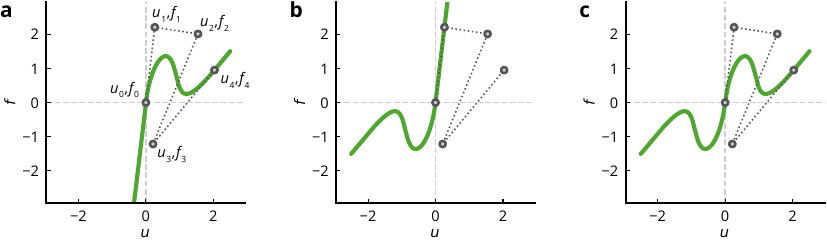}
    \caption{Bezier force-displacement curves for the tensile (a), compressive (b) and symmetric (c) modes.}
    \label{fig:bezier_uni}
\end{figure}

\paragraph{Bezier behavior}
A Bezier behavior is described by a Bezier curve of degree $n$, with control points positioned at $(u_0=0, f_0=0)$, $(u_1, f_1)$,$\dots$, $(u_{n}, f_{n})$, with linear extrapolation beyond the first and last control points that preserves $\mathcal{C}_1$ continuity. The nonlinear behavior defined by the Bezier curve can either describe the tensile behavior (a.k.a `tensile' mode, Fig.~\ref{fig:bezier_uni}a), the compressive behavior (a.k.a `compressive' mode, Fig.~\ref{fig:bezier_uni}b) or both simultaneously (a.k.a `symmetric' mode, Fig.~\ref{fig:bezier_uni}c).

For convenience, let us define the functional $\mathcal{F}$ as
\begin{equation}
\label{eq:mode_transform}
    \mathcal{F}(g)(x) = m(x)g(m(x)x),
\end{equation}
where $g$ is a function and
\begin{equation}
    m(x) =
    \begin{cases}
        +1&\text{if `tensile' mode},\\
        -1&\text{if `compressive' mode},\\
        \sgn{x}&\text{if `symmetric' mode}.
    \end{cases}
\end{equation}
Note that
\begin{align}
    \label{eq:mode_transform_der}
    \dfrac{\partial}{\partial x}\mathcal{F}(g)(x) &= g'(m(x)x):=\mathcal{F}'(g)(x),\\
    \label{eq:mode_transform_der2}
    \dfrac{\partial^2}{\partial x^2}\mathcal{F}(g)(x) &= m(x)g''(m(x)x):=\mathcal{F}''(g)(x).
\end{align}
Mathematically speaking, the force displacement curve $f(u)$ is defined as
\begin{equation}
\label{eq:bezier_fd_curve}
f(u) = \mathcal{F}(\bar{f})(u),
\end{equation}
where
\begin{equation}
\bar{f}(u)=
\begin{cases}
    (f_1/u_1)u&\text{if $u\le0$},\\
    ~&~\\
    b(a^{-1}(u))& \text{if $0<u \le u_n$}\\
    ~&~\\
    f_n + \dfrac{f_n - f_{n-1}}{u_n - u_{n-1}}(u - u_n)&\text{if $u > u_n$},
\end{cases}
\end{equation}
where the functions $a(.), b(.)$ describe the parametric Bezier curve:
\begin{equation}
\begin{cases}
    a(x) &= \sum_{i=0}^n u_i B_{i,n}(x)\\
    b(x) &= \sum_{i=0}^n f_i B_{i,n}(x),
\end{cases}
\end{equation}
where $x$ is the curve parameter that runs from 0 to 1 and $B_{i,n}$ are the Bernstein polynomials of degree $n$.
Note that the generalized force-displacement curve is only well defined if $a(x)$ is monotonic on the interval $[0, 1]$.
The stiffness $k(u)$ along the curve is given by the derivative of $f(u)$, i.e.:
\begin{align}
\label{eq:bezier_kd_curve}
k(u)=\dfrac{\mathrm{d}f}{\mathrm{d}u} &= \mathcal{F}'(\bar{f})(u),
\end{align}
where $\mathcal{F}'$ is defined in Eq.~(\ref{eq:mode_transform_der}) and
\begin{align}
\dfrac{\mathrm{d}\bar{f}}{\mathrm{d}u}=
\begin{cases}
     f_1/u_1&\text{if $u \le 0$},\\
     ~&~\\
     \dfrac{b'(a^{-1}(u))}{a'(a^{-1}(u))}&\text{if $0< u \le u_n$}\\
     ~&~\\
    \dfrac{f_n - f_{n-1}}{u_n - u_{n-1}}&\text{if $u > u_n$},\\
\end{cases}
\end{align}
where $a', b'$ are the derivatives of $a, b$:
\begin{equation}
a'(x)= n\sum_{i=0}^{n-1} (u_{i+1}-u_i)B_{i, n-1}(x) \quad\text{and}\quad b'(x)= n\sum_{i=0}^{n-1} (f_{i+1}-f_i)B_{i, n-1}(x).
\end{equation}
From the curve, the potential $v$, its gradient $\partial v / \partial \alpha$  and hessian $\partial^2 v / \partial \alpha^2$ can be constructed as follows:
\begin{align}
v(\alpha) &= \int_{0}^{\alpha-\alpha_0} f(u)\mathrm{d}u,\\
\dfrac{\partial v}{\partial\alpha}(\alpha) &=f(\alpha-\alpha_0),\\
\dfrac{\partial^2 v}{\partial\alpha^2}(\alpha) &=\dfrac{\mathrm{d}f}{\mathrm{d}u}(\alpha-\alpha_0)=k(\alpha-\alpha_0),
\end{align}
where $f$ and $k$ are described in Eqs.~(\ref{eq:bezier_fd_curve}, \ref{eq:bezier_kd_curve}).

\begin{figure}[ht]
    \centering
    \includegraphics{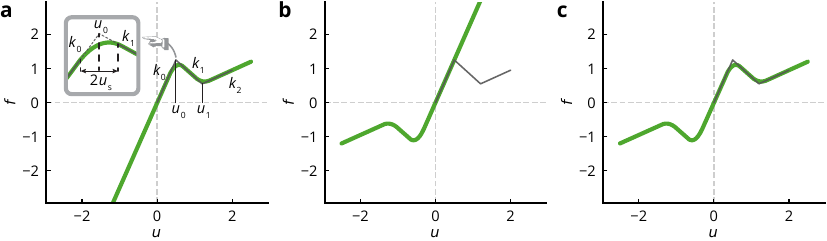}
    \caption{$\mathcal{C}_1$-piecewise linear force-displacement curves for the tensile (a), compressive (b) and symmetric (c) modes.}
    \label{fig:c1pw}
\end{figure}

\paragraph{$\mathcal{C}_1$-piecewise behavior}
A $\mathcal{C}_1$-piecewise behavior is defined by a piecewise linear function whose `corners' have been smoothed by quadratic functions, ensuring $\mathcal{C}_1$ continuity. Bilinear and trilinear piecewise linear force-displacement curves have been used to study softening, stiffening or bistable interactions in spring chains \cite{fermi1955studies,cohen_dynamics_2014, vainchtein_solitary_2024}. The piecewise curve presented here generalizes the concept to $n$ segments with control on the smoothness of the transitions. The curve is described by $n$ slopes $k_0, \dots, k_{n-1}$, $n-1$ displacement values $u_0, \dots, u_{n-2}$ at which the non-smoothed piecewise curve transitions from slope $k_i$ to $k_{i+1}$, and a smoothing parameter $u_\text{s}$ that controls the size of the region around the transition displacement $u_i$ where the curve is given by the quadratic function instead of the piecewise linear function. Similarly to the Bezier behavior, the nonlinear behavior defined by the smoothed piecewise linear curve can either describe the tensile behavior (a.k.a `tensile' mode, Fig.~\ref{fig:c1pw}a), the compressive behavior (a.k.a `compressive' mode, Fig.~\ref{fig:c1pw}b) or both simultaneously (a.k.a `symmetric' mode, Fig.~\ref{fig:c1pw}c).

Mathematically speaking,
\begin{equation}
\label{eq:c1pw_fd_curve}
f(u) = \mathcal{F}(\bar{f})(u),
\end{equation}
where $\mathcal{F}$ is defined in Eq.~(\ref{eq:mode_transform}) and
\begin{equation}
\bar{f}(u)=P(u; \bm{s}=[k_0, k_1, \dots, k_{n-1}]; \bm{\eta}=[u_0, u_1, \dots, u_{n-2}]; \eta_\text{s}=u_\text{s}),
\end{equation}
where
\begin{equation}
\label{eq:pw_definition}
 P(\eta;\bm{s};\bm{\eta};\eta_\text{s}) =
    \begin{cases}
    s_0 \eta&\text{if $\eta \le \eta_0 - \eta_\text{s}$}\\
    a_0 \eta^2 + b_0\eta + c_0&\text{if $\eta_{0} - \eta_\text{s} < \eta < \eta_0 + \eta_\text{s}$}\\
    s_1 \eta + p_1&\text{if $\eta_0 + \eta_\text{s} \le \eta \le \eta_1 - \eta_\text{s}$}\\
    \vdots&\vdots\\
    a_i \eta^2 + b_i\eta + c_i&\text{if $\eta_{i} - \eta_\text{s} < \eta < \eta_i + \eta_\text{s}$}\\
    s_i \eta + p_i&\text{if $\eta_{i-1}+\eta_\text{s} \le \eta \le \eta_i - \eta_\text{s}$}\\
    \vdots&\vdots\\
    a_{n-2} \eta^2 + b_{n-2}\eta + c_{n-2}&\text{if $\eta_{n-2} - \eta_\text{s} < \eta < \eta_{n-2} + \eta_\text{s}$}\\
    s_{n-1} \eta + p_{n-1}&\text{if $\eta_{n-2}+\eta_\text{s} \le \eta < +\infty $},\\
    \end{cases}
\end{equation}
with
\begin{align}
p_i &= p_{i-1} + \eta_{i-1} \left(s_{i-1} - s_i\right); \quad p_0=0,\\
\label{eq:a_i_pw}
a_i &= \dfrac{s_{i+1} - s_i}{4\eta_\text{s}},\\
b_i &= \dfrac{s_i( \eta_i + \eta_\text{s}) - s_{i+1}(\eta_i - \eta_\text{s})}{2\eta_\text{s}},\\
c_i&=\dfrac{(s_{i+1} - s_i) (\eta_i - \eta_\text{s})^ 2}{ 4 \eta_\text{s}} + p_i.
\end{align}
Note that for $P$ is well-defined only if $2\eta_\text{s}<\min \{2\eta_0, \eta_1-\eta_0, \dots, \eta_{n-2}-\eta_{n-3}\}$.
The stiffness $k(u)$ along the curve is given by the derivative of $f(u)$, i.e.:
\begin{align}
\label{eq:c1pw_kd_curve}
k(u)=\dfrac{\mathrm{d}f}{\mathrm{d}u} &= \mathcal{F}'(\bar{f})(u),
\end{align}
where $\mathcal{F}'$ is defined in Eq.~(\ref{eq:mode_transform_der}) and
\begin{align}
\dfrac{\mathrm{d}\bar{f}}{\mathrm{d}u}=\dfrac{\partial P}{\partial \eta}(u; [k_0, \dots, k_{n-1}]; [u_0, \dots, u_{n-2}];u_s),
\end{align}
where
\begin{equation}
    \label{eq:derivative_c1pw}
    \dfrac{\partial P}{\partial \eta}(\eta, \bm{s}, \bm{\eta}, \eta_s)=
    \begin{cases}
    s_0 &\text{if $\eta \le \eta_0 - \eta_\text{s}$}\\
    2a_0\eta + b_0&\text{if $\eta_{0} - \eta_\text{s} < \eta < \eta_0 + \eta_\text{s}$}\\
    s_1&\text{if $\eta_0 + \eta_\text{s} \le \eta \le \eta_1 - \eta_\text{s}$}\\
    \vdots&\vdots\\
    2a_i\eta + b_i&\text{if $\eta_{i} - \eta_\text{s} < \eta < \eta_i + \eta_\text{s}$}\\
    s_i&\text{if $\eta_{i-1}+\eta_\text{s} \le \eta \le \eta_i - \eta_\text{s}$}\\
    \vdots&\vdots\\
    2a_{n-2}\eta+ b_{n-2}&\text{if $\eta_{n-2} - \eta_\text{s} < \eta < \eta_{n-2} + \eta_\text{s}$}\\
    s_{n-1}&\text{if $\eta_{n-2}+\eta_\text{s} \le \eta < +\infty $}.\\
    \end{cases}
\end{equation}
From the curve, the potential $v$, its gradient $\partial v / \partial \alpha$  and hessian $\partial^2 v / \partial \alpha^2$ can be constructed as follows:
\begin{align}
v(\alpha) &= \int_{0}^{\alpha-\alpha_0} f(u)\mathrm{d}u,\\
\dfrac{\partial v}{\partial\alpha}(\alpha) &= f(\alpha-\alpha_0),\\
\dfrac{\partial^2 v}{\partial\alpha^2}(\alpha) &=
\dfrac{\mathrm{d}f}{\mathrm{d}u}(\alpha-\alpha_0)=k(\alpha-\alpha_0),
\end{align}
where $f$ and $k$ are described in Eqs.~(\ref{eq:c1pw_fd_curve}, \ref{eq:c1pw_kd_curve}).

\begin{figure}[ht]
    \centering
    \includegraphics{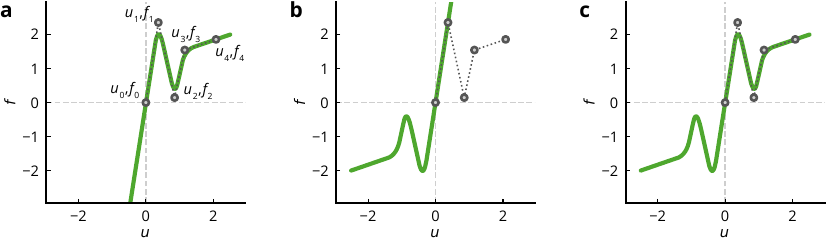}
    \caption{Zigzag force-displacement curves for the tensile (a), compressive (b) and symmetric (c) modes.}
    \label{fig:zz_uni}
\end{figure}

\paragraph{Zigzag behavior}
A zigzag behavior is described by a `smooth zigzag', that is, a polygonal chain with vertices positioned at $(u_0=0, f_0=0), (u_1, f_1), \dots, (u_n, f_n)$, whose `corners' have been smoothed. The amount of smoothing is governed by the parameter $\epsilon \in ]0, 1[$. The zigzag behavior serves a similar purpose as the $\mathcal{C}_1$-piecewise linear behavior, except that it can be generalized to a multi-valued curve as shown in subsection \ref{subsection:bivariate_behaviors}. Similarly to the Bezier behavior, the smooth zigzag can either describe the tensile behavior (Fig.~\ref{fig:zz_uni}a), the compressive behavior (Fig.~\ref{fig:zz_uni}b) or both simultaneously (Fig.~\ref{fig:zz_uni}c).

Mathematically speaking,
\begin{align}
\label{eq:zz_fd_curve}
    f(u) &= \mathcal{F}(\bar{f})(u)
\end{align}
where $\mathcal{F}$ is defined in Eq.~(\ref{eq:mode_transform}) and
\begin{equation}
\bar{f}(u)=b(a^{-1}(u)),
\end{equation}
where the functions $a(.)$ and $b(.)$ are defined by $\mathcal{C}_1$-piecewise linear functions:
\begin{align}
a(x)&=P(x;\bm{s}^a;\bm{\eta}^a;\eta^a_\text{s})\\
b(x)&=P(x;\bm{s}^b;\bm{\eta}^b;\eta^b_\text{s}),
\end{align}
with $P$ defined in Eq.~(\ref{eq:pw_definition}), $s_i^a=  (n-1)(u_{i+1}-u_i)$, $s^b_i=(n-1)(f_{i+1}-f_i)$, $\eta^a_i=\eta^b_i=i/(n-1)$, and $\eta_\text{s}^a=\eta_\text{s}^b=\epsilon / (2n-2)$.
Note that the generalized force-displacement curve is only well defined if $a(x)$ is monotonic on the interval $[0, +\infty [$, which is valid only if $0<u_1<\dots<u_{n-2}$.
The stiffness $k(u)$ along the curve is given by the derivative of $f(u)$, i.e.:
\begin{equation}
    \label{eq:zz_kd_curve}
    k(u) = \dfrac{\mathrm{d}f}{\mathrm{d}u}(u) = \mathcal{F}'(\bar{f})(u),
\end{equation}
where $\mathcal{F}'$ is defined in Eq.~(\ref{eq:mode_transform_der}) and
\begin{align}
    \dfrac{\mathrm{d}\bar{f}}{\mathrm{d}u}(u) &= b'(a^{-1}(u))/a'(a^{-1}(u)),
\end{align}
where $a' = \partial P/\partial \eta (x, \bm{s}^a, \bm{\eta}^a, \eta_\text{s}^a)$ and $b'=\partial P/\partial \eta (x, \bm{s}^b, \bm{\eta}^b, \eta_\text{s}^b)$ are the derivatives of $a$ and $b$ with respect to $x$, with  $\partial P/\partial \eta$ given in Eq.~(\ref{eq:derivative_c1pw}).
From the curve, the potential $v$, its gradient $\partial v / \partial \alpha$  and hessian $\partial^2 v / \partial \alpha^2$ can be constructed as follows:
\begin{align}
v(\alpha) &= \int_{0}^{\alpha-\alpha_0} f(u)\mathrm{d}u,\\
\dfrac{\partial v}{\partial\alpha}(\alpha) &=f(\alpha-\alpha_0),\\
\dfrac{\partial^2 v}{\partial\alpha^2}(\alpha) &=\dfrac{\mathrm{d}f}{\mathrm{d}u}(\alpha-\alpha_0)=k(\alpha-\alpha_0),
\end{align}
where $f$ and $k$ are described in Eqs.~(\ref{eq:zz_fd_curve}, \ref{eq:zz_kd_curve}).

\begin{figure}[ht]
    \centering
    \includegraphics{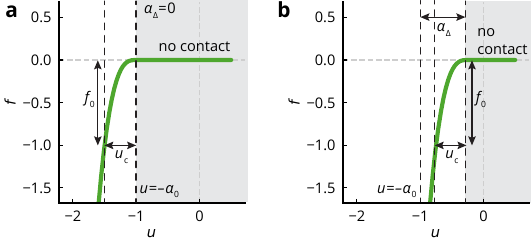}
    \caption{`Contact' force-displacement curves with $\alpha_0=1$ when $\alpha_\Delta=0$ (a) and $\alpha_\Delta=0.75$.}
    \label{fig:contact}
\end{figure}

\paragraph{Contact behavior}
A contact behavior mimics contact using a generalized force-displacement curve that produces a nonzero force only for measures $\alpha=u+\alpha_0$ below a certain threshold $\alpha_\Delta$ (Fig.~\ref{fig:contact}):

\begin{equation}
\label{eq:contact_fd_curve}
f(u)=
\begin{cases}
    0&\text{if $u \ge \alpha_\Delta - \alpha_0$}\\
    -f_0\left(\dfrac{\alpha_\Delta-\alpha_0-u}{u_\text{c}}\right)^3&\text{if $u < \alpha_\Delta - \alpha_0$},
\end{cases}
\end{equation}
where $f_0>0$ is the magnitude of the repulsive force when the measure is decreased by $u_\text{c}>0$ from $\alpha_\Delta$. The curve is $\mathcal{C}_2$ continuous.
From the curve the potential $v$, its gradient $\partial v / \partial \alpha$  and hessian $\partial^2 v / \partial \alpha^2$ can be constructed as follows:
\begin{align}
v(\alpha) &= \int_{\alpha_\Delta}^{\alpha} f(u)\mathrm{d}u=
\begin{cases}
0&\text{if $\alpha \ge \alpha_\Delta$}\\
\dfrac{f_0 u_\text{c}}{4}\left(\dfrac{\alpha_\Delta-\alpha}{u_\text{c}}\right)^4&\text{if $\alpha < \alpha_\Delta$},
\end{cases}\\
\dfrac{\partial v}{\partial\alpha}(\alpha) &=f(\alpha-\alpha_0)=
\begin{cases}
0&\text{if $\alpha \ge \alpha_\Delta$}\\
-f_0\left(\dfrac{\alpha_\Delta-\alpha}{u_\text{c}}\right)^3&\text{if $\alpha < \alpha_\Delta$},
\end{cases}\\
\dfrac{\partial^2 v}{\partial\alpha^2}(\alpha) &=\dfrac{\mathrm{d}f}{\mathrm{d}u}(\alpha-\alpha_0)=\begin{cases}
0&\text{if $\alpha \ge \alpha_\Delta$}\\
\dfrac{3f_0}{u_\text{c}}\left(\dfrac{\alpha_\Delta-\alpha}{u_\text{c}}\right)^2&\text{if $\alpha < \alpha_\Delta$}.
\end{cases}
\end{align}

\begin{figure}[ht]
    \centering
    \includegraphics{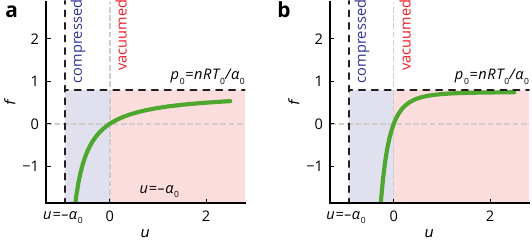}
    \caption{Force-displacement curves mimicking the ideal gas law. (a) Isothermal with $n=0.03, R=0.083, T_0=300$. (a) Isentropic with $n=0.03, R=0.083, T_0=300, \gamma=4$.}
    \label{fig:isothermal_isentropic}
\end{figure}

\paragraph{Isothermal behavior}
 An isothermal behavior is described by a generalized force-displacement curve that mimics the ideal gas law during an isothermal process (process where the temperature remains constant):
 \begin{equation}
     p-p_0 = nRT_0(1/V-1/V_0),
 \end{equation}
 where $p_0$ is the ambient pressure, $V_0$ is the volume of the gas at ambient pressure, $T_0$ is the temperature of the gas, $p$ the pressure, $V$ the volume, $n$ is the amount of substance, $R$ the gas constant. The generalized-force displacement curve is defined from Eq.~(\ref{eq:isothermal_curve}) by letting $f$ play the role of the pressure difference $p_0 -p$ and $u$ the role of volume change $V-V_0$, with $\alpha_0=V_0$ (Fig.~\ref{fig:isothermal_isentropic}a):
 \begin{equation}
 \label{eq:isothermal_curve}
 f(u) =\dfrac{nRT_0}{\alpha_0}\dfrac{u}{u+\alpha_0}.
 \end{equation}
 Note that a negative generalized force $f$ corresponds to a compressed state (the pressure $p$ is greater than the ambient pressure $p_0$), while a positive $f$ corresponds to a `vacuumed' state (the pressure $p$ is less than the ambient pressure $p_0$). From the curve, the potential $v$, its gradient $\partial v / \partial \alpha$  and hessian $\partial^2 v / \partial \alpha^2$ can be constructed as follows:
\begin{align}
v(\alpha) &= \int_{0}^{\alpha-\alpha_0} f(u)\mathrm{d}u = nRT_0(\alpha/\alpha_0 - 1 - \ln(\alpha/\alpha_0)),\\
\dfrac{\partial v}{\partial\alpha}(\alpha) &=f(\alpha-\alpha_0)=nRT_0\dfrac{\alpha-\alpha_0}{\alpha\alpha_0},\\
\dfrac{\partial^2 v}{\partial\alpha^2}(\alpha) &=\dfrac{\mathrm{d}f}{\mathrm{d}u}(\alpha-\alpha_0)=\dfrac{nRT_0}{\alpha^2}.
\end{align}

\paragraph{Isentropic behavior}
An isentropic behavior is described by a generalized force-displacement curve that mimics the ideal gas law during an isentropic process (process where the entropy remains constant):
\begin{equation}
    p-p_0 = nRT_0\left(\dfrac1V\left(\dfrac{V_0}{V}\right)^{\gamma-1} - \dfrac1{V_0}\right),
\end{equation}
where $p_0$ is the ambient pressure, $V_0$ the volume of the gas at ambient pressure, $T_0$ is the initial temperature, $p$ the pressure, $V$ the volume, $n$ the amount of substance, $R$ the gas constant and $\gamma$ the heat capacity ratio. The generalized-force displacement curve is defined from Eq. by letting $f$ play the role of the pressure difference $p_0 -p$ and $u$ the role of volume change $V-V_0$, with $\alpha_0=V_0$ (Fig.~\ref{fig:isothermal_isentropic}b):
\begin{equation}
 \label{eq:isentropic_curve}
f(u) = nRT_0\left(\dfrac1{\alpha_0} - \dfrac1{u+\alpha_0}\left(\dfrac{\alpha_0}{u+\alpha_0}\right)^{\gamma-1}\right),
\end{equation}

From the curve, the potential $v$, its gradient $\partial v / \partial \alpha$  and hessian $\partial^2 v / \partial \alpha^2$ can be constructed as follows:
\begin{align}
v(\alpha) &= \int_{0}^{\alpha-\alpha_0} f(u)\mathrm{d}u = nRT_0\left(\dfrac\alpha{\alpha_0} - 1\right) + \dfrac{nRT_0}{\gamma-1}\left(\left(\dfrac{\alpha_0}{\alpha}\right)^{\gamma-1} - 1\right),\\
\dfrac{\partial v}{\partial\alpha}(\alpha) &=f(\alpha-\alpha_0)=nRT_0\left(\dfrac1{\alpha_0} - \dfrac1{\alpha}\left(\dfrac{\alpha_0}{\alpha}\right)^{\gamma-1}\right),\\
\dfrac{\partial^2 v}{\partial\alpha^2}(\alpha) &=\dfrac{\mathrm{d}f}{\mathrm{d}u}(\alpha-\alpha_0)=\dfrac{nRT_0}{\alpha^2}\left(\dfrac{\alpha_0}{\alpha}\right)^{\gamma-1}.
\end{align}

\subsection{Bivariate behavior}
\label{subsection:bivariate_behaviors}
A bivariate behavior is characterized by a bivariate energy potential, which is constructed from a generalized force-displacement that can possibly be multi-valued, meaning that for a given generalized displacement $u$ can correspond to multiple generalized force values $f$. Such a curve is defined by the following parametric equations:
\begin{equation}
\label{eq:multi-valued_curve}
\begin{cases}
    u = a(x)\\
    f = b(x),
\end{cases}
\end{equation}
where $a(.)$ and $b(.)$ are two continuous functions, and $x$ is the curve parameter that monotonically increases as one moves along the curve. From the parametric curve $(u, f)=(a(x), b(x))$, the following bivariate potential, which takes two inputs, the measure $\alpha$ and an internal coordinate $t$, is constructed
\begin{equation}
    v(\alpha, t) = \dfrac12 k(t) \left(\alpha-\alpha_0 - a(t)\right)^2 + b(t)(\alpha - \alpha_0 - a(t)) + \int_{0}^{t}b(\tilde{t})a'(\tilde{t})\mathrm{d}\tilde{t}.
    \label{eq:bivariate_energy}
\end{equation}
Each point on the force-displacement curve $(a(x), b(x))$ corresponds to a state $(\alpha, t)=(\alpha_0+a(x), x)$ that is a static equilibrium state of the potential $v$ under force $b(x)$, as we will demonstrate through Eqs.~(\ref{eq:dpidalpha}, \ref{eq:dpidt}, \ref{eq:equi_u}, \ref{eq:equi_f}) (see colored box below). The function $k(.)$ can be defined so that the potential $v$ has the desired stability properties, as we will demonstrate afterwards. Its expression is given in Eq.~(\ref{eq:k_fun}). At equilibrium, the internal coordinate $t$ plays the role of the curve parameter. From the potential, we can compute its gradient and hessian components:
\begin{align}
\label{eq:dvda}
\dfrac{\partial v}{\partial \alpha} =&k(t)(\alpha-\alpha_0-a(t)) + b(t),\\
\label{eq:dvdt}
\dfrac{\partial v}{\partial t}=&(\alpha-\alpha_0-a(t))(b'(t) - k(t)a'(t)) + \dfrac{k'(t)}2 (\alpha-\alpha_0-a(t))^2,\\
\label{eq:d2vda2}
\dfrac{\partial^2 v}{\partial \alpha^2} =&k(t),\\
\label{eq:d2vdadt}
\dfrac{\partial ^2 v}{\partial \alpha \partial t }=&b'(t)-k(t)a'(t) + k'(t)(\alpha-\alpha_0-a(t)),\\
\label{eq:d2vdt2}
\dfrac{\partial^2 v}{ \partial t^2}=&(\alpha-\alpha_0-a(t))(b''(t) - k(t)a''(t)) + a'(t)(k(t)a'(t)-b'(t))\\
&+ (\alpha-\alpha_0 - a(t))\left(\dfrac{k''(t)}2 (\alpha-\alpha_0-a(t)) - 2k'(t)a'(t)\right). \nonumber
\end{align}
To define a bivariate behavior, it therefore suffices to provide the $a$ and $b$ functions (and their first and second derivatives) that describe the parametric generalized force-displacement curve. In \springable, the parametric curve can either be described as a Bezier curve (Eqs.~(\ref{eq:bivariate_bezier_a}, \ref{eq:bivariate_bezier_b})) or a smooth zigzag curve (Eqs.~(\ref{eq:bivariate_zz_a}, \ref{eq:bivariate_zz_b})) (see paragraphs below).

\begin{tcolorbox}[colback=bgc, title={\textbf{Proposition:} "Points on the curve $(a(t),b(t))$ correspond to static equilibria."}]
\paragraph{Proof}
States $(\alpha, t$) are equilibrium points if they are stationary points of the total potential energy $\pi(\alpha, t):=v(\alpha, t) - f\alpha$, meaning that they have to satisfy the two following equations,
\begin{align}
    0&=\dfrac{\partial \pi}{\partial \alpha} = \dfrac{\partial v}{\partial \alpha} - f \nonumber\\
    &= k(t)(\alpha-\alpha_0-a(t)) + b(t) - f, \label{eq:dpidalpha}\\
    ~&~ \nonumber\\
    0&=\dfrac{\partial \pi}{\partial t} = \dfrac{\partial v}{\partial t} \nonumber\\
    &= (\alpha-\alpha_0-a(t))\left(b'(t)-k(t)a'(t) + \dfrac{k'(t)}{2}(\alpha-\alpha_0-a(t))\right). \label{eq:dpidt}
\end{align} From (\ref{eq:dpidt}), we notice that
\begin{equation}
\alpha=\alpha_0 + a(t) \Leftrightarrow u = a(t)
\label{eq:equi_u}
\end{equation}
is a solution, which, when plugged back into Eq.~(\ref{eq:dpidalpha}), yields
\begin{equation}
f=b(t)\text{.}
\label{eq:equi_f}
\end{equation}
From Eqs.~(\ref{eq:equi_u}) and (\ref{eq:equi_f}), it can be noticed that each point on the force-displacement curve given in Eq.~(\ref{eq:multi-valued_curve}) is a stationary point of the total potential energy, which retrospectively explain why we define the elastic energy of the flexel using Eq.~(\ref{eq:bivariate_energy}).
\end{tcolorbox}

\paragraph{Defining $k(.)$ to satisfy stability conditions}
On the one hand, stability of equilibria are determined by the hessian matrix $\bm{h}$ of the total potential energy
\begin{equation}
    \bm{h}(\alpha, t) =
    \begin{bmatrix}
        \partial^2 \pi / \partial \alpha^2&\partial^2 \pi / \partial \alpha \partial t\\
        \partial^2 \pi / \partial \alpha \partial t&\partial^2 \pi / \partial t^2
    \end{bmatrix}.
\end{equation}
More precisely, stability of points under force-controlled conditions is governed by the number $\mu_f$ of negative eigenvalues in the matrix $\bm{h}$, each indicating an unstable deformation mode. Stable points under force-controlled conditions are therefore characterized by a hessian matrix with no negative eigenvalues ($\mu_f =0)$. Similarly, stability under displacement-controlled conditions is governed by the number $\mu_\alpha$ of negative eigenvalues in $\bm{h}$ from which the row and column corresponding to the variable $\alpha$ have been removed, that is, just the single component $\partial^2 \pi / \partial t^2$. Note that $\mu_f \ge \mu_\alpha$, with $2\ge\mu_f \ge 0$ and $1\ge \mu_\alpha \ge 0$, since the hessian matrix is 2-by-2.

\begin{figure}[ht]
    \centering
    \includegraphics{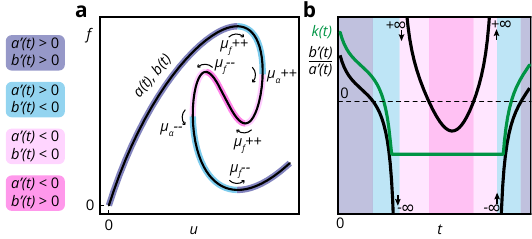}
    \caption{Example of multi-valued force-displacement curve and its slope. (a) Force-displacement curve described by parametric equations $u=a(t), f=b(t)$. (b) Slope of the tangent of the curve as a function of the curve parameter $t$ (black) and the associated $k(t)$ function (Eq.~(\ref{eq:k_fun})).}
    \label{fig:folds_and_k}
\end{figure}

On the other hand, the way that the curve folds (by reaching force and displacement extrema) should dictate the stability of equilibrium points along the $u-f$ curve \cite{maddocks_stability_1987}. This theory tells us that each time the curve folds clockwise (counter-clockwise) by reaching a force or displacement extrema, the number of negative unstable modes, indicated by $\mu_f$ or $\mu_\alpha$, under the respective loading condition should increase (decreased) by one (Fig.~\ref{fig:folds_and_k}a).

To ensure consistency between the folding rules and the numbers of negative eigenvalues, $\mu_f$ and $\mu_\alpha$, the following conditions must be satisfied:
\begin{align}
    \label{eq:cond1}
    &\text{if $(\nearrow) \equiv a'(t) > 0, b'(t) > 0 \Leftrightarrow \mu_f=0$ and $\mu_\alpha=0$;}\\
    \label{eq:cond2}
    &\text{if $(\searrow) \equiv a'(t) > 0, b'(t) < 0 \Leftrightarrow \mu_f=1$ and $\mu_\alpha=0$;}\\
    \label{eq:cond3}
    &\text{if $(\swarrow) \equiv a'(t) < 0, b'(t) < 0 \Leftrightarrow \mu_f=1$ and $\mu_\alpha=1$;}\\
    \label{eq:cond4}
    &\text{if $(\nwarrow) \equiv a'(t) < 0, b'(t) > 0\Leftrightarrow \mu_f=2$ and $\mu_\alpha=1$.}
\end{align}
Those conditions have been derived by first assuming that points along $(\nearrow$)-branches (that is, where $a'(t), b'(t) >0$) are stable under both force- and displacement-controlled conditions. Therefore,

\begin{itemize}
    \item when folding clockwise from a $(\nearrow)$-branch to a $(\searrow)$-branch (or inversely counter-clockwise), $\mu_f$ must be incremented (decremented) by one and $\mu_\alpha$ remains 0;
    \item when folding clockwise from a $(\searrow)$-branch to a $(\swarrow)$-branch (or inversely counter-clockwise), $\mu_f$ remains 1 and $\mu_\alpha$ is incremented (decremented) by one.
    \item when folding clockwise from a $(\swarrow)$-branch to a $(\nwarrow)$-branch (or inversely counter-clockwise), $\mu_f$ is incremented (decremented) by one and $\mu_\alpha$ remains 1.
\end{itemize}
Note that other folds such as $(\nearrow)-\text{to}-(\nwarrow)$ or its inverse $(\nwarrow)-\text{to}-(\nearrow)$ are not permitted, since the former would necessitate $0 \le \mu_\alpha (\nwarrow) = -1$ and the later would necessitate $1 \ge \mu_\alpha(\nearrow)=2$, which both are impossible. An example of generalized force-displacement showing valid folds is shown in Fig.~\ref{fig:folds_and_k}a.

To derive the conditions to satisfy the desired stability conditions (\ref{eq:cond1}),(\ref{eq:cond2}), (\ref{eq:cond3}) and (\ref{eq:cond4}) let us first express the hessian matrix when evaluated at an equilibrium point $(t, \alpha)=(t, \alpha_0+a(t))$:
\begin{equation}
\label{eq:hessian_matrix_at_eq}
\bm{h}(\alpha_0+a(t), t) =
\begin{bmatrix}
    k(t)&b'(t)-k(t)a'(t)\\
    b'(t)-k(t)a'(t)&a'(t)(k(t)a'(t)-b'(t))
\end{bmatrix}.
\end{equation} From conditions (\ref{eq:cond1}),(\ref{eq:cond2}), (\ref{eq:cond3}) and (\ref{eq:cond4}), notice that $\mu_\alpha=0$ if $a'>0$ else $1$, meaning that
\begin{equation}
\label{eq:ineq1}
\begin{cases}
    a'(ka'-b') > 0&\text{if $a'>0$}\\
    a'(ka'-b') < 0&\text{if $a'<0$}    
\end{cases} \Leftrightarrow
\begin{cases}
    k > b'/a'&\text{if $a'>0$}\\
    k < b'/a'&\text{if $a'<0$}.
\end{cases}
\end{equation}
To satisfy $\mu_f=0$ in condition (\ref{eq:cond1}), the hessian matrix (Eq.~(\ref{eq:hessian_matrix_at_eq})) must be positive-definite on ($\nearrow$)-branches, which, by applying Sylvester criterion, yields
\begin{align}
&k>0\quad\text{and}\quad\det{\bm{h}}=b'(ka'-b'))>0\quad\text{when $a', b'>0$}\\
&\Leftrightarrow \nonumber\\
\label{eq:ineq2}
&k > b'/a'\quad\text{when $a', b'>0$}.
\end{align}
To satisfy $\mu_f=1$ in condition (\ref{eq:cond2}), the hessian matrix must be indefinite on ($\searrow$)-branches, which, for a 2-by-2 matrix, yields
\begin{align}
&\det{\bm{h}}=b'(ka'-b')<0\quad\text{when $a'>0$, $b'<0$}\\
&\Leftrightarrow \nonumber\\
\label{eq:ineq3}
&k > b'/a'\quad\text{when $a'>0$, $b'<0$}.
\end{align}
To satisfy $\mu_f=1$ in condition (\ref{eq:cond3}), the hessian matrix must be indefinite on ($\swarrow$)-branches as well:
\begin{align}
&\det{\bm{h}}=b'(ka'-b')<0\quad\text{when $a', b'<0$}\\
&\Leftrightarrow \nonumber\\
\label{eq:ineq4}
&k < b'/a'\quad\text{when $a', b'<0$}.
\end{align}
To satisfy $\mu_f=2$ in condition (\ref{eq:cond4}), the hessian matrix must be negative-definite on ($\nwarrow$)-branches, which, by applying Sylvester criterion, yields
\begin{align}
&k<0\quad\text{and}\quad\det{\bm{h}}=b'(ka'-b'))>0\quad\text{when $a'<0$, $b'>0$}\\
&\Leftrightarrow \nonumber\\
\label{eq:ineq5}
&k < b'/a'\quad\text{when $a'<0$, $b'>0$}.
\end{align}
Inequalities (\ref{eq:ineq1}), (\ref{eq:ineq2}), (\ref{eq:ineq3}), (\ref{eq:ineq4}) and (\ref{eq:ineq5}) are consistent and can be equivalently summarized as follows
\begin{equation}
    \begin{cases}
        k(t) > b'(t) / a'(t)&\text{if $a'(t)>0$}\\
        k(t) < b'(t) / a'(t)&\text{if $a'(t)<0$},
    \end{cases}
\end{equation}
or, in other words, the $k$ function must be higher than the $b'/a'$ function when $a'>0$, else lower. For generalized force-displacement curves with valid folds, the $k$ function can always be constructed to satisfy those requirements, as illustrated in Fig.~\ref{fig:folds_and_k}b. In \springable, the $k$ function is constructed as follows
\begin{equation}
\label{eq:k_fun}
    k(t)=
    \begin{cases}
        k^\star&\text{if $k_{\min} - k_{\max}>2\delta$}\\
        \hat{k}(t)&\text{else},
    \end{cases}
\end{equation}
where
\begin{align}
    k_{\max} &=
        \displaystyle \max_{t:a'(t) > 0} b'(t)/a'(t)\\
    k_{\min} &=
    \begin{cases}
        \displaystyle \min_{t:a'(t) < 0} b'(t)/a'(t)&\text{if $\exists t: a'(t) <0$}\\
        +\infty&\text{else},
    \end{cases}\\
    \delta &= k_{\max} / 20,\\
    k^\star&=\min(k_{\min}-\delta, k_{\max}+\delta),\\
    \hat{k}(t)&=
    \begin{cases}
        \max\{b'(t)/a'(t) + \delta, k^\star\}&\forall t: a'(t) > 0\\
        k^\star&\forall t: a'(t) < 0.
    \end{cases}
\end{align}

\begin{figure}[ht]
    \centering
    \includegraphics{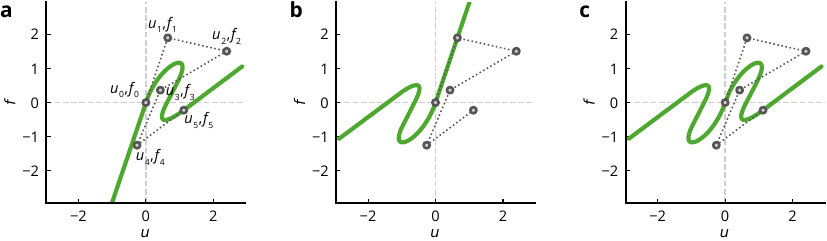}
    \caption{Multi-valued Bezier force-displacement curves for the tensile (a), compressive (b) and symmetric (c) modes.}
    \label{fig:bezier_bi}
\end{figure}

\paragraph{Bivariate Bezier behavior}
A bivariate Bezier behavior is a Bezier curve of degree $n$ with control points at $(u_0=0, f_0=0), (u_1, f_1) \dots (u_n, f_n)$, with linear extrapolation beyond the first and last control points that preserves $\mathcal{C}_1$-continuity. Unlike its univariate counterpart (subsection~\ref{subsection:univariate_behaviors}, paragraph `Bezier behavior'), its curve can be multi-valued. The nonlinear behavior defined by the Bezier curve can either describe the tensile behavior (a.k.a ‘tensile’ mode, Fig.~\ref{fig:bezier_bi}a), the compressive behavior (a.k.a ‘compressive’ mode, Fig.~\ref{fig:bezier_bi}b) or both simultaneously (a.k.a ‘symmetric’ mode, Fig.~\ref{fig:bezier_bi}c). Mathematically speaking, the force-displacement curve is defined by
\begin{align}
    \label{eq:bivariate_bezier_a}
    u &= a(t) = \mathcal{F}(\bar{a})(t)\\
    \label{eq:bivariate_bezier_b}
    f &= b(t) = \mathcal{F}(\bar{b})(t)
\end{align}
where the functional $\mathcal{F}$ is defined in Eq.~(\ref{eq:mode_transform}) and
\begin{align}
    \bar{a}(t) &=
    \begin{cases}
        nu_1t/t_\text{max}&\text{if $t \le 0$}\\
        \sum_{i=0}^n u_i B_{i,n}(t/t_\text{max}) & \text{if $0 < t \le t_\text{max}$}\\
        u_n + n(u_n-u_{n-1})(t/t_\text{max}-1)&\text{if $t > t_\text{max}$},
    \end{cases}\\
    \bar{b}(t) &=
    \begin{cases}
        nf_1t/t_\text{max}&\text{if $t \le 0$}\\
        \sum_{i=0}^n f_i B_{i,n}(t/t_\text{max}) & \text{if $0 < t \le t_\text{max}$}\\
        f_n + n(f_n-f_{n-1})(t/t_\text{max}-1)&\text{if $t > t_\text{max}$},
    \end{cases}
\end{align}
where $t_\text{max}=\sum_{i=1}^n|u_i-u_{i-1}|$ and $B_{i, n}$ are the Bernstein polynomials of degree $n$.

Functions $a$ and $b$ are used to construct an energy potential (Eq.~(\ref{eq:bivariate_energy})), and compute its gradient (Eqs.~(\ref{eq:dvda}, \ref{eq:dvdt})) and hessian (Eqs.~(\ref{eq:d2vda2}, \ref{eq:d2vdadt}, \ref{eq:d2vdt2})). To calculate those quantities, the first and second derivatives of $a$ and $b$ must also be provided:
\begin{align}
    a'(t) &= \mathcal{F}'(\bar{a})(t),\\
    b'(t) &= \mathcal{F}'(\bar{b})(t),\\
    a''(t) &= \mathcal{F}''(\bar{a})(t),\\
    b''(t) &= \mathcal{F}''(\bar{b})(t),
\end{align}
where the functionals $\mathcal{F}'$ and $\mathcal{F}''$ are defined in Eqs.~(\ref{eq:mode_transform_der}, \ref{eq:mode_transform_der2}) and
\begin{align} 
\bar{a}'(t)&=
    \begin{cases}
        nu_1/t_\text{max}&\text{if $t \le 0$}\\
        n\sum_{i=0}^{n-1} (u_{i+1}-u_i) B_{i,n-1}(t/t_\text{max})/t_\text{max} & \text{if $0 < t \le t_\text{max}$}\\
        n(u_n-u_{n-1})/t_\text{max}&\text{if $t > t_\text{max}$},
    \end{cases}\\
\bar{b}'(t)&=
    \begin{cases}
        nf_1/t_\text{max}&\text{if $t \le 0$}\\
        n\sum_{i=0}^{n-1} (f_{i+1}-f_i) B_{i,n-1}(t/t_\text{max})/t_\text{max} & \text{if $0 < t \le t_\text{max}$}\\
        n(f_n-f_{n-1})/t_\text{max}&\text{if $t > t_\text{max}$},
    \end{cases}\\
\bar{a}''(t) &=
    \begin{cases}
        n(n-1)\sum_{i=0}^{n-2} (u_{i+2}-2u_{i+1}+u_i) B_{i,n-2}(t/t_\text{max})/t^2_\text{max} & \text{if $0 < t \le t_\text{max}$}\\
        0&\text{else},
    \end{cases}\\
\bar{b}''(t) &=
    \begin{cases}
        n(n-1)\sum_{i=0}^{n-2} (f_{i+2}-2f_{i+1}+f_i) B_{i,n-2}(t/t_\text{max})/t^2_\text{max} & \text{if $0 < t \le t_\text{max}$}\\
        0&\text{else}.
    \end{cases}
\end{align}

\begin{figure}[ht]
    \centering
    \includegraphics{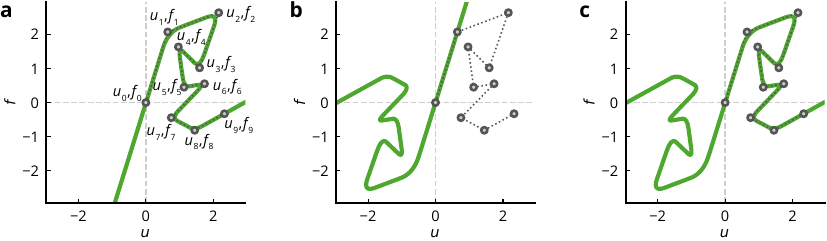}
    \caption{Multi-valued zigzag force-displacement curves for the tensile (a), compressive (b) and symmetric (c) modes.}
    \label{fig:zz_bi}
\end{figure}

\paragraph{Bivariate zigzag behavior}
A bivariate zigzag behavior is described by `smoothed zigzag', similar to its univariate counterpart (subsection~\ref{subsection:univariate_behaviors}, paragraph `zigzag behavior'). However, this bivariate version can defined multi-valued curves. The nonlinear behavior defined by the Bezier curve can either describe the tensile behavior (a.k.a ‘tensile’ mode, Fig.~\ref{fig:zz_bi}a), the compressive behavior (a.k.a ‘compressive’ mode, Fig.~\ref{fig:zz_bi}b) or both simultaneously (a.k.a ‘symmetric’ mode, Fig.~\ref{fig:zz_bi}c). Mathematically speaking, the force-displacement curve is defined by
\begin{align}
    \label{eq:bivariate_zz_a}
    u &= a(t) = \mathcal{F}(\bar{a})(t)\\
    \label{eq:bivariate_zz_b}
    f &= b(t) = \mathcal{F}(\bar{b})(t)
\end{align}
where the functional $\mathcal{F}$ is defined in Eq.~(\ref{eq:mode_transform}) and
\begin{align}
    \bar{a}(t) &= P(t/t_\text{max}, \bm{s}^a; \bm{\eta}^a;\eta_\text{s}^a),\\
    \bar{b}(t) &= P(t/t_\text{max}, \bm{s}^b; \bm{\eta}^b;\eta_\text{s}^b),
\end{align}
with $P$ defined in Eq.~(\ref{eq:pw_definition}), $t_\text{max}=\sum_{i=1}^n|u_i-u_{i-1}|$, $s_i^a= (n-1)(u_{i+1}-u_i)$, $s^b_i=(n-1)(f_{i+1}-f_i)$, $\eta^a_i=\eta^b_i=i/(n-1)$, and $\eta_\text{s}^a=\eta_\text{s}^b=\epsilon / (2n-2)$.

Functions $a$ and $b$ are used to construct an energy potential (Eq.~(\ref{eq:bivariate_energy})), and compute its gradient (Eqs.~(\ref{eq:dvda}, \ref{eq:dvdt})) and hessian (Eqs.~(\ref{eq:d2vda2}, \ref{eq:d2vdadt}, \ref{eq:d2vdt2})). To calculate those quantities, the first and second derivatives of $a$ and $b$ must also be provided:
\begin{align}
    a'(t) &= \mathcal{F}'(\bar{a})(t),\\
    b'(t) &= \mathcal{F}'(\bar{b})(t),\\
    a''(t) &= \mathcal{F}''(\bar{a})(t),\\
    b''(t) &= \mathcal{F}''(\bar{b})(t),
\end{align}
where the functionals $\mathcal{F}'$ and $\mathcal{F}''$ are defined in Eqs.~(\ref{eq:mode_transform_der}, \ref{eq:mode_transform_der2}) and
\begin{align} 
\bar{a}'(t)&=\dfrac{1}{t_\text{max}}\dfrac{\partial P}{\partial \eta}(t/t_\text{max}, \bm{s}^a; \bm{\eta}^a;\eta_\text{s}^a),\\
\bar{b}'(t)&=\dfrac{1}{t_\text{max}}\dfrac{\partial P}{\partial \eta}(t/t_\text{max}, \bm{s}^b; \bm{\eta}^b;\eta_\text{s}^b),\\
\bar{a}''(t) &= \dfrac{1}{t^2_\text{max}}\dfrac{\partial^2 P}{\partial \eta^2}(t/t_\text{max}, \bm{s}^a; \bm{\eta}^a;\eta_\text{s}^a),\\
\bar{b}''(t) &= \dfrac{1}{t^2_\text{max}}\dfrac{\partial^2 P}{\partial \eta^2}(t/t_\text{max}, \bm{s}^b; \bm{\eta}^b;\eta_\text{s}^b),
\end{align}
with $\partial P / \partial \eta$ defined in Eq.~(\ref{eq:derivative_c1pw}), and
\begin{equation}
\dfrac{\partial^2 P}{\partial \eta^2}(\eta, \bm{s}, \bm{\eta}, \eta_\text{s})=
\begin{cases}
    0 &\text{if $\eta \le \eta_0 - \eta_\text{s}$}\\
    2a_0&\text{if $\eta_{0} - \eta_\text{s} < \eta < \eta_0 + \eta_\text{s}$}\\
    0&\text{if $\eta_0 + \eta_\text{s} \le \eta \le \eta_1 - \eta_\text{s}$}\\
    \vdots&\vdots\\
    2a_i&\text{if $\eta_{i} - \eta_\text{s} < \eta < \eta_i + \eta_\text{s}$}\\
    0&\text{if $\eta_{i-1}+\eta_\text{s} \le \eta \le \eta_i - \eta_\text{s}$}\\
    \vdots&\vdots\\
    2a_{n-2}&\text{if $\eta_{n-2} - \eta_\text{s} < \eta < \eta_{n-2} + \eta_\text{s}$}\\
    0&\text{if $\eta_{n-2}+\eta_\text{s} \le \eta < +\infty $},\\
    \end{cases}
\end{equation}
with $a_i$ defined in Eq.~(\ref{eq:a_i_pw}).

\newpage
\section{Installation and simulation}
\subsection{Installation}
Simulations shown in this article have been produced with the version~1.0.1 of this toolkit. It is available on PyPI and can easily be installed by running
\begin{center}
    \texttt{python -m pip install springable==1.0.1}
\end{center} on Windows, or
\begin{center}
    \texttt{python3 -m pip install springable==1.0.1}
\end{center}
on MacOS and Linux. It is supported on Python versions~$\ge 3.10$.

\subsection{Simulating a model}
To simulate a model, you can execute the following Python script
\begin{tcolorbox}[title={\texttt{run\_simulation.py}}]
\begin{lstlisting}[language=Python,
  basicstyle=\ttfamily\small,
  keywordstyle=\color{blue},
  stringstyle=\color{darkgreen},
  commentstyle=\color{gray},
  showstringspaces=false,
  breaklines=true]
from springable.simulation import simulate_model

simulate_model('my_model.csv')
\end{lstlisting}
\end{tcolorbox}
\noindent by running
\begin{center}
    \texttt{python run\_simulation.py}
\end{center}
on Windows, or
\begin{center}
    \texttt{python3 run\_simulation.py}
\end{center}
on MacOS and Linux. The CSV file \texttt{my\_model.csv} describes the model (a flexel assembly subject to loading conditions). All the details on how to write such a file is described in section~\ref{section:model_file_spec}. The CSV model files used in the main article are provided in the section~\ref{section:model_descriptions}.

\newpage
\section{Model file specifications}
\label{section:model_file_spec}

A model is described in a CSV file (text file saved with extension \verb|.csv|). The CSV file is composed of three necessary parts.
\begin{itemize}
    \item The first part specifies the initial positions of the nodes and their boundary conditions (whether they are able to move horizontally and vertically). This is described in the \verb|NODES| section in the CSV file.
    \item The second part specifies how the nodes are coupled by flexels. This is described in multiple sections, one per type of flexels: \verb|LONGITUDINAL FLEXELS|, \verb|ANGULAR FLEXELS|, \verb|AREA FLEXELS|, \verb|DISTANCE FLEXELS|, \verb|X DISTANCE FLEXELS|, \verb|Y DISTANCE FLEXELS| and \verb|PATH FLEXELS|.
    \item The third part specifies how the structure is loaded, by defining one or multiple load steps wherein nodes are loaded along specific directions. This is described in the \verb|LOADING| section.
\end{itemize}

Many examples with the accompanying model depictions are provided in section~\ref{section:model_descriptions}.

For instructions on how to install the \texttt{springable}, you can refer to section~5.

\subsection{{The \texttt{NODES} section}}
The \verb|NODES| section start with a line named \verb|NODES| followed by multiple lines, one per node, with the following specification:
\begin{tcolorbox}
    \verb|<node index>, <x>, <y>, <fixed along x>, <fixed along y>|
\end{tcolorbox}
where
\begin{itemize}
    \item \verb|<node index>| is the index of the node (its label), it must be a  natural number (0, 1, 2, ...);
    \item  \verb|<x>| is the initial $x$-coordinate of the node (initial horizontal position), it can be any real number;
    \item  \verb|<y>| is the initial $y$-coordinate of the node (initial vertical position), it can be any real number;
    \item  \verb|<fixed along x>| is either `0' if the node is \emph{not} fixed along $x$ (that is, free to move horizontally), or `1' if the node is fixed along $x$;
    \item  \verb|<fixed along y>| is either `0' if the node is \emph{not} fixed along $y$ (that is, free to move vertically), or `1' if the node is fixed along $y$.
\end{itemize}

\paragraph{Example} Three nodes labeled `0', `1' and `2' are defined at initial positions $(0, 0)$, $(3, 2)$, $(6, 0)$. Nodes~0 and 2 are fixed both horizontally and vertically. Node~1 is free to move vertically, but not horizontally.
\begin{tcolorbox}
\begin{lstlisting}
NODES
0, 0.0, 0.0, 1, 1
1, 3.0, 2.0, 1, 0
2, 6.0, 0.0, 1, 1
\end{lstlisting}
\end{tcolorbox}
\paragraph{Notes}
\begin{itemize}
    \item Nodes can be defined in any order, as long as no node index is missing at the end of the section. If there are 6 nodes, node indices `0', `1', `2', `3', `4' and `5' must be used, but the order does not matter.
    \item Nodes that are not intended to be coupled by flexels can still be defined. They must be fixed along $x$ and $y$, as otherwise, zero modes (a.k.a. rigid body modes) will be present.
    \item When a node is defined, its initial coordinates can be used in math expressions, in order to, for example, facilitate the positioning of subsequent nodes. For example, once node `2' is defined, `X2' and `Y2' can be use as variables in the rest of the CSV file.
\end{itemize}

\subsection{{The \texttt{FLEXELS} sections}}
Flexels sharing the same type of geometric measure are grouped in separate sections. Flexels whose geometric measure is
\begin{itemize}
    \item a length (subsection~\ref{subsection:length}, Fig.~\ref{fig:geometric_measures}a) are grouped under section named \verb|LONGITUDINAL FLEXELS|;
    \item an angle (subsection~\ref{subsection:angle}, Fig.~\ref{fig:geometric_measures}b) under section named \verb|ANGULAR FLEXELS|;
    \item an area (subsections~\ref{subsection:area}, \ref{subsection:area_with_holes}, Fig.~\ref{fig:geometric_measures}c,g) under section named \verb|AREA FLEXELS|;
    \item a signed $x$-distance (subsection~\ref{subsection:signed_xy_distance}, Fig.~\ref{fig:geometric_measures}d) under section named \verb|X DISTANCE FLEXELS|;
    \item a signed $y$-distance (subsection~\ref{subsection:signed_xy_distance}, Fig.~\ref{fig:geometric_measures}d) under section \verb|Y DISTANCE FLEXELS|;
    \item a signed point-line distance (subsection~\ref{subsection:signed_distance}, Fig.~\ref{fig:geometric_measures}e) under section named \verb|DISTANCE FLEXELS|;
    \item a path length (subsection~\ref{subsection:path_length}, Fig.~\ref{fig:geometric_measures}f) under section named \verb|PATH FLEXELS|.
\end{itemize}
Under each section, a flexel is specified as follows:
\begin{tcolorbox}
    \verb|<list of node indices>, <nonlinear behavior>, <natural measure>|
\end{tcolorbox}
where
\begin{itemize}
    \item \verb|<list of nodes indices>| specifies the nodes coupled by the flexel (see paragraph~\ref{subsubsection:specifying_flexels_nodes} for more details);
    \item  \verb|<nonlinear behavior>| (together with \verb|<natural measure>|) specifies the intrinsic nonlinear of the flexel (see paragraph~\ref{subsubsection:specifying_flexels_behavior} for more details);
    \item  \verb|<natural measure>| describes $\alpha_0$: the natural measure of the flexel, that is its geometric measure at rest. It can be any real number. It is used with the \verb|<nonlinear behavior>| to construct the intrinsic nonlinear behavior of the flexel. It is an optional parameter. If not provided, the natural measure defaults to the the geometric measure computed from the initial positions of the nodes provided in the \verb|NODES| section.
\end{itemize}
\paragraph{Example} Two longitudinal flexels couples nodes `0' and `1' and nodes `1' and `2' respectively. One angular flexel is defined with an associated angle defined by nodes `0', `1' and `2' (with the vertex on node `1'). The longitudinal flexels are characterized by intrinsic Bezier behavior in compression. Their natural length (rest length) is the distance computed from the initial positions of their nodes. The angular flexel is characterized by a linear behavior (with spring constant 1.0) and a rest angle of $\pi$ radians (180$^\text{o}$).
\begin{tcolorbox}
\begin{lstlisting}[basicstyle=\ttfamily\color{gray}]
NODES
0, 0.0, 0.0, 1, 1
1, 3.0, 2.0, 1, 0
2, 6.0, 0.0, 1, 1
\end{lstlisting}
\begin{lstlisting}
LONGITUDINAL FLEXELS
0-1, BEZIER(u_i=[0.83; 0.74; 2.0]; f_i=[0.48; -0.83; 0.52]; mode=-1)
1-2, BEZIER2(u_i=[2.9; -2.3; 2.8]; f_i=[0.73; -1.0; 0.38]; mode=-1)
\end{lstlisting}
\begin{lstlisting}
ANGULAR FLEXELS
0-1-2, LINEAR(k=1.0), PI
\end{lstlisting}
\end{tcolorbox}

\subsubsection{Specifying the flexel's nodes}
\label{subsubsection:specifying_flexels_nodes}
\begin{itemize}
    \item For a longitudinal flexel, use
    \begin{tcolorbox}
\verb|<node N0 index>-<node N1 index>| 
    \end{tcolorbox}
    (see subsection~\ref{subsection:length}).
    \item For an angular flexel, use
    \begin{tcolorbox}
\verb|<node N0 index>-<node N1 index>-<node N2 index>| 
    \end{tcolorbox} (see subsection~\ref{subsection:angle}).
    \item For an area flexel, use
    \begin{tcolorbox}
\verb|<node N0 index>-<node N1 index>-...-<node N n-1 index>| 
    \end{tcolorbox} (see subsection~\ref{subsection:area}).
    \item For an $x$- (or a $y$-) distance flexel, use
    \begin{tcolorbox}
\verb|<node N0 index>-<node N1 index>| 
    \end{tcolorbox} (see subsection~\ref{subsection:signed_xy_distance}).
    \item For a distance flexel, use
    \begin{tcolorbox}
\verb|<node N0 index>-<node N1 index>-<node N2 index>| 
    \end{tcolorbox} (see subsection~\ref{subsection:signed_distance}).
    \item For a path flexel, use
    \begin{tcolorbox}
\verb|<node N0 index>-<node N1 index>-...-<node Np index>| 
    \end{tcolorbox} (see subsection~\ref{subsection:path_length}).
    \item For an area-with-holes flexel, use
    \begin{tcolorbox}
\small\verb|(<area0 node indices>)-(<area1 node indices>)-...-(<area p-1 node indices>)| 
    \end{tcolorbox} (see subsection~\ref{subsection:area_with_holes}).
    Example: \verb|(0-1-2-3-4)-(5-6-7)-(8-9-10-11)|.
    Note that flexels whose geometric measure is an area or an area with holes are grouped under the same section \verb|AREA FLEXELS|. Their actual geometric type will be determined from the way that their nodes are specified. 
\end{itemize}
\subsubsection{Specifying the flexel's intrinsic nonlinear behavior}
\label{subsubsection:specifying_flexels_behavior}
\begin{itemize}
\item For a linear behavior, use
\begin{tcolorbox}
    \verb|LINEAR(k=<k>)|
\end{tcolorbox}
where \verb|<k>| is the spring constant (see Eq.~(\ref{eq:linear_behavior})).
\item For a logarithmic behavior, use
\begin{tcolorbox}
    \verb|LOGARITHMIC(k=<k>)|
\end{tcolorbox}
(see Eq.~(\ref{eq:logarithmic_behavior})).
\item For a univariate Bezier behavior, use
\begin{tcolorbox}
    \small\verb|BEZIER(u_i=[<u1>; <u2>; ... ; <un>]; f_i=[<f1>; <f2>; ... ; <fn>]; mode=<mode>)|
\end{tcolorbox}
(see Eq.~(\ref{eq:bezier_fd_curve})). \verb|<mode>| is either `1' for `tensile' mode, `-1` for `compressive mode' or `0' for `symmetric mode'. It is an optional parameter; if not provided, it defaults to `0'.
\item For a $\mathcal{C}_1$-piecewise beahvior, use
\begin{tcolorbox}
    \footnotesize\verb|PIECEWISE(k_i=[<k0>; ...; <k n-1>]; u_i=[<u0>; ... ; <u n-2>]; us=<us>; mode=<mode>)|
\end{tcolorbox}
(see Eq.~(\ref{eq:c1pw_fd_curve})). \verb|<mode>| is either `1' for `tensile' mode, `-1` for `compressive mode' or `0' for `symmetric mode'. It is an optional parameter; if not provided, it defaults to `0'.
\item For a univariate zigzag behavior, use
\begin{tcolorbox}
    \footnotesize\verb|ZIGZAG(u_i=[<u1>; ... ; <un>]; f_i=[<f1>; ... ; <fn>]; epsilon=<epsilon>; mode=<mode>)|
\end{tcolorbox}
(see Eq.~(\ref{eq:zz_fd_curve})). \verb|<mode>| is either `1' for `tensile' mode, `-1` for `compressive mode' or `0' for `symmetric mode'. It is an optional parameter; if not provided, it defaults to `0'.
\item For a contact behavior, use
\begin{tcolorbox}
    \verb|CONTACT(f0=<f0>; uc=<uc>; delta=<alpha_delta>)|
\end{tcolorbox}
(see Eq.~(\ref{eq:contact_fd_curve})).
\item For an isothermal behavior, use
\begin{tcolorbox}
    \verb|ISOTHERMAL(n=<n>; R=<R>; T0=<T0>)|
\end{tcolorbox}
(see Eq.~(\ref{eq:isothermal_curve})).
\item For an isentropic behavior, use
\begin{tcolorbox}
    \verb|ISENTROPIC(n=<n>; R=<R>; T0=<T0>; gamma=<gamma>)|
\end{tcolorbox}
(see Eq.~(\ref{eq:isentropic_curve})).
\item For a bivariate Bezier behavior, use
\begin{tcolorbox}
    \small\verb|BEZIER2(u_i=[<u1>; <u2>; ... ; <un>]; f_i=[<f1>; <f2>; ... ; <fn>]; mode=<mode>)|
\end{tcolorbox}
(see Eqs.~(\ref{eq:bivariate_bezier_a}, \ref{eq:bivariate_bezier_b})). \verb|<mode>| is either `1' for `tensile' mode, `-1` for `compressive mode' or `0' for `symmetric mode'. It is an optional parameter; if not provided, it defaults to `0'.
\item For a bivariate zigzag behavior, use
\begin{tcolorbox}
    \footnotesize\verb|ZIGZAG2(u_i=[<u1>; ... ; <un>]; f_i=[<f1>; ... ; <fn>]; epsilon=<epsilon>; mode=<mode>)|
\end{tcolorbox}
(see Eqs.~(\ref{eq:bivariate_zz_a}, \ref{eq:bivariate_zz_b})). \verb|<mode>| is either `1' for `tensile' mode, `-1` for `compressive mode' or `0' for `symmetric mode'. It is an optional parameter; if not provided, it defaults to `0'.
\end{itemize}

\begin{figure}[ht]
    \centering
    \includegraphics[width=0.8\textwidth]{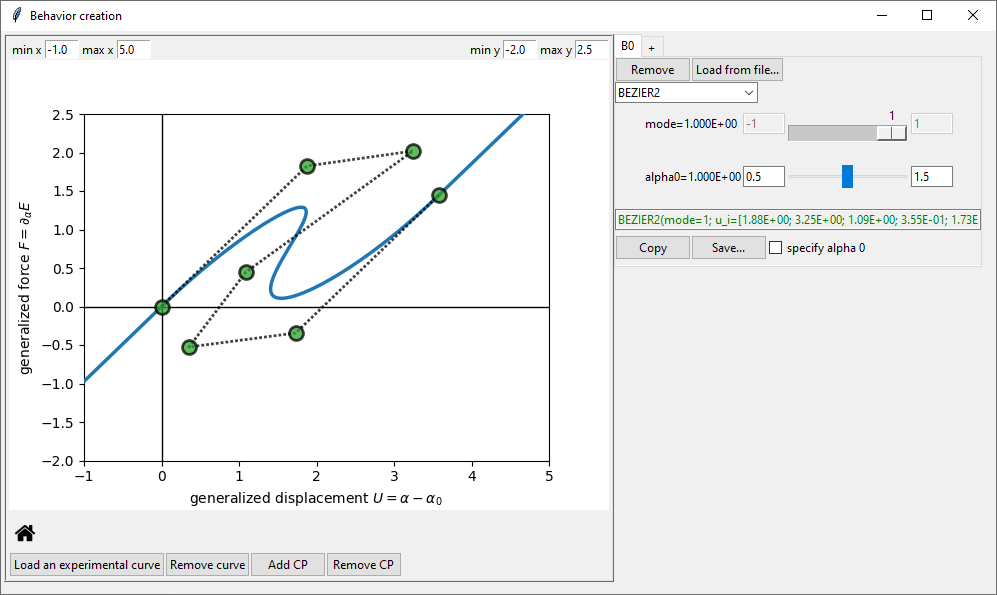}
    \caption{Screenshot of the behavior creation graphical interface, which can be used to make custom nonlinear behavior and generate the corresponding code to be pasted or imported in the model CSV file.}
    \label{fig:screenshot_gui}
\end{figure}

\paragraph{Notes}
\begin{itemize}
    \item A nonlinear behavior can be saved in a separate CSV file and used in a model file using
\begin{tcolorbox}
    \verb|FROMFILE(<nonlinear behavior csv file>)|
\end{tcolorbox}
Example:
\begin{tcolorbox}
\begin{lstlisting}[basicstyle=\ttfamily\color{gray}]
NODES
0, 0.0, 0.0, 1, 1
1, 3.0, 2.0, 1, 0
2, 6.0, 0.0, 1, 1
\end{lstlisting}
\begin{lstlisting}[basicstyle=\ttfamily\color{gray}]
LONGITUDINAL FLEXELS
0-1, BEZIER(u_i=[0.83; 0.74; 2.0]; f_i=[0.48; -0.83; 0.52]; mode=-1)
1-2, BEZIER2(u_i=[2.9; -2.3; 2.8]; f_i=[0.73; -1.0; 0.38]; mode=-1)
\end{lstlisting}
\begin{lstlisting}
ANGULAR FLEXELS
0-1-2, FROMFILE('custom_nonlinear_behavior.csv'), PI
\end{lstlisting}
\end{tcolorbox}
where the \verb|custom_nonlinear_behavior.csv| is for example:
\begin{tcolorbox}[title=\texttt{custom\_nonlinear\_behavior.csv}]
\begin{lstlisting}
BEZIER2(u_i=[0.21; -0.1; 3.14]; f_i=[1.0; -2.0; +3.0]; mode=0)
\end{lstlisting}
\end{tcolorbox}
The file path to the behavior is relative to the working directory, that is, the directory from where the script is run. If the CSV behavior file lives in a subdirectory \texttt{path/to/behavior.csv} relative to the working directory, then we would use
\begin{tcolorbox}
    \verb|FROMFILE('path'; 'to'; 'behavior.csv')|
\end{tcolorbox}
To specify a CSV behavior file that would live in a subdirectory \texttt{relative/path/to/behavior.csv} relative to the CSV model file instead, we can use the keyword \texttt{HERE} that encodes the directory where the CSV model file lives (relative to the working directory) as follows:
\begin{tcolorbox}
    \verb|FROMFILE(HERE; 'relative; 'path'; 'to'; 'behavior.csv')|
\end{tcolorbox}

\item Nonlinear behaviors can be interactively tuned and created using the `behavior creation' graphical interface (Fig.~\ref{fig:screenshot_gui}), which can be started by running the following Python script
\begin{tcolorbox}
\begin{lstlisting}[language=Python,
  basicstyle=\ttfamily\small,
  keywordstyle=\color{blue},
  stringstyle=\color{darkgreen},
  commentstyle=\color{gray},
  showstringspaces=false,
  breaklines=true]
from springable.behavior_creation import start

start()
\end{lstlisting}
\end{tcolorbox} 
\end{itemize}

\subsection{The \texttt{LOADING} section}
The \verb|LOADING| section is composed of one or multiple load steps. Each load step describes an additional force load acting on the nodes, relative to the end state reached at the previous load step. When multiple nodes are loaded within the same load step, each additional force component will evolve in proportion to a single parameter. In other words, each load step is characterized by a step load vector direction $\tilde{\bm{F}}^\text{dir}$ that remains constant during the entire load step (Eq.~(\ref{eq:parametrized_sys})). Before each load step, some nodes that were free to move during the previous load step can be blocked (along $x$, $y$ or both), so that they remain fixed (along $x$, $y$ or both) in the state reached at the end of the previous load step. The starting state of the first load step starts in the state reached by minimizing the total elastic energy $E$ (Eq.~(\ref{eq:total_elastic_energy})) using the configuration described in the \verb|NODES| section as initial guess. The \verb|LOADING| section is structured as follows:
\begin{tcolorbox}
\begin{lstlisting}
<load step>
then
<load step>
then
...
\end{lstlisting}
\end{tcolorbox}
where each \verb|<load step>| is composed of one of multiple lines, one per loaded coordinate, as follows:
\begin{tcolorbox}
\begin{lstlisting}
<node index>, <direction>, <force>, <max displacement>
\end{lstlisting}
\end{tcolorbox}
with
\begin{itemize}
\item \verb|<node index>| is the index of the node on which the force is applied,
\item \verb|<direction>| is either `X' or `Y' depending on whether the force is applied along the $x$ or $y$ coordinate of the node,
\item \verb|<force>| is the signed magnitude of the additional force applied on node \verb|<node index>| along \verb|<direction>| (negative or positive real number),
\item \verb|<max displacement>| is the signed magnitude of the additional displacement of node \verb|<node index>| along \verb|<direction>| beyond which the load step is considered completed (positive of negative real number). It is an optional parameter; if not provided, the load step ends when the applied force is reached.
\end{itemize}
In addition to these `nodal load lines', the \verb|<load step>| can optionally start with a list of coordinates that were free that need to be blocked before applying the load step, using the following specification:
\begin{tcolorbox}
\begin{lstlisting}
block
<node index>, <direction>
<node index>, <direction>
...
\end{lstlisting}
\end{tcolorbox}
where each \verb|<load index>| is the index of the node that will be blocked along the direction specified in \verb|<direction>|.

\paragraph{Example} A load of $-3.0$ is first applied along the vertical direction of node 3 (so, a downward force of magnitude 3.0). When the prescribed force load is reached or when node 3 is moved downward by more than 4.0 (whichever comes first), the load step is completed. Then, node 3 is blocked vertically and node 1 is block horizontally, followed by the application of a vertical force of magnitude 6.0 on node 1.
\begin{tcolorbox}
\begin{lstlisting}[basicstyle=\ttfamily\color{gray}]
NODES
0, 0.0, 0.0, 1, 1
1, 3.0, 2.0, 0, 0
2, 6.0, 0.0, 1, 1
3, 0.0, -1.0, 1, 0
\end{lstlisting}
\begin{lstlisting}[basicstyle=\ttfamily\color{gray}]
LONGITUDINAL FLEXELS
0-1, BEZIER(u_i=[0.83; 0.74; 2.0]; f_i=[0.48; -0.83; 0.52]; mode=-1)
1-2, BEZIER(u_i=[0.81; 0.72; 2.0]; f_i=[0.45; -0.81; 0.52]; mode=-1)
1-3, LINEAR(k=2.0)
\end{lstlisting}
\begin{lstlisting}[basicstyle=\ttfamily\color{gray}]
ANGULAR FLEXELS
0-1-2, LINEAR(k=1.0), PI
\end{lstlisting}
\begin{lstlisting}
LOADING
3, Y, -3.0, -4.0
then
block
3, Y
1, X
1, Y, 6.0
\end{lstlisting}
\end{tcolorbox}
\paragraph{Notes}
\begin{itemize}
    \item When multiple loads are applied within the same load step, they will evolve proportionally in \emph{force}, not \emph{displacement}.
    \item Each load is always relative to the state reached at the previous load step. So, when multiple load steps are specified, each load (force and max displacement) is \emph{not absolute}, but relative. 
    \item Within a single load step, the same node can be loaded multiple times. If the direction is the same, the loads will be simply added together to make an equivalent load. If the directions are different, the node will be effectively loaded in a specific direction that is not purely horizontal or vertical.
\end{itemize}

\subsection{The \texttt{PARAMETERS} section}
Optionally, parameters can be defined at the start of the CSV model file, under a separate section starting with \verb|PARAMETERS| using the following specification for each parameter:
\begin{tcolorbox}
\begin{lstlisting}
<parameter name>, <parameter value>
\end{lstlisting}
\end{tcolorbox}
where
\begin{itemize}
    \item \verb|<parameter name>| is the name of the parameter (which should only contains letters, numbers and underscores `\verb|_|', no space or other special characters), it is also recommended to use lowercase only, as uppercase are reserved for keywords;
    \item \verb|<parameter value>| is the value of the parameter, it can be any real number, or a string of text between single quotes (which can be useful if the parameter is the name of the custom nonlinear behavior file).
\end{itemize}
\paragraph{Example} Multiple parameters are defined: \verb|width|=6, \verb|height|=3, \verb|angle|=45,\\\verb|nonlinear_behavior0|=\verb|'curve0.csv'|,
\verb|nonlinear_behavior1|=\verb|'curve1.csv'|,\\
\verb|stiffness|=1, \verb|f|=3.
\begin{tcolorbox}
\begin{lstlisting}
PARAMETERS
width, 6.0
height, 3.0
angle, 45
nonlinear_behavior0, 'curve0.csv'
nonlinear_behavior1, 'curve1.csv'
stiffness, 1.0
f, 3.0
\end{lstlisting}
\begin{lstlisting}[basicstyle=\ttfamily\color{gray}]
NODES
0, 0.0, 0.0, 1, 1
1, width/2, width/2*TAN(angle/180*PI), 1, 0
2, width, 0.0, 1, 1
3, 0.0, -width/3, 1, 0
\end{lstlisting}
\begin{lstlisting}[basicstyle=\ttfamily\color{gray}]
LONGITUDINAL FLEXELS
0-1, FROMFILE(nonlinear_behavior0)
1-2, FROMFILE(nonlinear_behavior0)
1-3, FROMFILE(nonlinear_behavior1)
\end{lstlisting}
\begin{lstlisting}[basicstyle=\ttfamily\color{gray}]
ANGULAR FLEXELS
0-1-2, LINEAR(k=stiffness), PI
\end{lstlisting}
\begin{lstlisting}[basicstyle=\ttfamily\color{gray}]
LOADING
3, Y, -f, -4.0
\end{lstlisting}
\end{tcolorbox}
\paragraph{Notes}
\begin{itemize}
    \item Numerical parameters can be combined in mathematical expressions using operators \verb|+|, \verb|-|, \verb|*|, \verb|/|, \verb|**| (exponentiation) and functions \verb|SIN()|, \verb|COS()|, \verb|TAN()|, \verb|ARCSIN()|, \verb|ARCCOS()|, \verb|ARCTAN()|, \verb|SQRT()|, anywhere in the model file \emph{outside the parameter section}, where a real number is expected (so numerical parameters cannot be used to describe a node index).
    \item A parameter cannot be defined as a function of other parameters.
    \item The parameter \verb|PI| can be used directly without having to define it to represent the number $\pi$.
    \item Once a node is defined (with, let's say, node index = 4), parameters \verb|X4| and \verb|Y4| are automatically available to describe the initial $x$ and $y$ positions of node~4 (as described in the \verb|NODES| section).
\end{itemize}

\newpage
\section{Model descriptions}
\label{section:model_descriptions}
The model CSV files used in the main manuscript are provided in this section. To simulate all those models, the following Python script can be run.
\begin{tcolorbox}
\begin{lstlisting}[language=Python,
  basicstyle=\ttfamily\small,
  keywordstyle=\color{blue},
  stringstyle=\color{darkgreen},
  commentstyle=\color{gray},
  showstringspaces=false,
  breaklines=true]
from springable.discover import run_simulations_from_article

run_simulations_from_article()
\end{lstlisting}
\end{tcolorbox}

To simulate the model of each individual figure, you can refer to the Python scripts specified at the end of each subsection 7.x (see herein below).

For instructions on how to install the \texttt{springable}, you can refer to section~5.

\subsection{Figure 1}
\subsubsection{Nonmonotonic Von-Mises truss (Fig.~1a)}
\begin{tcolorbox}[title=Model CSV file for Fig.~1a: \texttt{fig1a\_model.csv}]
\begin{lstlisting}
PARAMETERS
alpha, 45.0
l, 1.0
NODES
0, 0.0, 0.0, 1, 1
1, l*COS(alpha/180*PI), l*SIN(alpha/180*PI), 0, 0
2, 2*l*COS(alpha/180*PI), 0.0, 1, 1
3, l*COS(alpha/180*PI), Y1-l, 1, 0
LONGITUDINAL FLEXELS
0-1, LINEAR(k=0.6)
1-2, LINEAR(k=0.6)
1-3, LINEAR(k=20.0)
LOADING
3, Y, -0.5, -1.2*2*l*SIN(alpha/180*PI)

\end{lstlisting}
\begin{center}
\includegraphics[]{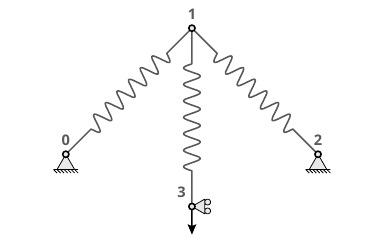}
\end{center}
\end{tcolorbox}

\subsubsection{Multi-valued Von-Mises truss (Fig.~1b)}
\begin{tcolorbox}[title=Model CSV file for Fig.~1b: \texttt{fig1b\_model.csv}]
\begin{lstlisting}
PARAMETERS
alpha, 45.0
l, 1.0
NODES
0, 0.0, 0.0, 1, 1
1, l*COS(alpha/180*PI), l*SIN(alpha/180*PI), 0, 0
2, 2*l*COS(alpha/180*PI), 0.0, 1, 1
3, l*COS(alpha/180*PI), Y1-l, 1, 0
LONGITUDINAL FLEXELS
0-1, LINEAR(k=1.0)
1-2, LINEAR(k=1.0)
1-3, LINEAR(k=0.33)
LOADING
3, Y, -0.5, -1.2*2*l*SIN(alpha/180*PI)
\end{lstlisting}
\begin{center}
\includegraphics[]{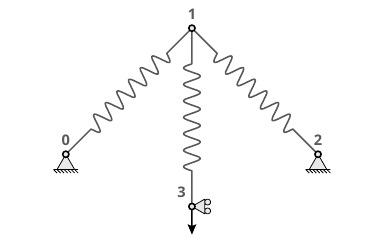}
\end{center}
\end{tcolorbox}

\subsubsection{Nonlinear flexels in series (Fig.~1e)}
\begin{tcolorbox}[title=Model CSV file for Fig.~1e: \texttt{fig1e\_model.csv}]
\begin{lstlisting}
NODES
0, 0.0, 0.0, 1, 1
1, 4.0, 0.0, 0, 1
2, 8.0, 0.0, 0, 1
LONGITUDINAL FLEXELS
0-1, FROMFILE('fig1c_behavior.csv')
2-1, FROMFILE('fig1d_behavior.csv')
LOADING
2, X, -0.325, -3.5
\end{lstlisting}
\begin{center}
\includegraphics[]{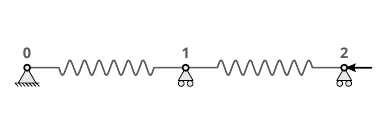}
\end{center}
\end{tcolorbox}

\subsubsection{Nonlinear flexels connected at an angle (Fig.~1f)}
\begin{tcolorbox}[title=Model CSV file for Fig.~1f: \texttt{fig1f\_model.csv}]
\begin{lstlisting}
NODES
0, 0.0, 0.0, 1, 1
1, 2.5, 3.0, 0, 0
2, 5.0, 0.0, 1, 1
LONGITUDINAL FLEXELS
0-1, FROMFILE('fig1c_behavior.csv')
1-2, FROMFILE('fig1d_behavior.csv')
LOADING
1, Y, -2.0, -5.5
\end{lstlisting}
\begin{center}
\includegraphics[]{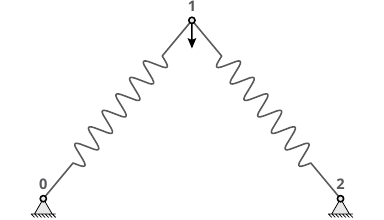}
\end{center}
\end{tcolorbox}

\subsubsection{Nonlinear behaviors (Fig.~1c,d,g,h)}
\begin{tcolorbox}[title=Nonlinear behavior of Fig.~1c: \texttt{fig1c\_behavior.csv}]
\begin{lstlisting}[basicstyle=\scriptsize\ttfamily]
BEZIER(u_i=[8.323E-01; 7.419E-01; 2.019E+00]; f_i=[4.784E-01; -8.377E-01; 5.216E-01]; mode=-1)
\end{lstlisting}
\end{tcolorbox}
\begin{tcolorbox}[title=Nonlinear behavior of Fig.~1d: \texttt{fig1d\_behavior.csv}]
\begin{lstlisting}[basicstyle=\scriptsize\ttfamily]
BEZIER2(u_i=[2.931E+00; -2.323E+00; 2.841E+00]; f_i=[7.294E-01; -1.045E+00; 3.831E-01]; mode=-1)
\end{lstlisting}
\end{tcolorbox}
\begin{tcolorbox}[title=Nonlinear behavior of Fig.~1g: \texttt{fig1g\_behavior.csv}]
\begin{lstlisting}[basicstyle=\scriptsize\ttfamily]
ZIGZAG2(u_i=[9.355E-01; 4.065E-01; 2.845E+00; 1.490E+00; 1.335E+00; -1.484E-01; 2.381E+00; 1.839E+00; 3.503E+00]; f_i=[1.395E-01; -1.633E-01; 2.386E-01; -1.652E-01; 1.605E-01; -2.490E-01; 1.624E-01; -1.576E-01; 1.243E-01]; epsilon=7.500E-01; mode=-1)
\end{lstlisting}
\end{tcolorbox}
\begin{tcolorbox}[title=Nonlinear behavior of Fig.~1h: \texttt{fig1h\_behavior.csv}]
\begin{lstlisting}[basicstyle=\scriptsize\ttfamily]
ZIGZAG2(u_i=[6.742E-01; 3.806E-01; 1.649E+00; 2.477E+00; 1.765E+00; 1.032E+00; 9.484E-01; -1.086E-01; 1.589E+00; 1.996E+00; 1.303E+00; 2.206E+00; 3.682E+00; 4.690E+00; 4.028E+00; 4.735E+00; 5.533E+00]; f_i=[2.085E-01; -2.071E-01; 1.530E-01; 9.885E-02; 5.267E-02; -1.911E-01; 2.460E-01; -3.543E-01; 1.190E-01; 1.017E-01; -1.465E-01; -1.371E-02; 1.227E-02; 1.553E-01; -1.356E-01; -5.714E-02; 1.830E-01]; epsilon=7.500E-01; mode=-1)
\end{lstlisting}
\end{tcolorbox}

\subsubsection{Simulation script}
To simulate the models related to Fig.~1, the following script can be run. The CSV behavior files must live in the same (working) directory as the CSV model files.
\begin{tcolorbox}
\begin{lstlisting}[language=Python,
  basicstyle=\ttfamily\small,
  keywordstyle=\color{blue},
  stringstyle=\color{darkgreen},
  commentstyle=\color{gray},
  showstringspaces=false,
  breaklines=true]
from springable.simulation import simulate_model

simulate_model('fig1a_model.csv')
simulate_model('fig1b_model.csv')
simulate_model('fig1e_model.csv')
simulate_model('fig1f_model.csv')
\end{lstlisting}
\end{tcolorbox}

\subsection{Figure 3}
\subsubsection{Example of nonlinear angular flexel (Fig.~3a)}
\begin{tcolorbox}[title=Model CSV file of Fig.~3a: \texttt{fig3a\_model.csv}]
\begin{lstlisting}
NODES
0, -1.0, 0.0, 1, 1
1, 1.0, 0.0, 0, 1
2, 0.0, 0.15, 0, 0
ANGULAR FLEXELS
1-2-0, FROMFILE('fig3a_behavior.csv')
LONGITUDINAL FLEXELS
0-2, LINEAR(k=1.0)
2-1, LINEAR(k=1.0)
LOADING
2, Y, -5.0
\end{lstlisting}
\begin{center}
\includegraphics[]{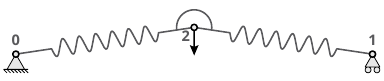}
\end{center}
\end{tcolorbox}
\begin{tcolorbox}[title=Nonlinear behavior used in model of Fig.~3a: \texttt{fig3a\_behavior.csv}]
\begin{lstlisting}[basicstyle=\scriptsize\ttfamily]
BEZIER2(u_i=[2.0E-01; 10E-01; 10E-01; -13.33E-02; -2E-01; 3.3E-01; 6.774E-01]; f_i=[2.749E+00; 3.297E+00; 1.515E-01; 1.623E+00; 1.190E+00; -2.648E+00; 1.364E+00])
\end{lstlisting}
\end{tcolorbox}

\subsubsection{Example of nonlinear area flexel (Fig.~3b)}
\begin{tcolorbox}[title=Model CSV file of Fig.~3b: \texttt{fig3b\_model.csv}]
\begin{lstlisting}[basicstyle=\scriptsize\ttfamily]
PARAMETERS
diameter, 10.0
alpha0, 0.675
kl, 2.0
kr, 0.075
NODES
0, diameter/2 * COS(alpha0 + (PI-2*alpha0) / 4 * 0), diameter/2 * SIN(alpha0 + (PI-2*alpha0) / 4 * 0), 1, 1
1, diameter/2 * COS(alpha0 + (PI-2*alpha0) / 4 * 1), diameter/2 * SIN(alpha0 + (PI-2*alpha0) / 4 * 1), 0, 0
2, diameter/2 * COS(alpha0 + (PI-2*alpha0) / 4 * 2), diameter/2 * SIN(alpha0 + (PI-2*alpha0) / 4 * 2), 0, 0
3, diameter/2 * COS(alpha0 + (PI-2*alpha0) / 4 * 3), diameter/2 * SIN(alpha0 + (PI-2*alpha0) / 4 * 3), 0, 0
4, diameter/2 * COS(alpha0 + (PI-2*alpha0) / 4 * 4), diameter/2 * SIN(alpha0 + (PI-2*alpha0) / 4 * 4), 1, 1
5, X4, Y4-diameter/7, 1, 0
6, (X0+X4)/2, Y4-diameter/7, 1, 0
7, X0, Y4-diameter/7, 1, 0
LONGITUDINAL FLEXELS
0-1, LINEAR(k=kl)
1-2, LINEAR(k=kl)
2-3, LINEAR(k=kl)
3-4, LINEAR(k=kl)
ANGULAR FLEXELS
7-0-1, LINEAR(k=0.99*kr)
0-1-2, LINEAR(k=0.98*kr)
1-2-3, LINEAR(k=1.01*kr)
2-3-4, LINEAR(k=1.00*kr)
3-4-5, LINEAR(k=1.02*kr)
6-5-4, LINEAR(k=40.0)
7-6-5, LINEAR(k=40.0)
0-7-6, LINEAR(k=40.0)
AREA FLEXELS
0-1-2-3-4-5-6-7, ISOTHERMAL(n=0.14; R=1.0; T0=4.0)
LOADING
6, Y, -0.25, -diameter*1.2
\end{lstlisting}
\begin{center}
\includegraphics[]{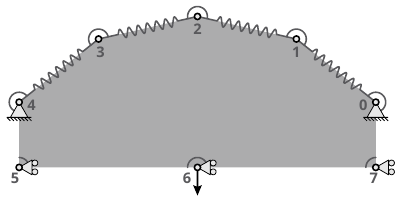}
\end{center}
\end{tcolorbox}

\subsubsection{Example of nonlinear path flexel (Fig.~3c)}
\begin{tcolorbox}[title=Model CSV file of Fig.~3c: \texttt{fig3c\_model.csv}]
\begin{lstlisting}
NODES
0, 0.0, 0.0, 1, 1
1, 0.5, 0.5, 0, 0
2, -1.5, 0.0, 1, 1
3, -2.0, 0.0, 0, 1
LONGITUDINAL FLEXELS
0-1, LINEAR(k=20.0)
ANGULAR FLEXELS
1-0-2, LINEAR(k=0.2)
PATH FLEXELS
1-2-3, PIECEWISE(k_i=[0.02;10.0];u_i=[0.5];us=0.01)
LOADING
3, X, -1.25
\end{lstlisting}
\begin{center}
\includegraphics[]{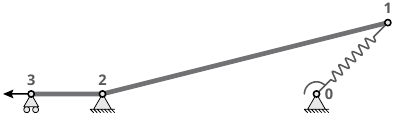}
\end{center}
\end{tcolorbox}

\subsubsection{Example of nonlinear distance flexel (Fig.~3d)}
\begin{tcolorbox}[title=Model CSV file of Fig.~3d: \texttt{fig3d\_model.csv}]
\begin{lstlisting}
NODES
0, 0.0, 0.0, 1, 1
1, 1.0, 0.35, 0, 0
2, 2.0, 0.0, 0, 1
3, 0.0, 0.5, 1, 1
4, 2.0, 1.0, 1, 1
LONGITUDINAL FLEXELS
0-1, LINEAR(k=20.0)
1-2, LINEAR(k=20.0)
ANGULAR FLEXELS
0-1-2, LINEAR(k=2.0)
DISTANCE FLEXELS
1-4-3, CONTACT(f0=3.0;uc=0.05;delta=0.0)
LOADING
2, X, -10.0
\end{lstlisting}
\begin{center}
\includegraphics[]{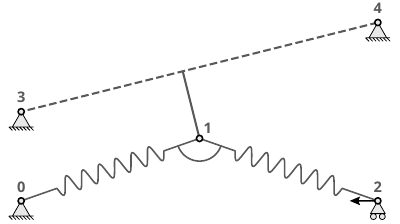}
\end{center}
\end{tcolorbox}

\subsubsection{Simulation script}
To simulate the models related to Fig.~3, the following script can be run. The CSV behavior files must live in the same (working) directory as the CSV model files.
\begin{tcolorbox}
\begin{lstlisting}[language=Python,
  basicstyle=\ttfamily\small,
  keywordstyle=\color{blue},
  stringstyle=\color{darkgreen},
  commentstyle=\color{gray},
  showstringspaces=false,
  breaklines=true]
from springable.simulation import simulate_model

simulate_model('fig3a_model.csv')
simulate_model('fig3b_model.csv')
simulate_model('fig3c_model.csv')
simulate_model('fig3d_model.csv')
\end{lstlisting}
\end{tcolorbox}

\subsection{Figure 4}
\subsubsection{Model of the two nonlinear building blocks in series (Fig.~4d)}
\begin{tcolorbox}[title=Model CSV file of Fig.~4d: \texttt{fig4d\_model.csv}]
\begin{lstlisting}
NODES
0, 0.0,  0.0, 1, 1
1, 25.0, 0.0, 0, 1
2, 50.0, 0.0, 0, 1
LONGITUDINAL FLEXELS
0-1, FROMFILE('block2_fig4c_behavior.csv')
1-2, FROMFILE('block1_fig4c_behavior.csv')
LOADING
1, X, -2.0*9.81/1000
then
2, X, 1.0, 25
\end{lstlisting}
\begin{center}
\includegraphics[]{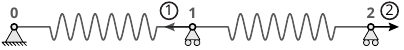}
\end{center}
\end{tcolorbox}

\begin{tcolorbox}[title=Nonlinear behavior of Fig.~4c-top: \texttt{block1\_fig4c\_behavior.csv}]
\begin{lstlisting}[basicstyle=\scriptsize\ttfamily]
BEZIER(u_i=[3.5435E+00; 4.6615E+00; 9.1484E+00; 1.2964E+01]; f_i=[1.4147E+00; 4.6408E-01; -2.8637E-01; 8.8110E-01])
\end{lstlisting}
\end{tcolorbox}

\begin{tcolorbox}[title=Nonlinear behavior of Fig.~4c-bottom: \texttt{block2\_fig4c\_behavior.csv}]
\begin{lstlisting}[basicstyle=\scriptsize\ttfamily]
BEZIER(u_i=[4.6731E+00; 4.7645E+00; 1.1955E+01; 1.5443E+01]; f_i=[1.6709E+00; 2.4704E-01; -1.8301E-01; 3.8442E-01])
\end{lstlisting}
\end{tcolorbox}

\subsubsection{Simulation script}
To simulate the model related to Fig.~4, the following script can be run. The CSV behavior files must live in the same (working) directory as the CSV model file.
\begin{tcolorbox}
\begin{lstlisting}[language=Python,
  basicstyle=\ttfamily\small,
  keywordstyle=\color{blue},
  stringstyle=\color{darkgreen},
  commentstyle=\color{gray},
  showstringspaces=false,
  breaklines=true]
from springable.simulation import simulate_model

simulate_model('fig4d_model.csv')
\end{lstlisting}
\end{tcolorbox}

\subsection{Figure 5}
\subsubsection{Tensegrity (Fig.~5a)}
\begin{tcolorbox}[title=Model CSV file of Fig.~5a (no prestress): \texttt{fig5atop\_model.csv}]
\begin{lstlisting}
PARAMETERS
k, 0.5
a, 1.0
b, 1.0
c, 1.5
prestress, 1e-5
cable_behavior, 'tensegrity_cable_behavior.csv'
NODES
0, 0.0, 0.0, 1, 1
1, X0+COS(ARCSIN(c/2/a))*a, Y0+c/2, 0, 0
2, X1+b, Y1, 0, 0
3, X2+COS(ARCSIN(c/2/a))*a, Y0, 1, 1
4, X2, -Y2, 0, 0
5, X1, -Y1, 0, 0
LONGITUDINAL FLEXELS
# cables
0-1, FROMFILE(cable_behavior), a * (1-prestress)
1-2, FROMFILE(cable_behavior), b * (1-prestress)
2-3, FROMFILE(cable_behavior), a * (1-prestress)
3-4, FROMFILE(cable_behavior), a * (1-prestress)
4-5, FROMFILE(cable_behavior), b * (1-prestress)
5-0, FROMFILE(cable_behavior), a * (1-prestress)
# struts
1-5, LINEAR(k=k), c * (1+prestress)
2-4, LINEAR(k=k), c * (1+prestress)
LOADING
5, X, +0.1, -0.001
\end{lstlisting}
\begin{center}
\includegraphics[]{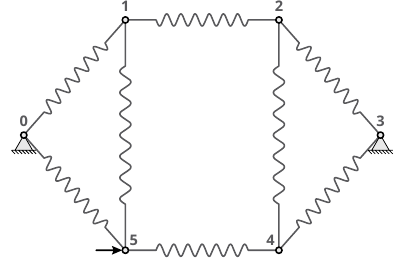}
\end{center}
\end{tcolorbox}
\begin{tcolorbox}[title=Model CSV file of Fig.~5a (with prestress): \texttt{fig5abottom\_model.csv}]
\begin{lstlisting}
PARAMETERS
k, 0.5
a, 1.0
b, 1.0
c, 1.5
prestress, 0.3
cable_behavior, 'tensegrity_cable_behavior.csv'
NODES
0, 0.0, 0.0, 1, 1
1, X0+COS(ARCSIN(c/2/a))*a, Y0+c/2, 0, 0
2, X1+b, Y1, 0, 0
3, X2+COS(ARCSIN(c/2/a))*a, Y0, 1, 1
4, X2, -Y2, 0, 0
5, X1, -Y1, 0, 0
LONGITUDINAL FLEXELS
# cables
0-1, FROMFILE(cable_behavior), a * (1-prestress)
1-2, FROMFILE(cable_behavior), b * (1-prestress)
2-3, FROMFILE(cable_behavior), a * (1-prestress)
3-4, FROMFILE(cable_behavior), a * (1-prestress)
4-5, FROMFILE(cable_behavior), b * (1-prestress)
5-0, FROMFILE(cable_behavior), a * (1-prestress)
# struts
1-5, LINEAR(k=k), c * (1+prestress)
2-4, LINEAR(k=k), c * (1+prestress)
LOADING
5, X, +0.1, -0.001
\end{lstlisting}
\begin{center}
\includegraphics[]{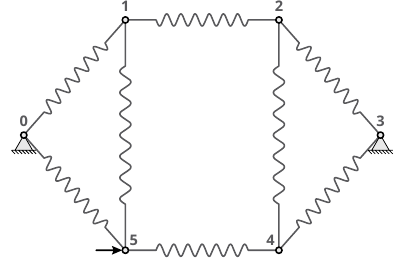}
\end{center}
\end{tcolorbox}
\begin{tcolorbox}[title=Nonlinear behavior used in model of Fig.~5a: \texttt{tensegrity\_cable\_behavior.csv}]
\begin{lstlisting}[basicstyle=\scriptsize\ttfamily]
PIECEWISE(k_i=[1e-6; 1.0];u_i=[1e-3]; us=0.99e-3;mode=1)
\end{lstlisting}
\end{tcolorbox}

\subsubsection{Tape spring (Fig.~5b)}
\begin{tcolorbox}[title=Model CSV file of Fig.~5b: \texttt{fig5b\_model.csv}]
\begin{lstlisting}
NODES
0, 0.0, 0.0, 1, 1
1, 1.0, 0.0, 0, 0
2, 2.0, 0.0, 1, 1
3, -2.0, 0.0, 0, 1
PATH FLEXELS
3-0-1-2, LINEAR(k=1000.0)
ANGULAR FLEXELS
2-1-0, FROMFILE('tape_behavior.csv'), PI
LOADING
1, Y, -1e-3
then
3, X, +50.0, +0.75
then
block
3, X
1, X, 0.5, 1.0
\end{lstlisting}
\begin{center}
\includegraphics[width=0.6\textwidth]{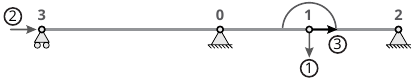}
\end{center}
\end{tcolorbox}
\begin{tcolorbox}[title=Nonlinear behavior used in model of Fig.~5b: \texttt{tape\_behavior.csv}]
\begin{lstlisting}[basicstyle=\scriptsize\ttfamily]
ZIGZAG2(u_i=[1.553E-01; 3.548E-01; 5.613E-01; 3.419E-01; 1.418E-01; 1.524E+00]; f_i=[1.227E+00; 1.448E+00; 1.292E+00; 1.123E+00; 1.234E-01; 1.893E-01]; epsilon=9.000E-01)
\end{lstlisting}
\end{tcolorbox}

\subsubsection{Bumping buckled beams (Fig.~5c)}
\begin{tcolorbox}[title={Model CSV file of Fig.~5c $(d=0.25)$: \texttt{fig5cright\_model.csv}}]
\begin{minipage}{0.55\textwidth}
\begin{lstlisting}
PARAMETERS
kr0, 1.0
kr1, 27.0
d, 0.25
h, 1.0
eps, 1e-4
NODES
0, 0.0, h, 1, 0
1, X0+d, Y0, 1, 0
2, X0+eps, Y0 - h/2, 0, 0
3, X1-eps, Y1 - h/2, 0, 0
4, X0, Y0 - h, 1, 1
5, X1, Y1 - h, 1, 1
6, (X0+X1)/2, (Y0+Y1)/2+0.1, 1, 0
LONGITUDINAL FLEXELS
0-2, LINEAR(k=1e3)
2-4, LINEAR(k=1e3)
1-3, LINEAR(k=3e3)
3-5, LINEAR(k=3e3)
ANGULAR FLEXELS
0-2-4, LINEAR(k=kr0)
1-3-5, LINEAR(k=kr1)
X DISTANCE FLEXELS
3-2, CONTACT(f0=5.0;uc=0.001;delta=0.0)
Y DISTANCE FLEXELS
6-0, LINEAR(k=1e5)
6-1, LINEAR(k=1e5)
LOADING
6, Y, -500.0, -h/3.0


\end{lstlisting}
\end{minipage}
\begin{minipage}{0.4\textwidth}
\begin{center}
\includegraphics[width=0.5\textwidth]{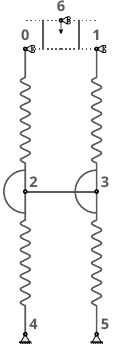}
\end{center}
\end{minipage}
\end{tcolorbox}
\begin{tcolorbox}[title={Model CSV file of Fig.~5c $(d=0.30)$: \texttt{fig5cleft\_model.csv}}]
\begin{minipage}{0.55\textwidth}
\begin{lstlisting}
PARAMETERS
kr0, 1.0
kr1, 27.0
d, 0.3
h, 1.0
eps, 1e-4
NODES
0, 0.0, h, 1, 0
1, X0+d, Y0, 1, 0
2, X0+eps, Y0 - h/2, 0, 0
3, X1-eps, Y1 - h/2, 0, 0
4, X0, Y0 - h, 1, 1
5, X1, Y1 - h, 1, 1
6, (X0+X1)/2, (Y0+Y1)/2+0.1, 1, 0
LONGITUDINAL FLEXELS
0-2, LINEAR(k=1e3)
2-4, LINEAR(k=1e3)
1-3, LINEAR(k=3e3)
3-5, LINEAR(k=3e3)
ANGULAR FLEXELS
0-2-4, LINEAR(k=kr0)
1-3-5, LINEAR(k=kr1)
X DISTANCE FLEXELS
3-2, CONTACT(f0=5.0;uc=0.001;delta=0.0)
Y DISTANCE FLEXELS
6-0, LINEAR(k=1e5)
6-1, LINEAR(k=1e5)
LOADING
6, Y, -500.0, -h/3.0

\end{lstlisting}
\end{minipage}
\begin{minipage}{0.4\textwidth}
\begin{center}
\includegraphics[width=0.5\textwidth]{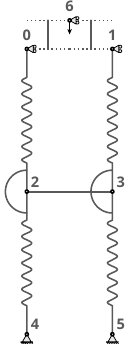}
\end{center}
\end{minipage}
\end{tcolorbox}

\newpage
\subsubsection{Metafluidic gripper (Fig.~5d)}
\begin{tcolorbox}[title={Model CSV file of Fig.~5d: \texttt{fig5d\_model.csv}},enhanced, breakable,before lower={%
    \ifnumequal{\tcbbreakpart}{\tcbtotalparts}{}{%
      \par\smallskip\hfill\textit{\tiny continued on next page}%
    }
  }]
\begin{minipage}[t]{0.48\textwidth}
\begin{lstlisting}[basicstyle=\scriptsize\ttfamily, firstline=1, lastline=44]
PARAMETERS
dx, 0.1
k1, 40
k0, 0.6
kv, 2.8
kr, 0.002
delta, 0.1
NODES
0 , 0*dx, 3*dx, 1, 1
1 , 1*dx, 3*dx, 0, 0
2 , 2*dx, 3*dx, 0, 0
3 , 3*dx, 3*dx, 0, 0
4 , 4*dx, 3*dx, 0, 0
5 , 4*dx, 2*dx, 0, 0
6 , 3*dx, 2*dx, 0, 0
7 , 2*dx, 2*dx, 0, 0
8 , 1*dx, 2*dx, 0, 0
9 , 0*dx, 2*dx, 1, 1
10, 0*dx, -1*dx, 1, 1
11, 1*dx, -1*dx, 0, 0
12, 2*dx, -1*dx, 0, 0
13, 3*dx, -1*dx, 0, 0
14, 4*dx, -1*dx, 0, 0
15, 4*dx, -2*dx, 0, 0
16, 3*dx, -2*dx, 0, 0
17, 2*dx, -2*dx, 0, 0
18, 1*dx, -2*dx, 0, 0
19, 0*dx, -2*dx, 1, 1
20, -4*dx, -2*dx, 0, 1
21, -4*dx, 3*dx, 0, 1
22, -4.5*dx, 0.5*dx, 0, 1
LONGITUDINAL FLEXELS
0-1, LINEAR(k=k0)
1-2, LINEAR(k=k0)
2-3, LINEAR(k=k0)
3-4, LINEAR(k=k0)
5-6, LINEAR(k=k1)
6-7, LINEAR(k=k1)
7-8, LINEAR(k=k1)
8-9, LINEAR(k=k1)
1-8, LINEAR(k=kv)
2-7, LINEAR(k=kv)
3-6, LINEAR(k=kv)
4-5, LINEAR(k=kv)
\end{lstlisting}
\emph{(to be continued on the rightside)}
\end{minipage}
\hfill
\vrule
\hfill
\begin{minipage}[t]{0.48\textwidth}
\emph{(continued)}
\begin{lstlisting}[basicstyle=\scriptsize\ttfamily, firstline=45]
10-11, LINEAR(k=k1)
11-12, LINEAR(k=k1)
12-13, LINEAR(k=k1)
13-14, LINEAR(k=k1)
15-16, LINEAR(k=k0)
16-17, LINEAR(k=k0)
17-18, LINEAR(k=k0)
18-19, LINEAR(k=k0)
11-18, LINEAR(k=kv)
12-17, LINEAR(k=kv)
13-16, LINEAR(k=kv)
14-15, LINEAR(k=kv)
ANGULAR FLEXELS
0-1-2, LINEAR(k=kr)
1-2-3, LINEAR(k=kr)
2-3-4, LINEAR(k=kr)
3-4-5, LINEAR(k=kr)
4-5-6, LINEAR(k=kr)
5-6-7, LINEAR(k=kr)
6-7-8, LINEAR(k=kr)
7-8-9, LINEAR(k=kr)
8-9-0, LINEAR(k=kr)
9-0-1, LINEAR(k=kr)
10-11-12, LINEAR(k=kr)
11-12-13, LINEAR(k=kr)
12-13-14, LINEAR(k=kr)
13-14-15, LINEAR(k=kr)
14-15-16, LINEAR(k=kr)
15-16-17, LINEAR(k=kr)
16-17-18, LINEAR(k=kr)
17-18-19, LINEAR(k=kr)
18-19-10, LINEAR(k=kr)
19-10-11, LINEAR(k=kr)
X DISTANCE FLEXELS
21-22, LINEAR(k=100.0)
20-22, LINEAR(k=100.0)
Y DISTANCE FLEXELS
5-14, CONTACT(f0=5.0; uc=0.0025; delta=delta)
AREA FLEXELS
0-1-2-3-4-5-6-7-8-9-10-11-12-13-14-15-16-17-18-19-20-21, FROMFILE('metafluid_behavior.csv')
LOADING
22, X, 1.0, 3.95*dx

\end{lstlisting}
\end{minipage}
\begin{center}
\includegraphics[width=0.5\textwidth]{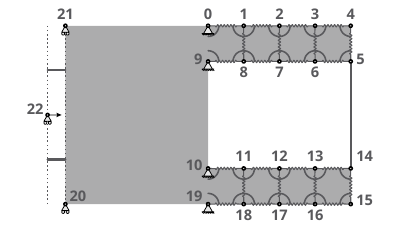}
\end{center}
\end{tcolorbox}

\begin{tcolorbox}[title=Nonlinear behavior used in model of Fig.~5d: \texttt{metafluid\_behavior.csv}]
\begin{lstlisting}[basicstyle=\scriptsize\ttfamily]
ZIGZAG2(u_i=[6.000E-02; 5.000E-03; 7.500E-02; 1.000E-02; 9.000E-02; 1.500E-02; 1.050E-01; 2.000E-02; 1.200E-01; 2.500E-02; 1.350E-01; 3.000E-02; 1.500E-01]; f_i=[1.000E+00; 0.000E+00; 1.010E+00; 1.000E-02; 1.020E+00; 2.000E-02; 1.030E+00; 3.000E-02; 1.040E+00; 4.000E-02; 1.050E+00; 5.000E-02; 1.060E+00]; epsilon=5.000E-01)
\end{lstlisting}
\end{tcolorbox}

\subsubsection{Simulation script}
To simulate the models related to Fig.~5, the following script can be run. The CSV behavior files must live in the same (working) directory as the CSV model files.
\begin{tcolorbox}
\begin{lstlisting}[language=Python,
  basicstyle=\ttfamily\small,
  keywordstyle=\color{blue},
  stringstyle=\color{darkgreen},
  commentstyle=\color{gray},
  showstringspaces=false,
  breaklines=true]
from springable.simulation import simulate_model

settings = {'radius': 0.005, 'convergence_value': 1e-8, 'detect_mechanism': False}

simulate_model('fig5abottom_model.csv', solver_settings=solver_settings)
simulate_model('fig5atop_model.csv', solver_settings=settings)
simulate_model('fig5b_model.csv', solver_settings=settings)
simulate_model('fig5cleft_model.csv', solver_settings=settings)
simulate_model('fig5cright_model.csv', solver_settings=settings)
simulate_model('fig5d_model.csv', solver_settings=settings)
\end{lstlisting}
\end{tcolorbox}